\documentclass[a4paper,11pt]{article}
\pdfoutput=1 

\usepackage{jheppub} 

\usepackage[T1]{fontenc} 

\usepackage{latexsym, bbm, color,appendix,multirow}

\newcommand{\mpcac}{m_{\mbox{\tiny{PCAC}}}}
\newcommand{\SU}{\mathrm{SU}}
\newcommand{\U}{\mathrm{U}}
\newcommand{\re}{{\rm{Re}}}

\newcommand{\dd}{{\rm{d}}}
\newcommand{\Tr}{{\rm Tr\,}}

\newcommand{\nconf}{n_{\mbox{\tiny{conf}}}}

\newcommand{\be}{\begin{equation}}
\newcommand{\eq}{\begin{equation}}
\newcommand{\en}{\end{equation}}
\newcommand{\ee}{\end{equation}}
\newcommand{\eqar}{\begin{eqnarray}}
\newcommand{\bea}{\begin{eqnarray}}
\newcommand{\enar}{\end{eqnarray}}
\newcommand{\eea}{\end{eqnarray}}

\title{\boldmath Mesons in large-$N$ QCD}

\author[a,b]{Gunnar~Bali,}
\author[c]{Francis~Bursa,}
\author[a]{Luca~Castagnini,}
\author[a]{Sara~Collins,}
\author[d]{Luigi~Del~Debbio,}
\author[e]{Biagio~Lucini}
\author[f]{and Marco~Panero}

\affiliation[a]{Institute for Theoretical Physics,  University of Regensburg,\\D-93040 Regensburg, Germany}
\affiliation[b]{Tata Institute of Fundamental Research,\\Homi Bhabha Road, Mumbai 400005, India}
\affiliation[c]{SUPA, School of Physics and Astronomy, University of Glasgow,\\Glasgow, G12 8QQ, UK}
\affiliation[d]{SUPA and The Tait Institute, School of Physics and Astronomy, University of Edinburgh,\\Edinburgh EH9 3JZ, UK}
\affiliation[e]{College of Science, Swansea University,\\Singleton Park, Swansea SA2 8PP, UK}
\affiliation[f]{Department of Physics and Helsinki Institute of Physics, University of Helsinki,\\FIN-00560 Helsinki, Finland}

\emailAdd{gunnar.bali@ur.de}
\emailAdd{francis.bursa@glasgow.ac.uk}
\emailAdd{luca.castagnini@physik.uni-regensburg.de}
\emailAdd{sara.collins@physik.uni-regensburg.de}
\emailAdd{luigi.del.debbio@ed.ac.uk}
\emailAdd{b.lucini@swansea.ac.uk}
\emailAdd{marco.panero@helsinki.fi}

\abstract{We present the results of a systematic, first-principles study of the spectrum and decay constants of mesons for different numbers of color charges $N$, via lattice computations. We restrict our attention to states in the non-zero isospin sector, evaluating the masses associated with the ground-state and first excitation in the pseudoscalar, vector, scalar, and axial vector channels. Our results are based on a new set of simulations of four dimensional $\SU(N)$ Yang-Mills theories with the number of colors ranging from $N=2$ to $N=17$; the spectra and the decay constants are computed in the quenched approximation (which becomes exact in the 't~Hooft limit) using Wilson fermions. After discussing the extrapolations to the chiral and large-$N$ limits, we present a comparison of our results to some of the numerical computations and analytical predictions available in the literature---including, in particular, those from holographic computations.}

\arxivnumber{1304.4437}

\begin{document}
\maketitle
\flushbottom

\section{Introduction} 
\label{sec:introduction} 

The Standard Model of elementary particles describes the strong interaction by Quantum Chromodynamics (QCD): a non-Abelian gauge theory characterized by local $\SU(3)$ color invariance, and with flavors of fermionic matter fields (quarks) in the fundamental representation of the gauge group. Due to asymptotic freedom, the predictions of QCD for processes at high energies can reliably be worked out through perturbative calculations. At low energies, however, the running coupling becomes large; in this regime many features of QCD are determined by non-perturbative phenomena: confinement and chiral symmetry breaking. For this reason, the theoretical derivation of low energy QCD properties must rely on non-perturbative methods.

The standard non-perturbative definition of QCD is based on lattice regularization~\cite{Wilson:1974sk}, which makes the theory mathematically well-defined and amenable to analytical as well as to numerical studies. Thanks to  theoretical, algorithmic and computer-power progress, during the last decade many large-scale dynamical lattice QCD computations have been performed at realistic values of the physical parameters, allowing one to numerically obtain predictions in energy regimes otherwise inaccessible to a first-principles approach, see, e.g., ref.~\cite{Kronfeld:2012uk} for a recent review.

A different non-perturbative approach to QCD is based on an expansion in powers of $1/N$, i.e. in powers of the inverse of the number of color charges~\cite{largeN}. When the number of colors $N$ is taken to infinity, and the coupling $g$ is sent to zero, keeping the product $g^2 N$ fixed, the theory reveals striking mathematical simplifications---see refs.~\cite{reviews,reviews-end} for recent reviews. One may study the physical $N=3$ case expanding around the $1/N \to 0$ limit. In particular, in the so-called 't~Hooft limit (in which the number of flavors $n_f$ is held fixed), one finds that the amplitudes for physical processes are determined by a particular subset of Feynman diagrams (planar diagrams), the low-energy spectrum consists of stable meson and glueball states, and the scattering matrix becomes trivial. Examples of other interesting implications of the large-$N$ limit include those discussed, e.g., in refs.~\cite{analytical_results_at_large_N,analytical_results_at_large_N_1,analytical_results_at_large_N_2,analytical_results_at_large_N_3,analytical_results_at_large_N_4}.

Another non-perturbative approach to low-energy properties of strongly coupled non-Abelian gauge theories is based on the conjectured correspondence between gauge and string theories~\cite{Maldacena:1997re,Gubser:1998bc,Witten:1998qj}. During the last decade, many studies have employed techniques based on this correspondence, to build models which reproduce (at least qualitatively or semi-quantitatively) the main features of the mesonic spectrum of QCD~\cite{Erdmenger:2007cm}. Remarkably, the large-$N$ limit plays a technically crucial role also in the context of these holographic computations: the correspondence relates the strongly coupled regime of a gauge theory with an infinite number of colors to the classical gravity limit of a dual string model in an anti-de~Sitter spacetime, a setup that can be studied with analytical or semi-analytical methods.

In order to understand whether predictions derived from approaches relying on the large-$N$ limit can be relevant also for the physical case of QCD with $N=3$ colors, it is crucial to estimate the quantitative impact of finite-$N$ corrections. For this purpose, recently several lattice studies investigated the dependence on $N$ of various quantities of interest---including string tensions for sources in different representations, the glueball spectrum and its dependence on the $\theta$-term, the topological susceptibility or the finite-temperature equation of state---in different $\SU(N)$ Yang-Mills theories~\cite{Lucini:2001ej,Lucini:2003zr,Lucini:2004yh,Lucini:2004my,Lucini:2008vi,DelDebbio:2001sj,DelDebbio:2002xa,Del Debbio:2004rw,Del Debbio:2006df,Bursa:2005yv,largeN4D_1,largeN4D_2,Bringoltz:2005rr,Panero:2009tv,Panero:2008mg,Datta:2010sq,Mykkanen:2012ri,largeN4D_3,Bonati:2013tt, Lucini:2012wq}. These works revealed precocious scaling to the large-$N$ limit: up to modest $\mathcal{O}(N^{-2})$ corrections even the $\SU(3)$ and $\SU(2)$ theories appear to be ``close'' to the large-$N$ limit. Similar results have also been found in $D=2+1$ dimensions~\cite{Teper:1998te,Liddle:2008kk,largeN3D_1,largeN3D_2,Bursa:2005tk,largeN3D_2a,largeN3D_3,largeN3D_4,largeN3D_5}.

The purpose of the present paper is to further expand this line of research, by studying the light mesons in gauge theories based on different $\SU(N)$ gauge groups to explore the connections between the different non-perturbative approaches outlined above. More precisely, we extract the meson masses from numerical simulations and investigate their dependence on the number of colors. We discuss the approach to the large-$N$ limit and compare our results to the theoretical expectations from large-$N$ expansions and holographic computations. 

All the results that we present in this paper are obtained from quenched simulations. A technical remark about this is in order: in theories including dynamical quarks, virtual fermion loops are suppressed by powers of $n_f/N$, so, in general, corrections with respect to the large-$N$ limit are expected to be larger than for the Yang-Mills case. However, when $N$ is infinite, the theory exactly reduces to a (unitary) quenched model: this justifies studying the large-$N$ meson spectrum neglecting dynamical fermions and, in fact, it also provides an intuitive explanation of why the quenched approximation performs rather well even for the $N=3$ case~\cite{Aoki:1999yr}. From the practical point of view, lattice computations of meson masses in the large-$N$ limit in the quenched setup offer two major advantages:
\begin{enumerate}
\item neglecting the contribution from virtual fermion loops allows one to bypass the major computational overhead associated with the inclusion of the Dirac operator determinant in the system dynamics;
\item the convergence to the large-$N$ limit is faster since the leading corrections due to the finiteness of the number of colors scale quadratically (rather than linearly) in $1/N$, enabling reliable extrapolations, even from simulations at rather small values of $N$.
\end{enumerate}

In the past, related studies have been reported in refs.~\cite{DelDebbio:2007wk, Bali:2008an,Bali:2007kt, DeGrand:2012hd, Hietanen:2009tu}: one of the goals of our present work consists of improving and extending these results, by going to lighter quark masses, larger $N$ and larger volumes, and by increasing the statistics, the number of interpolators included in the variational basis and of states that we investigate.
We also aim at clarifying a discrepancy between the results of refs.~\cite{DelDebbio:2007wk, Bali:2008an,Bali:2007kt,DeGrand:2012hd}, which at large $N$ found a value of the vector meson mass close to the one of real-world QCD, and those obtained in ref.~\cite{Hietanen:2009tu}, which, on the contrary, reported a value approximately twice as large. 

In section~\ref{sec:latticesetup}, we define the setup of our lattice computation and, in section~\ref{sec:numericalresults}, we present our numerical results. In section \ref{sec:rho} we discuss discrepancies with ref.~\cite{Hietanen:2009tu}. In section~\ref{sec:postdictions} we compare our results to  analytical predictions and finally, in section~\ref{sec:conclusions}, we summarize our findings and conclude.

Preliminary results of this work have been presented in ref.~\cite{Bali:2013fpo}.

\section{Setup of the lattice computation}
\label{sec:latticesetup}

In this work, we non-perturbatively study theories with $\SU(N)$ internal color symmetry with $N=2$, $3$, $4$, $5$, $6$, $7$ and $17$ color charges, regularized on a finite, isotropic hypercubic lattice $\Lambda$ of spacing $a$, in Euclidean spacetime. In the following, we denote the lattice hypervolume as $L_s^3 \times L_t= (N_s^3  \times N_t )a^4$. Expectation values of observables are obtained as statistical averages over quenched configuration ensembles, using the Wilson discretization of the generating functional of the continuum theory detailed below:
\eq
\label{lattice_partition_function}
Z = \int \prod_{x \in \Lambda} \prod_{\mu=1}^4 {\dd}U_\mu(x) e^{-S},
\en
where ${\dd}U_\mu(x)$ denotes the Haar measure for each $U_\mu(x) \in \SU(N)$ link matrix, $\mu=1,\dots,4$ and $S$ is the ``plaquette'' discretization of the Yang-Mills action:
\eq
\label{Wilson_lattice_gauge_action}
S = \beta \sum_{x \in \Lambda} \sum_{ \mu < \nu } \left[1 - \frac{1}{N} \re\,\Tr U_{\mu\nu}(x)\right].
\en
$\beta$ is related to the bare lattice gauge coupling $g_0$ via $\beta=2N/g_0^2$ and 
\eq
\label{plaquette}
U_{\mu\nu}(x) = U_\mu(x) U_\nu(x+a\hat\mu) U^\dagger_\mu(x+a\hat\nu) U^\dagger_\nu(x).
\en
The expectation values of gauge-invariant physical observables $O$ are defined as:
\eq
\label{expectation_value}
\langle O \rangle = \frac{1}{Z} \int \prod_{x \in \Lambda} \prod_{\mu=1}^4 {\dd}U_\mu(x) \; O \; e^{-S}
\en
and are numerically estimated via Monte Carlo sampling. The numerical results presented in this work are obtained from sets of configurations generated by code based on the Chroma suite~\cite{Edwards:2004sx}, using standard local updates~\cite{Creutz:1980zw,Fabricius:1984wp,Kennedy:1985nu,Adler:1981sn,Brown:1987rra,Cabibbo:1982zn}, which we have adapted to work for a generic $N$. In the following, we denote the number of configurations used in our computations (for each set of parameters) as $\nconf$.

Our lattice implementation of quark propagators is based on the Wilson discretization of the continuum Dirac operator~\cite{Wilson:1974sk}:
\begin{equation}
a\mathcal{M}(x,y)[U]=\delta_{x,y} - \kappa \sum_{\mu=1}^4 \left[ (1-\gamma_\mu) U_\mu (x)\delta_{x+a\hat\mu,y} + (1+\gamma_\mu) U_\mu^\dagger (y) \delta_{x-a\hat\mu,y} \right] \,,
\end{equation}
where the hopping parameter $\kappa$ is related to the bare quark mass $m_q$ via:
\begin{equation}
\label{bare mass}
am_q=\frac{1}{2}\left(\frac{1}{\kappa}-\frac{1}{\kappa_c}\right)
\end{equation}
and $\kappa_c$ denotes the critical value, corresponding to a massless quark. The additive constant is given by $\kappa_c^{-1}=8+{\mathcal O}(\beta^{-1})$ and its
 non-perturbative determination is discussed in subsection \ref{subsec:pcacMass}.

To set the scale, we use the string tension calculations of ref.~\cite{Lucini:2005vg,Allton:2008ty} for $N=2$, $3$, $4$ and $6$, and those of ref.~\cite{Lucini:2012wq} for $N=5$ and $7$. In each case we choose the coupling
\begin{equation}
\beta=\frac{2N}{g^2}=\frac{2N^2}{\lambda},
\end{equation}
such that the (square root of the) string tension in lattice units
$a\sqrt\sigma \simeq 0.2093$ is the same for each $N$. 

Using the \emph{ad hoc} value $\sigma=1$~GeV/fm, 
our lattice spacing corresponds to $a\approx 0.093$~fm
or $a^{-1}\approx 2.1$~GeV.
Strictly speaking, we can only predict ratios of dimensionful quantities.
In the real world where experiments are performed,
$n_f>0$, $N=3\neq\infty$ and even the string tension is
not well defined. This means that any absolute scale
setting in physical units
will be arbitrary and is just meant as a rough guide.

\begin{figure}
\begin{center}
\includegraphics[width=0.68\textwidth]{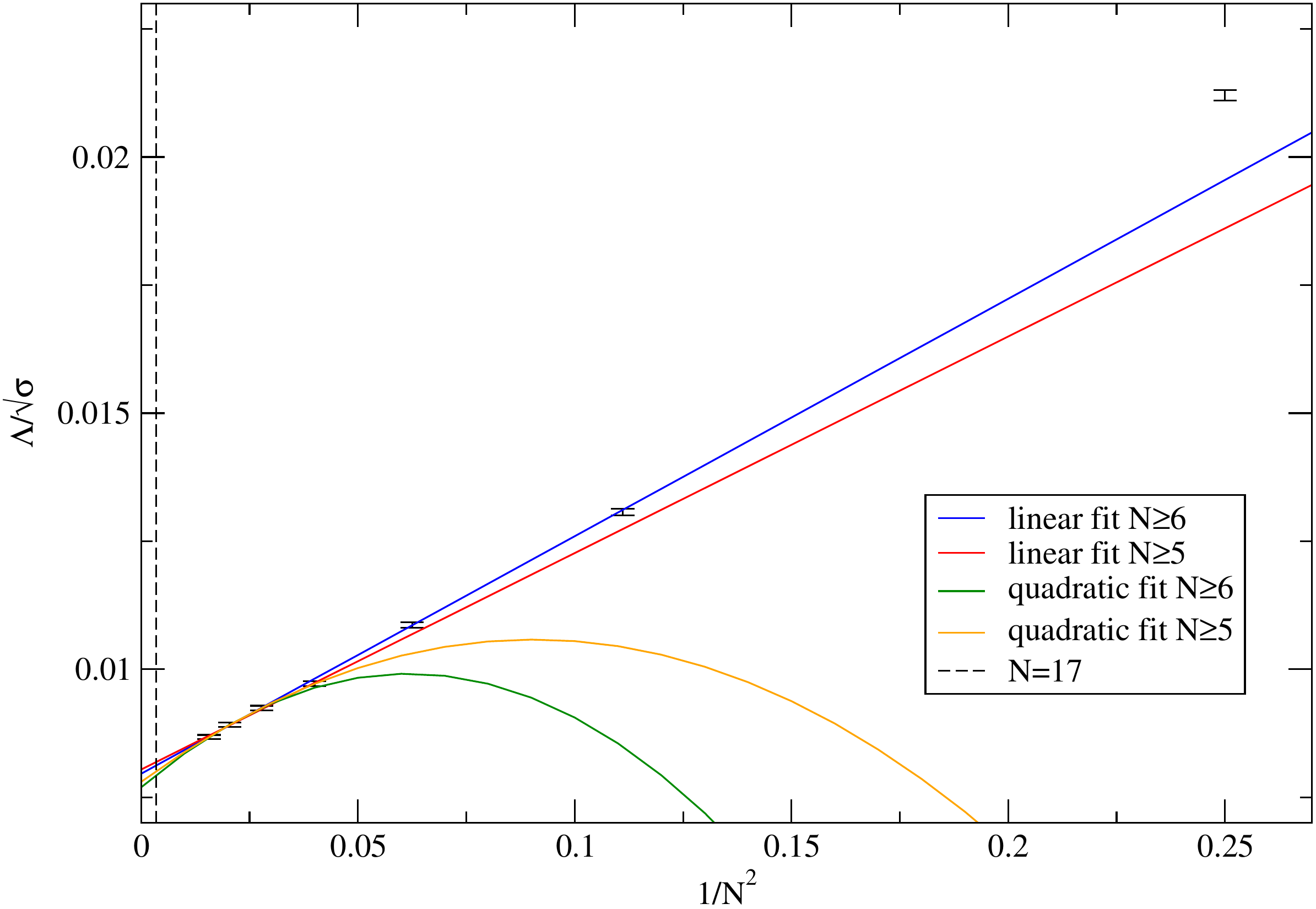}
\end{center}
\caption{$\Lambda$-parameter estimates of eq.~(\ref{lambdalattice}), in units of
the square root of the string tension. The errors shown are propagated
from those of $a\sqrt{\sigma}$.}
\label{fig:lambdaLat}
\end{figure}

For the theory with $\SU(17)$ gauge group, there are no string tension calculations available, so we extracted a $\beta$-value  from a fit of the QCD $\Lambda$-parameter in the lattice scheme:
\begin{equation}
\label{lambdalattice}
\Lambda \approx a^{-1}   \exp\left[-\frac{1}{2 b_0
\lambda(a^{-1} )}\right] \cdot \left[ b_0 \lambda(a^{-1} ) \right]^{-\frac{b_1}{2 b^2_0}} \cdot \left[1 + \frac{1}{2 b_0^3}\left( b_1 ^2 - b_2^L \beta_0 \right) \lambda(a^{-1} ) + \dots\right] ,
\end{equation}
with \cite{Bode:2001uz,Luscher:1995np}
\begin{equation}
b_0 = \frac{11}{3 \left(4 \pi \right)^2} ,\quad b_1 = \frac{34}{3  \left(4 \pi \right)^4 }, \quad b_2^L =\frac{1}{ \left(4 \pi \right)^6}  \left( -366.2 + \frac{1433.8}{N^2} - \frac{2143}{N^4} \right). 
\end{equation}
$\Lambda/\sqrt{\sigma}$ was calculated from the data presented in~\cite{Lucini:2005vg,Allton:2008ty,Lucini:2012wq} for $\SU(2 \le N  \le 8)$ and is shown in figure~\ref{fig:lambdaLat}. Using the data for $N=6$, $7$ and $8$ and a linear fit in $1/{N^2}$, we obtained $\beta=208.45$ for $N=17$. Adding further values of $N\ge4$ or using a quadratic fit in $1/N^2$ changed this value by less than 0.3\% (see table~\ref{tab:betaFits}).

\begin{table}
\begin{center}

\begin{tabular}{|c|c|c|}
\hline
Fit & 3 points & 4 points \\ \hline
linear & 208.45 & 208.16 \\
quadratic & 209.04 & 208.77 \\ \hline
\end{tabular}
\end{center}
\caption{Fit results for $\beta$ at $N=17$.  }
\label{tab:betaFits}
\end{table}

\begin{table}
\begin{center}
\begin{tabular}{|c|c|c|c|c|c|}
\hline
$N$  & $N_s^3 \times N_t$ & $\beta$ & $\lambda$&$10^5\kappa$ & $n_{\mathrm{conf}}$ \\
\hline
$2$ & $16^3 \times 32$ & \multirow{3}{*}{2.4645}  &\multirow{3}{*}{3.246}& 14581,  14827,  15008,  15096 &  400 \\ 
 & $24^3 \times 48$ &  & &14581,  14827,  15008,  15096,  15195.9 ,15249.6   &  200 \\ 
  & $32^3 \times 64$ &  & & 14581,  14827,  15008,  15096,  15195.9 ,15249.6   &  100 \\ 
  \hline
  $3$ & $16^3 \times 32$ &\multirow{3}{*}{6.0175} &\multirow{3}{*}{2.991}&  15002,  15220, 15380, 15458   &  200\\ 
 & $24^3 \times 48$ &  &&  15002,  15220, 15380, 15458, 15563.8, 15613   & 200 \\ 
  & $32^3 \times 64$ &  &&   15002,  15220, 15380, 15458, 15563.8, 15613  &  100 \\ 
\hline
  $4$ & $16^3 \times 32$ &\multirow{2}{*}{11.028} &\multirow{2}{*}{2.902}& 15184, 15400, 15559, 15635   &  200\\ 
 & $24^3 \times 48$ &  &&   15184, 15400, 15559, 15635, 15717.3, 15764    &  200\\ 
 \hline
  $5$ & $16^3 \times 32$ &\multirow{2}{*}{17.535}&\multirow{2}{*}{2.851}& 15205, 15426, 15592, 15658   & 200 \\ 
 & $24^3 \times 48$ &  & &   15205, 15426, 15592, 15658, 15754.8, 15835.5  &  200\\ 
\hline
  $6$ & $16^3 \times 32$ &\multirow{2}{*}{25.452} &\multirow{2}{*}{2.829}&15264, 15479, 15636, 15712     & 200 \\ 
 & $24^3 \times 48$ &  &&  15264, 15479, 15636, 15712, 15805.1, 15884.5     &  200 \\ 
\hline
  $7$ & $16^3 \times 32$ &\multirow{2}{*}{34.8343}&\multirow{2}{*}{2.813}&  15281.6, 15496.7, 15654.7, 15733.9  &  200\\ 
 & $24^3 \times 48$ &  & & 15281.6, 15496.7, 15654.7, 15733.9, 15827.3, 15906.2    &  200 \\ 
\hline
  $17$ & $12^3 \times 24$ & $208.45$ &2.773&  15298, 15521, 15684, 15755,  15853.1,  15931 &  80\\ 
\hline
\end{tabular}
\end{center}
\caption{Parameters of the main set of lattice simulations used in this work,  
$\beta$ denotes the gauge action parameter, while $\kappa$ is the hopping parameter appearing in the quark propagator. 
All configurations were separated by 200
combined heatbath and overrelaxation Monte Carlo sweeps and found to
be effectively statistically independent. For orientation we also include the bare 't~Hooft parameter.}
\label{parameters} 
\end{table}

Table~\ref{parameters} summarizes the essential technical information of our computations. The $N\leq 7$
results presented in the fits and plots here are obtained from the $24^3 \times 48$  lattices,
corresponding to $L_s\approx 2.2$~fm. In order to study finite size effects (FSE), we also performed additional simulations, both using smaller
and (for $N=2$ and $3$) larger volumes---see the discussion in subsection~\ref{subsec:FSE}.
For $\SU(17)$ we employed a smaller $12^3\times 24$ volume. The corresponding extent
$L_s=12a\approx 2.512/\sqrt{\sigma}$ is well above the inverse critical
temperature~\cite{Lucini:2012wq} $T_c^{-1}\approx 1.681/\sqrt{\sigma}\lesssim L_c$. 

The $\kappa$-values were selected
to keep one set of six pion masses approximately constant across the
different $\mathrm{SU}(N)$ theories. To achieve this, we combined the results reported in ref.~\cite{Bali:2008an} for the groups studied therein with initial estimates for the groups that had not been studied before. We vary the ``pion'' mass
down to $m_\pi/\sqrt{\sigma}  \approx 0.5$ for groups with $N \geq 5$,
and to $m_\pi/\sqrt{\sigma}  \approx 0.75$ for $N<5$. We also simulated
a smaller quark mass for $\mathrm{SU}(N<5)$ but found significant numbers
of ``exceptional configurations"~\cite{Bardeen:1997gv} (up to 15~\% of
the total); we leave these data out of this work.
For $N = 5$, at the lowest quark mass, only two
exceptional configurations were encountered that we removed from the analysis.

\subsection{Smearing and operators used}

The meson spectrum can be extracted from zero-momentum correlators of interpolating operators of the form:
\begin{equation}
O_k (\mathbf{x},t) = \overline\psi(\mathbf{x},t) \Gamma_k \psi (\mathbf{x},t),
\end{equation}
where different choices of $\Gamma_k$ correspond to different physical states. We used degenerate quark masses for the quark fields $u$ and $d$, which allowed us to study the spectrum of the particles $\pi,\rho,a_0,a_1,b_1$ (see table~\ref{bilinear}). A generic meson correlator can then be computed as:
\begin{equation}
C_{\Gamma\Gamma}(t) =\sum_{\mathbf{x}}\left \langle  O (\mathbf{x},t)  \overline{O} (0) \right\rangle =  - \sum_{\mathbf{x}} \mathrm{Tr}\left \langle  \Gamma G(\mathbf{x},t) \Gamma \gamma_5 G(\mathbf{x},t) \gamma_5\right\rangle,
\end{equation}
where the propagator $G(\mathbf{x},t)$ is obtained by inverting the Dirac operator with the stabilized biconjugate gradient (BiCGStab) algorithm with even/odd preconditioning. The trace indicates a sum over spin and color. In the rare cases, when the BiCGStab routine failed to converge (which only occurred at the lowest quark masses), we reverted to the standard Conjugate Gradient (CG) algorithm.

In our correlators, we used both point-like and extended sources and sinks, employing several steps of Wuppertal smearing~\cite{Gusken:1989qx} which iteratively modifies a fermion field as:
\begin{equation}
\label{wuppsmear}
\psi^{n+1} = \frac{1}{1 + 6\, \omega}\left ( \psi^{n}  + \omega \sum_{j=\pm 1}^{\pm 3} U'_{j}(x) \psi^n(x + a\hat{j} ) \right ),
\end{equation}
where $n$ denotes the number of iterations, $\omega$ is the smearing parameter (we used $ \omega=0.25$) and $U'$ is the gauge field smeared by 10 iterations of the spatial APE smearing routine~\cite{Falcioni:1984ei}:
\begin{eqnarray}
\label{apesmear}
&& U'_i(x) = \mathrm{Proj}_{\SU(N)} \Big[ \alpha \, U_i(x) + \sum_{i \neq j}  U_{j}(x)U_{i}(x +a \hat{j})U^\dagger_j(x+ a\hat{i})  \nonumber \\
&& \qquad \qquad \qquad + U^\dagger_{j}(x - a\hat{i})U_{i}(x -a \hat{j})U_j(x+ a\hat{i} - a\hat{j}) \Big],
\end{eqnarray}
with smearing parameter $\alpha=2.5$.

\begin{table}
\begin{center}

\begin{tabular}{|c|c|c|c|c|c|}
\hline
Particle & $\pi$ & $\rho$ & $a_0$ & $a_1$ &$b_1$\\ \hline
Bilinear & $\bar{u} \gamma_5 d$  & $\bar{u} \gamma_i d$  & $\bar{u} d$ & $\bar{u}\gamma_5 \gamma_i d$  & $\frac {1}{2} \epsilon_{ijk} \bar{u} \gamma_i \gamma_j  d$ \\
\hline
$J^{PC}$ & $0^{-+}$& $1^{--}$& $0^{++}$ &  $1^{++}$&$1^{+-}$\\\hline
\end{tabular}
\end{center}
\caption{List of the studied channels and their bilinear operators used in the correlation functions.  }
\label{bilinear}
\end{table}

\subsection{Variational method}

We extracted the ground state and the first excited level using the variational analysis discussed in ref.~\cite{Michael:1985ne,Luscher:1990ck,Burch:2004he}. For each channel, we computed the cross-correlation matrix $C_{ij}(t) =\langle  O_i (t) \overline{O}_j (0)  \rangle$, where $i$ and $j$ correspond to the number of iterations ($0$, $20$, $80$ or $180$ steps) of Wuppertal smearing, for the sources and the sinks. 

Then we solved the generalized eigenvalue problem:
\begin{equation}
C(t)  	\mathbf{v}^\alpha = \lambda^\alpha(t) C(t_0) \mathbf{v}^\alpha
\end{equation}
and extracted the mass $m$ performing hyperbolic-cosine fits of the largest and second largest eigenvalues, $\lambda^0$ and $\lambda^1$, in $[t_\mathrm{min},t_\mathrm{max}]$ ranges ($t_\mathrm{max} \leq a N_t/2$):
\begin{equation}\label{eq:coshfit}
\lambda(t)  = A \left(e^{-m t} + e^{-m(a N_t - t)}\right).
\end{equation}
All statistical uncertainties were estimated using a jackknife procedure. With four different operators $O_i$ in many cases we were able to extract the first three states, however we regard the second excited states as unreliable at the present statistics. Using a subset of three operators out of the four mentioned above leads to compatible mass values within errors. 

Varying $t_0$ in the range $[0,2a]$ gives compatible results, so we used $t_0 =a$. To select the best fit ranges, for each particle we first studied the effective mass defined as:
\begin{equation}
a\,m_{\mathrm{eff}} =\mathrm{arccosh} \left[ \frac{ \lambda(t+a) +  \lambda(t-a)   }{2  \lambda(t) } \right].
\end{equation}
We determined $t_\mathrm{min}$ as the Euclidean time separation at which $m_{\mathrm{eff}}$ reaches a plateau (within statistical uncertainties), so that the contribution from higher states is negligible, while $t_\mathrm{max}$ is the value where $m_{\mathrm{eff}}$ becomes too noisy for stable fits. The signals become more precise at larger $N$ and lower $\kappa$-values. Typically we fit $\lambda(t)$ in the range $[5a,a N_t/2]$ for the ground states and in the range $[5 a ,10 a]$ for the first excited states, adjusting those ranges (by one or two lattice spacings) on a case-by-case basis. Fitting to eq.~(\ref{eq:coshfit}) with this procedure leads to reduced $\chi^2$-values which are well below one. 

The cost of inverting the propagator increases at lower quark masses, with the signal becoming noisier at the same time. For this reason we focused only on the ground states for the lowest two quark masses of each $\SU(N)$ group and instead of using the variational method we computed the two point functions using only $80$ steps of smearing for the sources/sinks. We then applied the same analysis for $\lambda(t)$ directly to the correlator $C(t)$.

\section{Numerical results}
\label{sec:numericalresults}

\subsection{The PCAC mass and the critical hopping parameter $\kappa_c$}
\label{subsec:pcacMass}

\begin{figure}
\begin{center}
\includegraphics[width=0.48\textwidth]{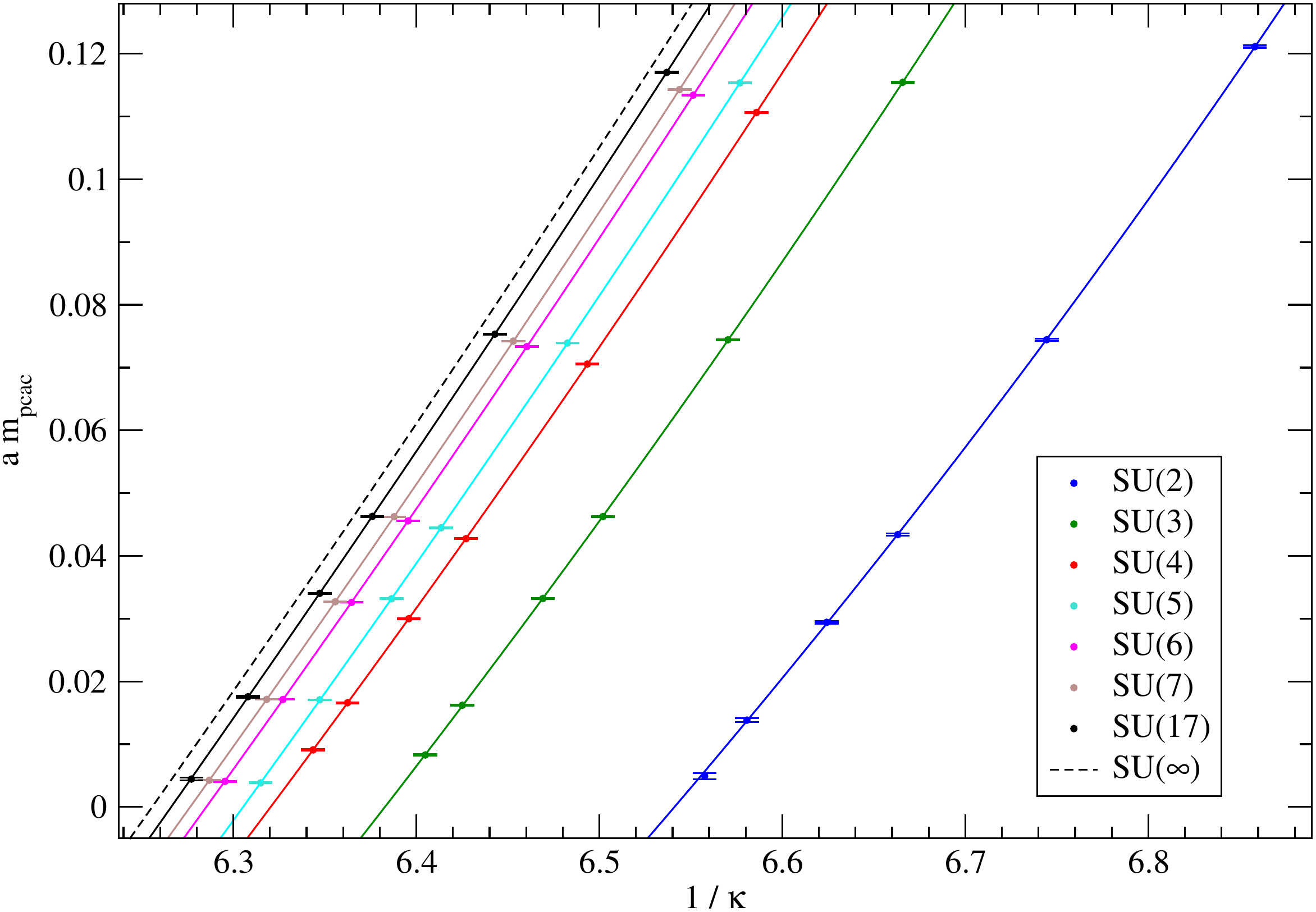} \includegraphics[width=0.48\textwidth]{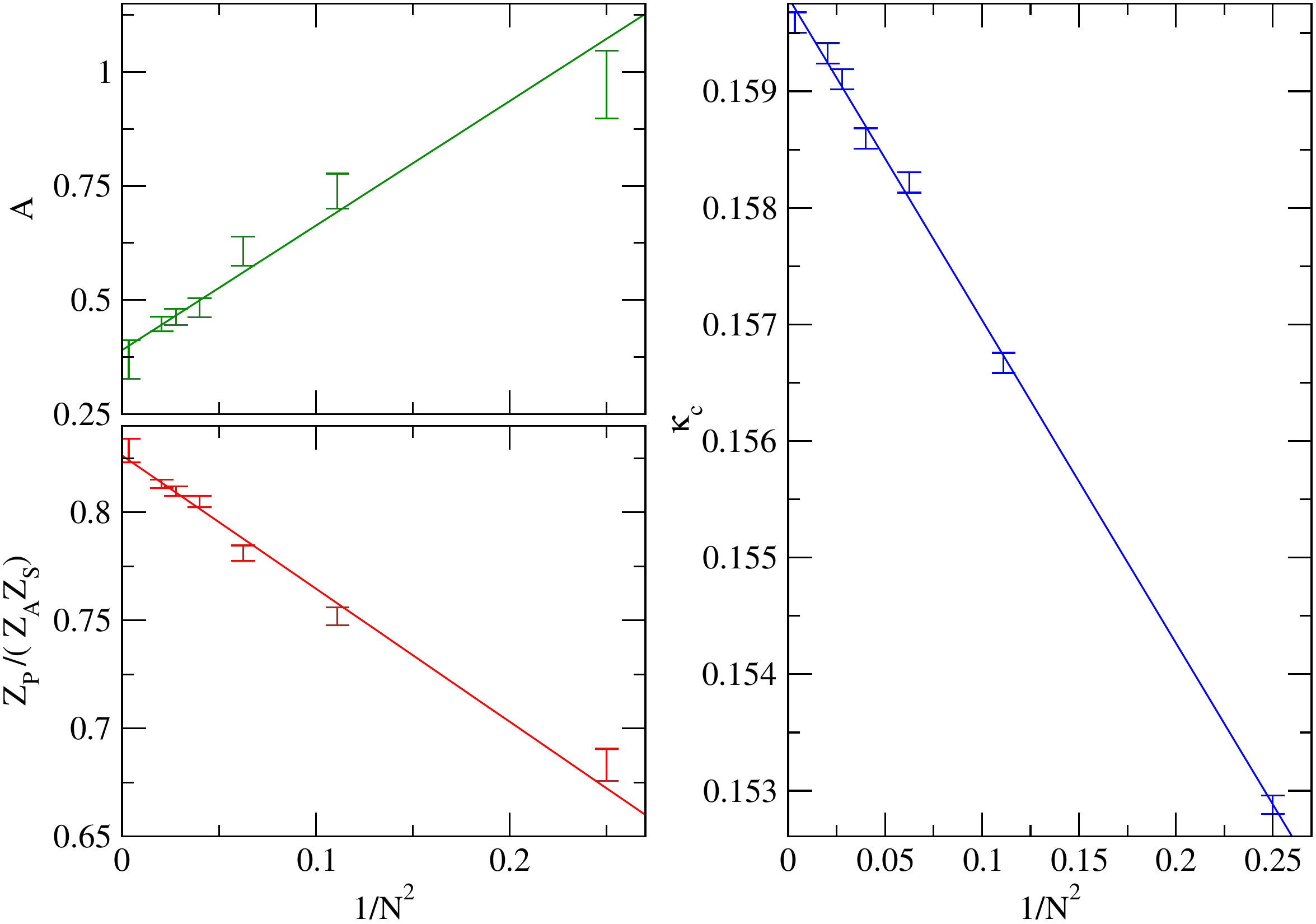}

\end{center}
\caption{Fit of the PCAC mass to eq.~(\ref{eq:pcacMass}) (left), together with the $1/N^2$ fit of the parameters $A$, $Z_P \left(Z_A Z_S\right)^{-1}$ and $\kappa_c$ (right). We plot systematic errors for $\kappa_c$.}
\label{pcac}
\end{figure}

Our non-perturbative determination of $\kappa_c$ is based on the partially conserved axial current (PCAC) relation, 
\begin{equation}
\sum_{\vec{r}} \langle 0 | \partial_\mu A^\mu (\vec{r},t) | \pi \rangle= 2 \mpcac (t) \, \sum_{\vec{r}}  \langle 0 | P(\vec{r},t) | \pi \rangle ,
\end{equation}
where $ A^\mu(\vec{r},t) = \bar{u} (\vec{r},t) \gamma_\mu \gamma_5 d(\vec{r},t) $, $P(\vec{r},t) = \bar{u} (\vec{r},t)  \gamma_5 d(\vec{r},t) $
and $\mpcac=\lim_{t\rightarrow\infty}\mpcac(t)$. On the lattice we compute $\mpcac$ as:
\begin{equation}
a\,\mpcac (t) =  \frac{C_{\gamma_0\gamma_5,\gamma_5}(t+a) - C_{\gamma_0\gamma_5,\gamma_5}(t-a)   }{4C_{\gamma_5,\gamma_5}(t) } ,
\end{equation}
where the pion sources are smeared, and fit this to a constant. This quantity, which is not affected by chiral logarithms, is related to the vector quark mass $a m_q$ defined in eq.~(\ref{bare mass}) by renormalization constants:
\begin{equation}
\mpcac  =  \frac{Z_P} {Z_A Z_S} m_q.
\end{equation}
Taking into account the leading lattice corrections $(1+b_Xa\,m)$ to renormalization constants $Z_X$, we can fit our lattice results to the expression: 
\begin{equation}\label{eq:pcacMass}
a\,\mpcac= \frac{Z_P}{Z_A Z_S} \left( 1 + A \,a\,\mpcac + \dots \right) \frac{1}{2} \left(\frac{1}{\kappa}-\frac{1}{\kappa_c}\right),
\end{equation}
from which we extract $\kappa_c$, ${Z_P}/({Z_A Z_S})$ and $A$ (see table~\ref{tab:pcacParams} of appendix~\ref{sec:addtabfig}
). The fits are plotted in figure~\ref{pcac}.

The (unrenormalized) PCAC mass can be determined very precisely (see tables \ref{tab:su2table1}-\ref{tab:su17table1}), so we will expand every meson mass as a function of this variable. 

We expect the parameters $A,{Z_P}/({Z_A Z_S})$ and $\kappa_c$ to have $\mathcal{O}(1/N^2)$ corrections, hence we fit them to
\begin{equation}
\alpha_1 + \frac{\alpha_2}{N^2},
\end{equation}
as shown in figure~\ref{pcac}. With this analysis, we obtain good fits for $A$ and ${Z_P}/({Z_A Z_S})$, with values of the reduced $\chi^2$ close to $1$, while for $\kappa_c$, although qualitatively the behavior looks very promising, we get a $\chi^2$ per d.o.f. of 300, indicating that uncertainties in our data are underestimated. In fact $\kappa_c$ can be considered a function of $\beta$ only, which was chosen to match the string tension among the different groups. This process introduces a systematic error which propagates to $\kappa_c$.  A qualitative way to estimate this propagation is to consider finite differences between the values of $\kappa_c, \beta$ and $a\sqrt{\sigma}$ from refs.~\cite{DelDebbio:2007wk, Lucini:2005vg,Allton:2008ty, Lucini:2012wq} and to compute
\begin{equation}\label{eq:systemerr}
 \delta\kappa_c= \frac{\Delta\kappa_c}{\Delta\beta} \frac{\Delta\beta}{\Delta(a\sqrt{\sigma})} \delta{(a\sqrt{\sigma})} .
\end{equation}
The r.h.s factors of the equation above are listed in table~\ref{tab:kappacsyserr}, together with $\delta\kappa_c$ for the groups available in refs.~\cite{DelDebbio:2007wk, Bali:2008an}. These systematic uncertainties due to the matching of $\beta$ are ten to twenty times larger than the statistical ones and, since they are approximately constant across the $N$-values, we used the same errors for the remaining groups. Taking this into account, the reduced $\chi^2$-value of the $\kappa_c$ fit becomes 1.6. 

The $1/N^2$ fit results are:
\begin{eqnarray}\label{eq:pcacParamEq}
 \frac{Z_P}{Z_A Z_S}  & = & 0.8291(20)  -\frac{0.699(45)}{N^2},\\
A &= &0.390(13)+  \frac{2.73(26)}{N^2},\\
\kappa_c &=& 0.1598555(33)(447)-\frac{0.028242(68)(394)}{N^2},
\end{eqnarray}

where in the $\kappa_c$-case the second error is the systematic one, due to the slight mismatch in the string tension, detailed above. 
We find the ratio ${Z_P}/({Z_A Z_S})$ to
vary between
0.68 ($N=2$) and 0.83 ($N=17$), with the $\mathrm{SU}(3)$-value 0.75,
which is consistent with the non-perturbative result
0.82(11)~\cite{Gimenez:1998ue} obtained at $\beta=6.0$, close
to our value $\beta=6.0175$. 

\subsection{The pion mass}
The pion masses are shown in figure~\ref{fig:pionDelta} and presented in tables \ref{tab:su2table1}--\ref{tab:su17table1} of appendix~\ref{sec:addtabfig} as dimensionless ratios, dividing them by the square root of the string tension $\sqrt{\sigma}$. Quenched chiral perturbation theory~\cite{Sharpe:1992ft} predicts
\begin{equation}
\label{piondelta1}
\left(am_\pi\right)^2= A \left(  am_q \right)^{\frac{1}{1+\delta } } + \dots,
\end{equation}
where the exponent $\delta$ is due to the presence of chiral logarithms. The theory predicts $\delta$ to be positive, $\mathcal{O}(10^{-1})$ for $\SU(3)$, and suppressed as $1/N$ at large $N$. However, it is known that data in the region where $m_\pi/\sqrt{\sigma} > 1$ are not very sensitive to chiral logarithms~\cite{Chen:2003im}, and fitting larger pion masses according to eq.~(\ref{piondelta1}) would lead to values of $\delta$ with even the wrong sign. For this reason, we included a subleading term of the quark mass expansion, performing fits according to:
\begin{equation}\label{piondelta2}
\frac{m_\pi^2}{\sigma} = A \left( \frac{\mpcac}{\sqrt{\sigma}} \right)^{\frac{1}{1+\delta } }  + B  \frac{\mpcac^2}{\sigma}.
\end{equation}
In these fits, the $\delta$ exponent is, essentially, determined by the lowest pion masses---which, unfortunately, are the points with the largest uncertainties. This leads to rather large relative errors for $\delta$. Nevertheless, we found clear evidence that $\delta$ gets smaller when $N$ is increasing. 
Within our precision limits, $\delta$ is found to be consistent with zero for all $N\geq6$. In fact, for larger $N$ one can omit $\delta$ completely from the formula and still obtain a good fit. Conversely, for $N\leq 3$ we find $\mathcal{O}(10^{-1})$ values of $\delta$, where for $\SU(3)$ we get results consistent with those reported in ref.~\cite{Bardeen:2000cz}, in which Wilson and clover actions were used, and also consistent with table 3 of ref.~\cite{Chen:2003im}, where different values of $\delta$ were calculated, using different actions. This suggests a small $1/N$ coefficient and thus we include a higher order term into the $1/N$ fit. 

The $N$ dependence of $\delta$ and the expansion of $A$ and $B$ in powers of $1/N^2$ (see figure~\ref{fig:pionDelta}) give a $\chi^2$ per degree of freedom close to $ 2 $ and read:
\begin{eqnarray}\label{eq:pionDeltaAandB}
A&=& 11.99( 0.10 )  - \frac{8.7(1.6)}{N^2},\\
B &=& 2.05 ( 0.13 ) +  \frac{5.0(2.2)}{N^2},\\
\delta &=& \frac{0.056(19)}{N} + \frac{0.94(21)}{N^3}.\label{eq:delta1}
\end{eqnarray}

In order to assess the systematics on the exponent $\delta$, we performed a combined fit $(\mpcac,N)$ of our data to eq.~(\ref{piondelta2}), using all the $N$-values at once and excluding the two highest masses for each group. Since the Wilson action explicitly breaks the chiral symmetry, we include in the fit also a constant term which however is found to be consistent with zero. The resulting curve has $\chi^2/\mathrm{d.o.f.}=1.6$ and reads,
\begin{eqnarray}
\label{piondelta3}
\frac{m_\pi^2}{\sigma} &=&  0.0015(36) + \left(  11.67(15 ) -  \frac{8.1(5.4 ) }{N^2}  \right) \left( \frac{\mpcac}{\sqrt{\sigma}} \right)^{\frac{1}{1+\delta } }  +  \nonumber \\ & &+\left(  2.95( 42 ) -  \frac{1( 15 ) }{N^2} \right) \frac{\mpcac^2}{\sigma}  \\
\delta &=& \frac{0.093(27)}{N} + \frac{1.00(52)}{N^3}.
\end{eqnarray}

Note that the $\mpcac^2$ term is now less well determined, due to the exclusion of high mass points. However, the $\delta$-parametrisation is consistent with the one of eq.~(\ref{eq:delta1}).

\begin{figure}
\begin{center}
\includegraphics[width=0.68\textwidth]{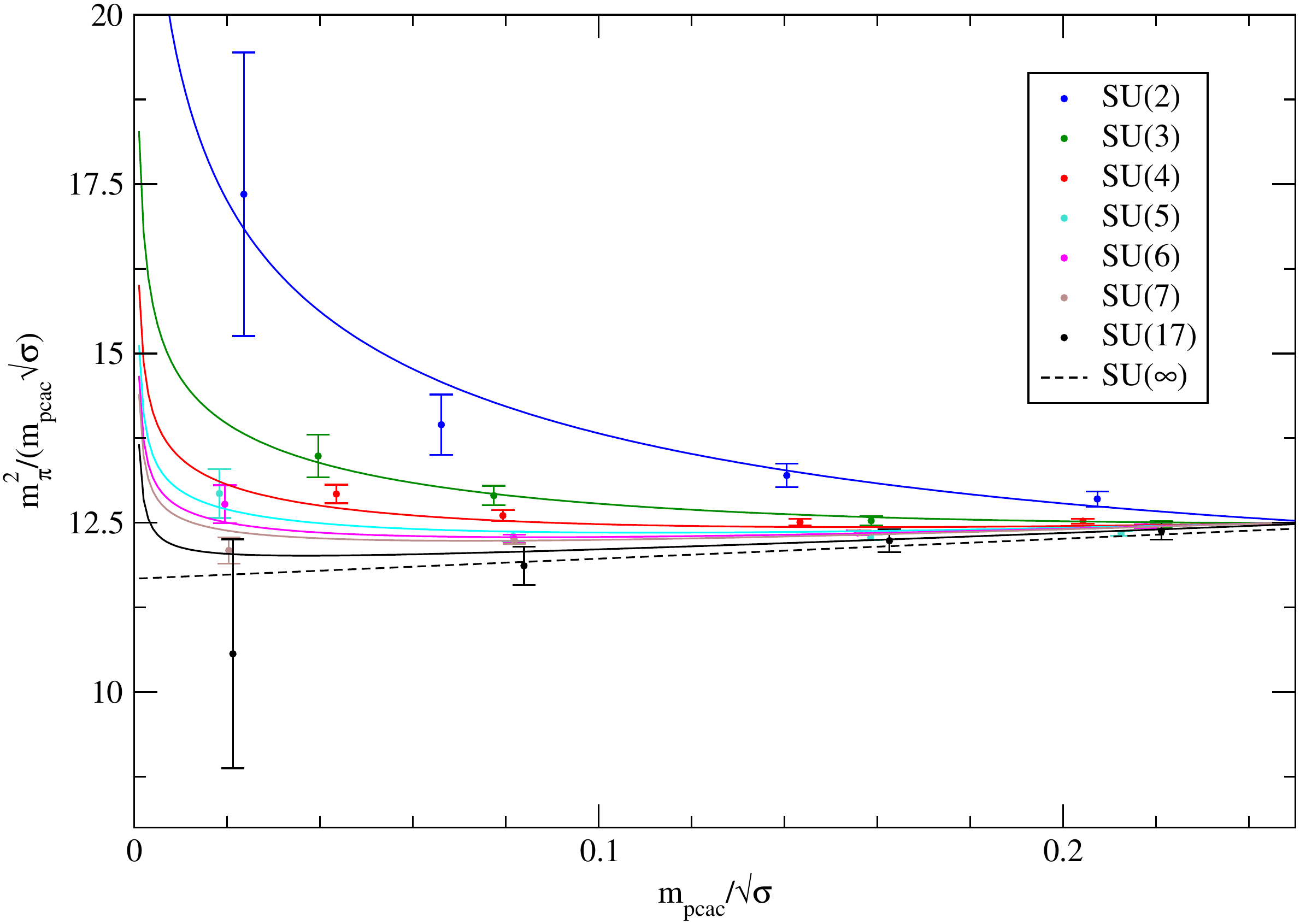} 

\end{center}
\caption{Combined fit of $m^2_\pi/\left(\mpcac \sqrt{\sigma}\right)$. The solid curves are calculated using eq.~(\ref{piondelta3}), for each $N$. }
\label{fig:pionDelta2}
\end{figure}

In figure~\ref{fig:pionDelta2} we plot for each group the pion mass according to eq.~(\ref{piondelta3}) divided by the PCAC mass, in order to emphasise the deviations from a linear behaviour, due to the exponent $\delta$.

Below, we expand all the remaining meson masses as functions of $\mpcac$. These can easily be translated into dependencies on $m_\pi^2$ through eq.~(\ref{piondelta2}) above.

\begin{figure}
\begin{center}
\includegraphics[width=0.48\textwidth]{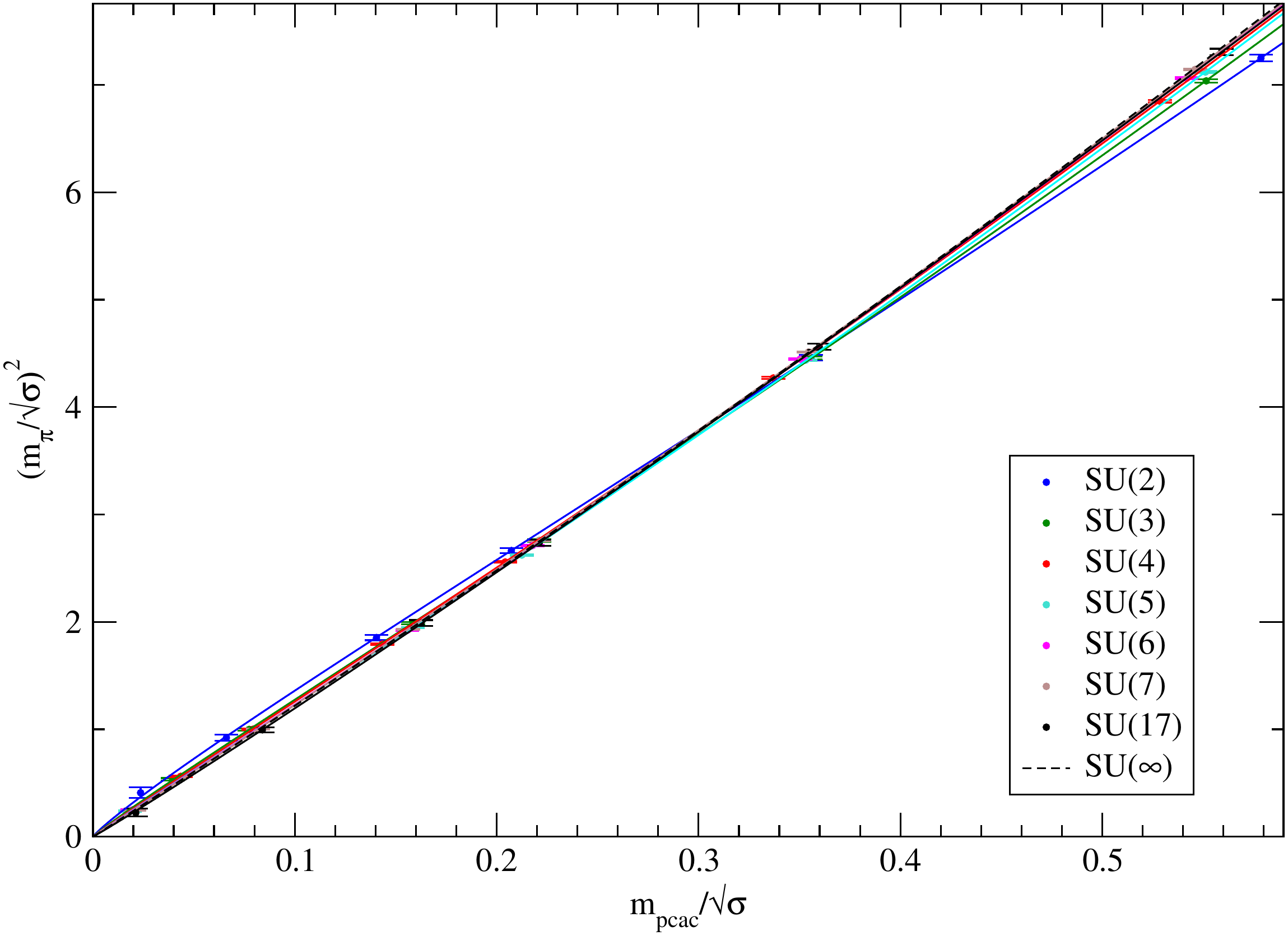} \includegraphics[width=0.48\textwidth]{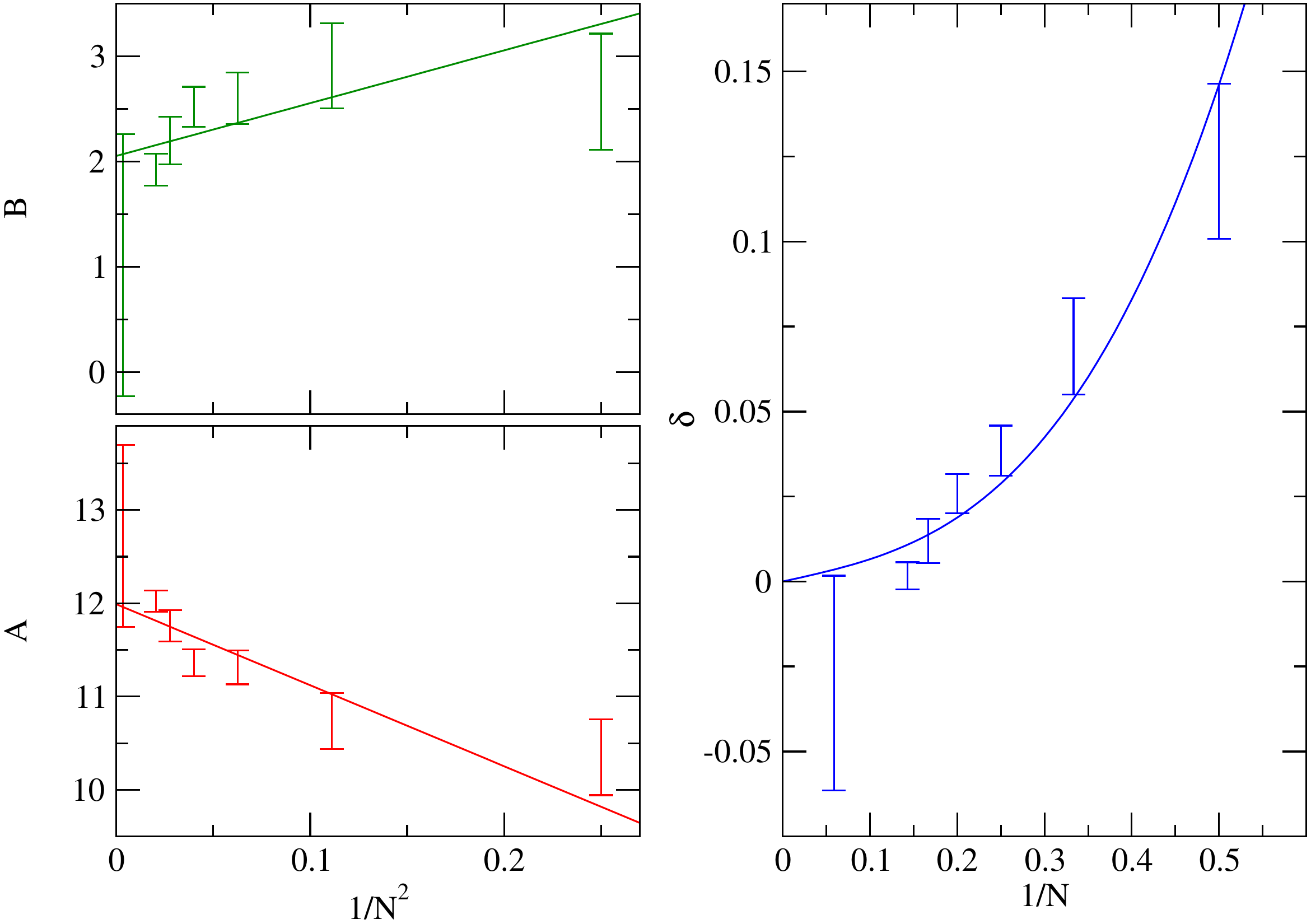}

\end{center}
\caption{Fit of the squared pion mass, in units of the string tension, to eq.~(\ref{piondelta2}) (left panel), $N$ dependence of the
fit parameters~eq.~(\ref{eq:pionDeltaAandB})--(\ref{eq:delta1}) (right).}
\label{fig:pionDelta}
\end{figure}

\subsection{The $\rho$ mass}
\label{subsec:rhomass}
Quenched chiral perturbation theory predicts a dependence of $m_\rho$ on the square root of the quark mass $m_q$~\cite{Booth:1996hk}, in contrast to the unquenched theory, where the leading behavior is linear in $m_q$. Thus the expansion of $m_\rho$ takes the form:
\begin{equation}\label{rhoVsPionFull}
m_\rho = m_{\rho,0} + C_{1/2} m_q^{1/2} + C_{1} m_q  + C_{3/2} m_q^{{3/2}} + \dots,
\end{equation}
where the $C_{1/2}$ coefficient is expected to be negative and to vanish as $1/N$ in the large-$N$ limit~\cite{Booth:1996hk}, restoring the linear behavior.

\begin{figure}
\begin{center}
\includegraphics[width=0.48\textwidth]{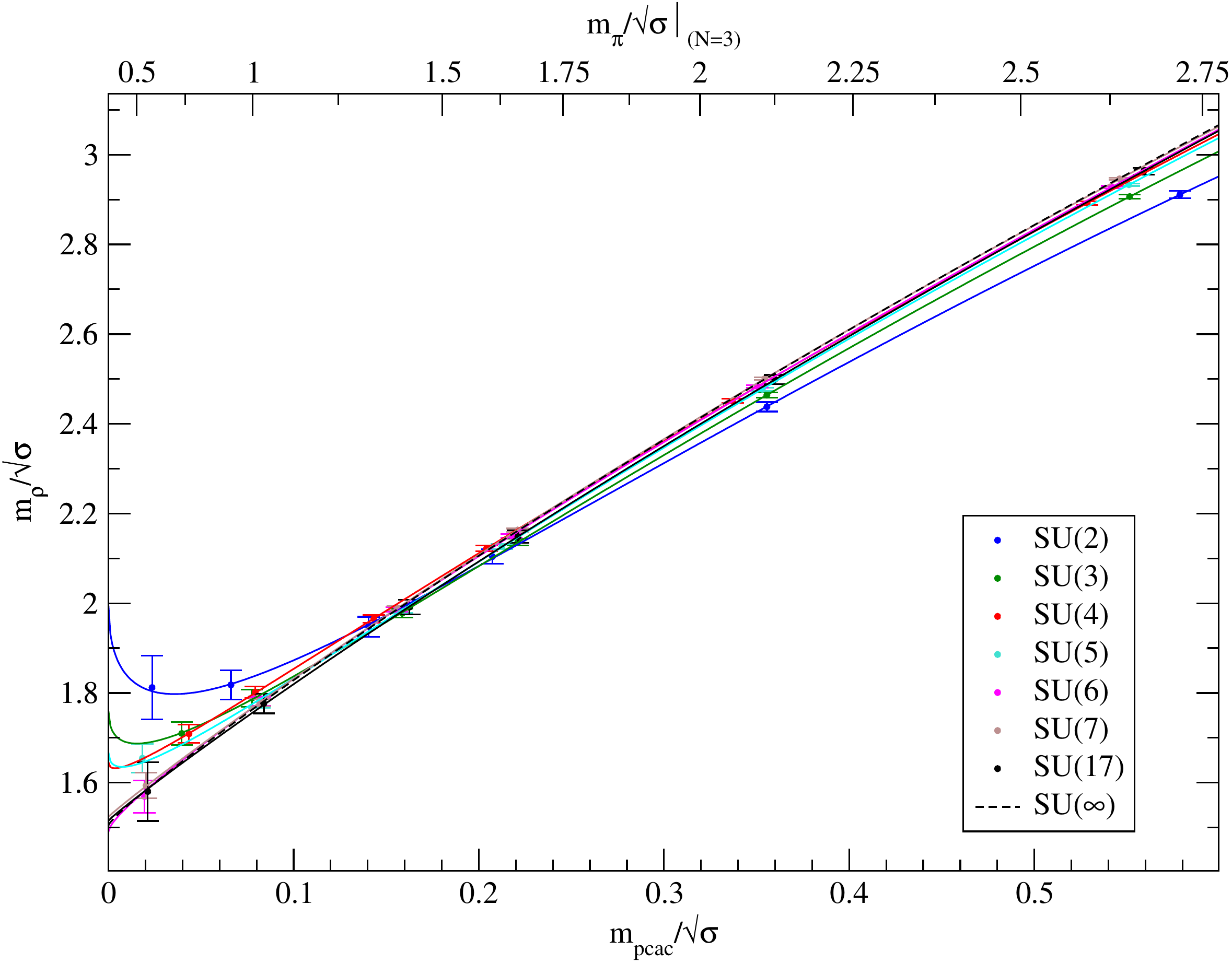} \includegraphics[width=0.48\textwidth]{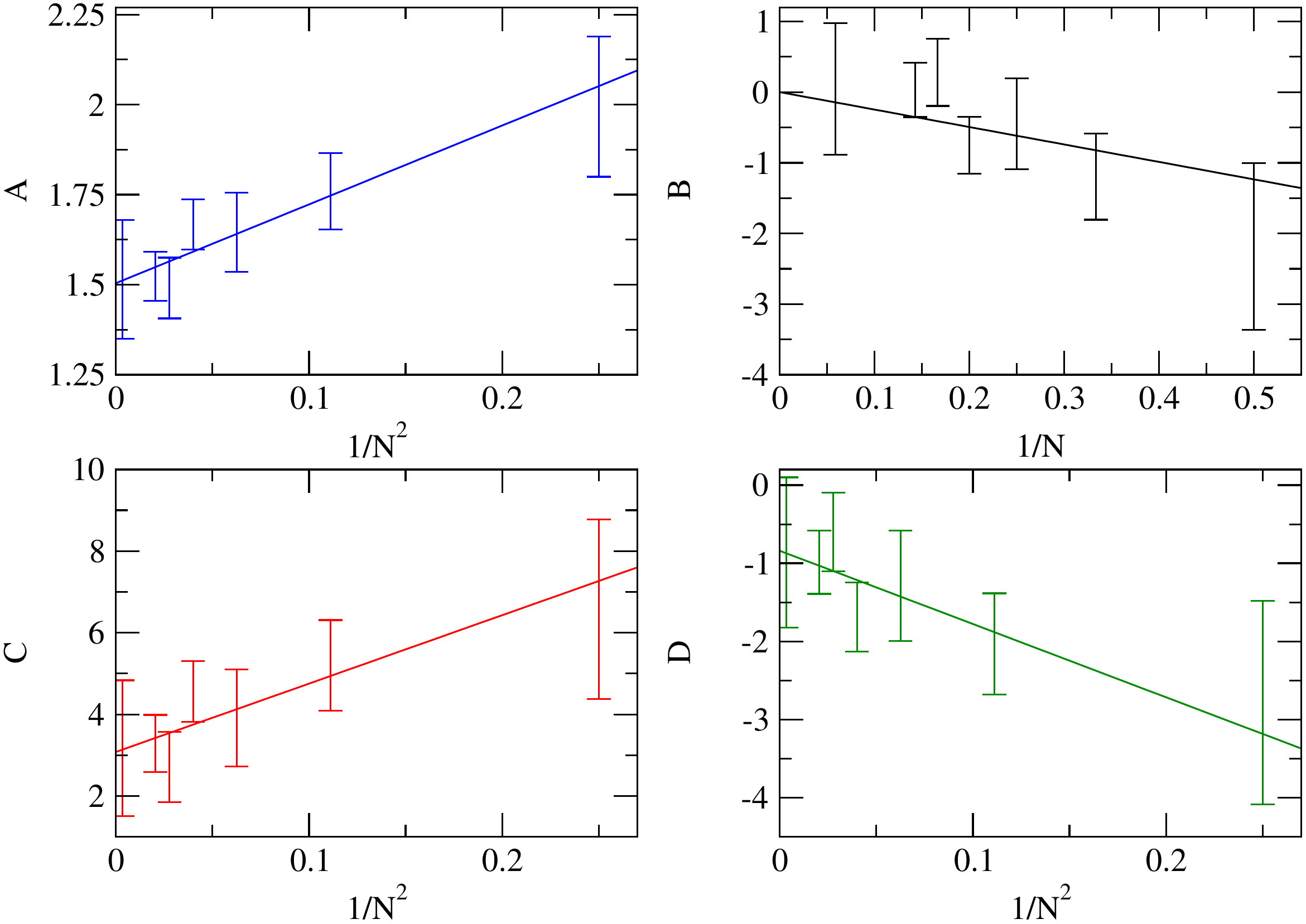}

\end{center}
\caption{Fit of the $\rho$ mass to: $m_{\rho}/\sqrt{\sigma} = A + B \cdot (\mpcac/\sqrt{\sigma})^{1/2} + C \cdot \mpcac/\sqrt{\sigma}+ D \cdot (\mpcac/\sqrt{\sigma})^{3/2} $ (left). $1/N^2$ fits of the parameters $A$, $C$ and $D$ and $1/N$ fit of $B$ (right). }
\label{fig:rhoPcac}
\end{figure}

The masses of the $\rho$ states are listed in tables~\ref{tab:su2table1} to~\ref{tab:su17table1} and plotted in figure~\ref{fig:rhoPcac} against the PCAC mass. For each group we fit:
\begin{equation}\label{eq:rhoVsPcac0}
\frac{m_{\rho}}{\sqrt{\sigma}} = A + B \left( \frac{\mpcac}{\sqrt{\sigma}} \right )^{1/2}+ C \frac{\mpcac}{\sqrt{\sigma}}  + D\left( \frac{\mpcac}{\sqrt{\sigma}}\right)^{3/2} ,
\end{equation}
and then expand $A$, $C$ and $D$ in powers of $1/N^2$, while fitting $B$ linearly in $1/N$. Note that $D$ should vanish like $1/N^2$ in the large-$N$ limit. The parameter $B$ is found to be compatible with zero for $N > 5$ while the parameter $D$ has unexpectedly a finite value: this effect is due to contamination from higher orders in the quark mass, as we show below. 

The large-$N$ expansion of the parameters reads:
\begin{eqnarray}\label{eq:rhoVsPcac1}
\frac{m_{ \rho }}{\sqrt{\sigma}} &=& \left(  1.504( 51 ) +  \frac{2.19( 75 ) }{N^2} \right) -\frac{2.47(94)}{N}  \left(\frac{\mpcac}{\sqrt{\sigma}} \right )^{1/2}  \nonumber \\
&&+ \left( 3.08( 53 ) +  \frac{16.8( 8.2)}{N^2} \right) \left(\frac{\mpcac}{\sqrt{\sigma}}\right)  +   \left(  -0.84(31)    -\frac{9.4( 4.8)}{N^2} \right)  \left(\frac{\mpcac}{\sqrt{\sigma}} \right )^{3/2}.
\end{eqnarray}

\begin{figure}
\begin{center}
\includegraphics[width=0.48\textwidth]{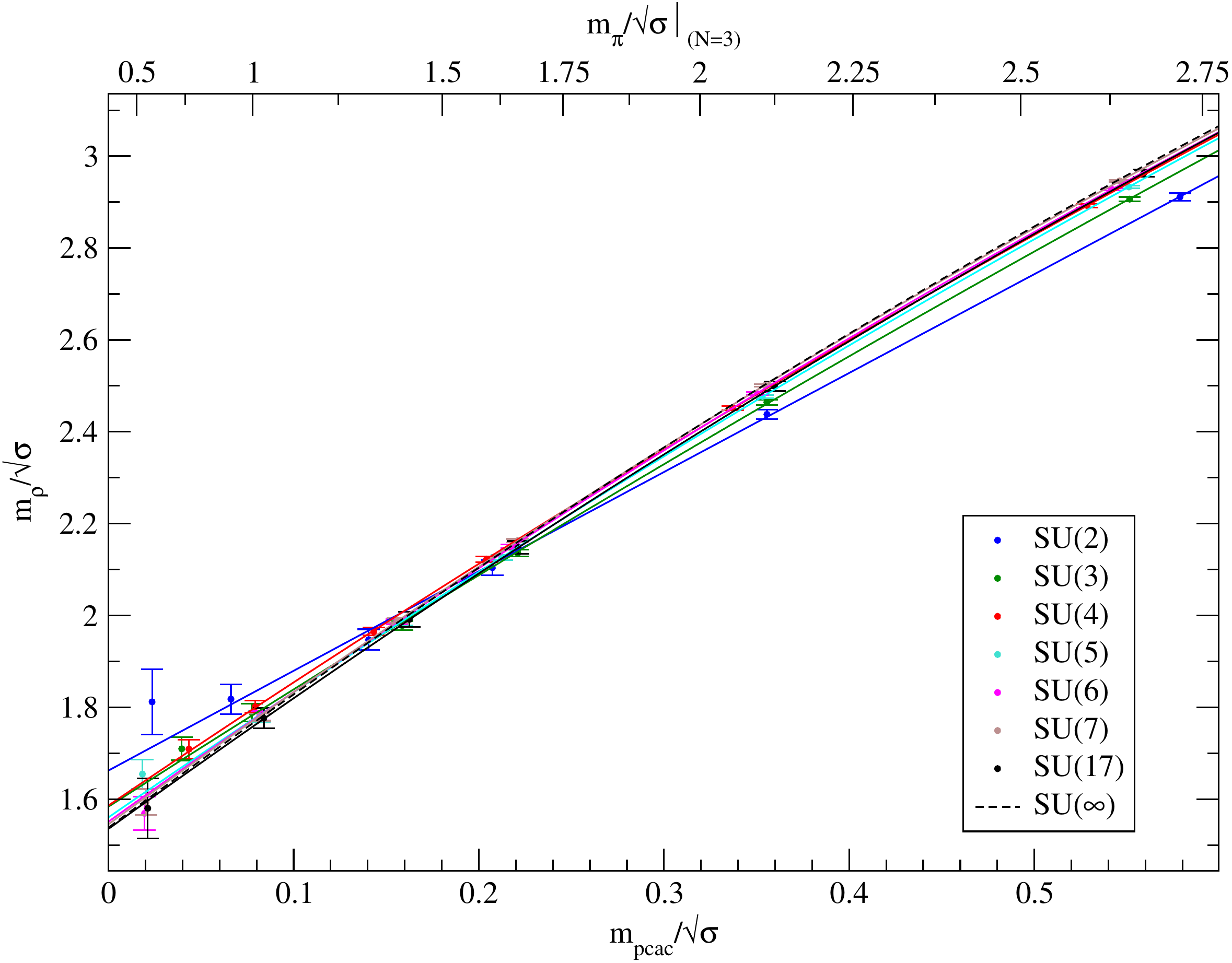} \includegraphics[width=0.48\textwidth]{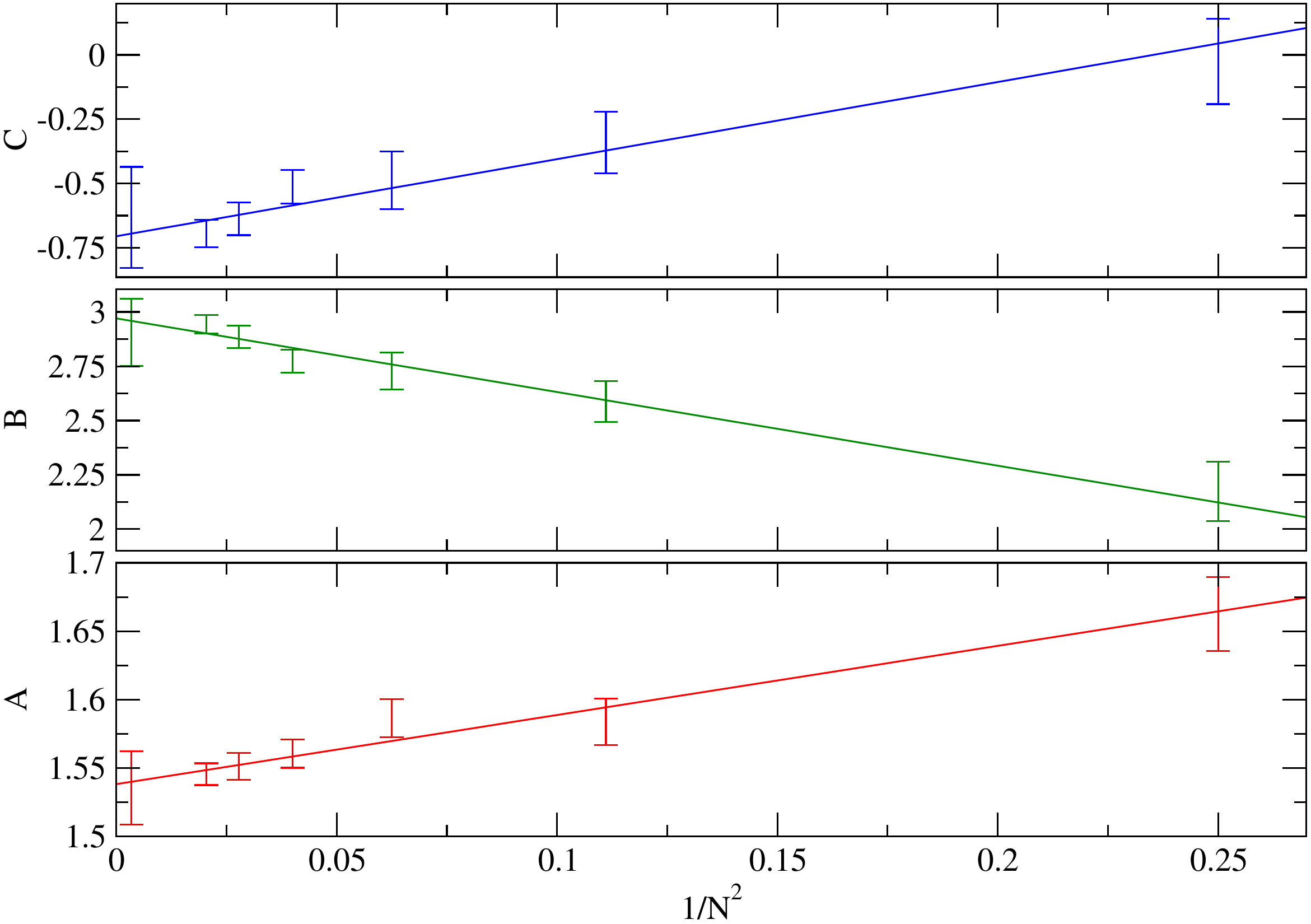}

\end{center}
\caption{Quadratic fit of the $\rho$ mass in units of the square root of the string tension, according to eq.~(\ref{eq:rhoVsPcac2}).}
\label{fig:rhoPcacWrong}
\end{figure}

To address the question of the non-vanishing large-$N$ value of $D$, we interpolated the data with an alternative fit (figure~\ref{fig:rhoPcacWrong}) of the form: 
\begin{eqnarray}\label{eq:rhoVsPcac2}
\frac{m_{ \rho }}{\sqrt{\sigma}} &=& \left( 1.5382(65) +  \frac{0.51(11)}{N^2} \right)+ \left(2.970(34)  -  \frac{3.39(55)}{N^2} \right) \left(\frac{\mpcac}{\sqrt{\sigma}}\right)   \nonumber \\
&&+   \left(  -0.706(43)    +\frac{3.00(68)}{N^2} \right)  \left(\frac{\mpcac}{\sqrt{\sigma}} \right )^{2}.
\end{eqnarray}
Although the $1/N$ counting of eq.~(\ref{eq:rhoVsPcac2}) is not consistent, because of the missing square root term, one should notice that  eq.~(\ref{eq:rhoVsPcac1}) and eq.~(\ref{eq:rhoVsPcac2}) share the same $N\rightarrow\infty$ behaviour. In fact the $N=\infty$ coefficients of eq.~(\ref{eq:rhoVsPcac1}) agree within the errors with those of eq.~(\ref{eq:rhoVsPcac2}). In particular, we stress that the non-vanishing term $0.84(31)  \left({\mpcac}/{\sqrt{\sigma}}\right )^{3/2}$ is consistent with $0.706(43)  \left({\mpcac}/{\sqrt{\sigma}}\right )^{2}$, for the range of $\mpcac$ studied. In principle one should introduce a quadratic term into eq.~(\ref{eq:rhoVsPcac1}), in practice however, such a fit to six points with five free parameters becomes unstable and does not allow us to study the $1/N$ behaviour of the coefficients. 

\begin{figure}
\begin{center}
\includegraphics[width=0.48\textwidth]{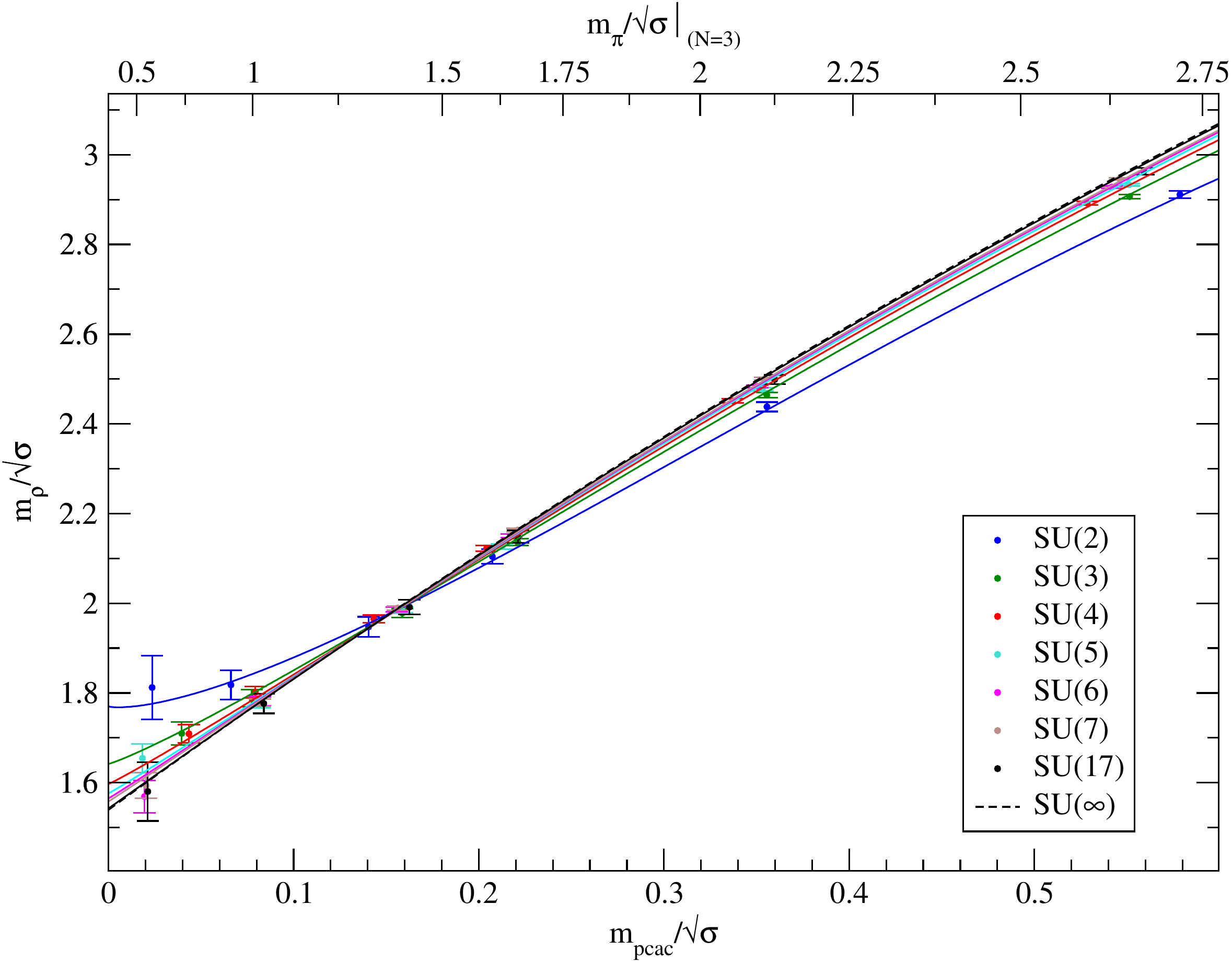}

\end{center}
\caption{Combined fit of the $\rho$ mass in units of the square root of the string tension. The solid curves are calculated from eq.~(\ref{eq:rhoVsPcac3}) for the different $N$.}
\label{fig:rhoCombined}
\end{figure}

A possible solution is to fit the data for all groups at once using a combined fit, where we fix the $(\mpcac,N)$ functional form up to the second order term in the quark mass (figure~\ref{fig:rhoCombined}). With this approach we obtain:
\begin{eqnarray}\label{eq:rhoVsPcac3}
\frac{m_{ \rho }}{\sqrt{\sigma}} &=& \left(  1.5395(83) +  \frac{0.92(21)}{N^2} \right) -\frac{0.06(14)}{N}  \left(\frac{\mpcac}{\sqrt{\sigma}} \right )^{1/2}  + \left(2.994(44)  -  \frac{13.9(48)}{N^2} \right) \left(\frac{\mpcac}{\sqrt{\sigma}}\right)   \nonumber \\
&&+ \left(   \frac{27( 11)}{N^2} \right)  \left(\frac{\mpcac}{\sqrt{\sigma}} \right )^{3/2} +  \left(  -0.739(50)    -\frac{15.1(73)}{N^2} \right)  \left(\frac{\mpcac}{\sqrt{\sigma}} \right )^{2},
\end{eqnarray}
with a $\chi^2/\mathrm{d.o.f.}=2$. The drawback of this approach is that data at small $(m_q,N)$ have less weight in the combined fit, leading to a smaller coefficient for the $\mpcac^{1/2}$ term.

In order to compare our results with the holographic prediction (see section \ref{subsec:holography}) it is useful to study  $m_\rho$ as a function of the pion mass, paying particular attention to the linear term of $m_\rho(m_\pi^2)$. Since the results of eqs.~(\ref{eq:rhoVsPcac1})--(\ref{eq:rhoVsPcac3}) tend to agree in the large-$N$ limit and considering that our extrapolation relies on small $N$, which might be affected by quenched deviations, we quote eq.~(\ref{eq:rhoVsPcac1}) as our best phenomenological parametrization of the data. To avoid further propagation of errors, we fit for each $N$ the $\rho$ data directly to the $\pi$ masses using the fit form of eq.~(\ref{eq:rhoVsPcac1}), with the substitution $\mpcac \rightarrow m_\pi^2$. Then we extrapolate the fitted slope to $N\rightarrow\infty$,  obtaining:
\begin{equation}\label{eq:slopeRho1}
\frac{m_{ \rho} (m_\pi) }{m_\rho(0) } =  1+ 0.360(64)\left( \frac{m_\pi}{ m_\rho(0) }\right)^2  +\dots .
\end{equation}
As a consistency check, we can omit the data corresponding to the largest quark masses and repeat the analysis using only the smallest three masses and a simple linear fit. This leads to:
\begin{equation}
\label{eq:slopeRho2}
\frac{m_\rho^{\mbox{\tiny{lin}}}  (m_\pi) }{m_\rho(0) } =  1+ 0.317(2)\left( \frac{m_\pi}{ m_\rho(0) }\right)^2  +\dots,
\end{equation}
where the smaller error may be unreliable because the fit has two free parameters for only three data points. We quote eq.~(\ref{eq:slopeRho1}) as our final result.
As already pointed out in ref.~\cite{Bali:2008an}, the results for the slope are close to the holographic prediction obtained in the context of the model presented in ref.~\cite{Constable:1999ch} (see ref.~\cite{Erdmenger:2007cm} for a detailed discussion).

\subsection{The scalar particle}
\begin{figure}
\begin{center}
\includegraphics[width=0.48\textwidth]{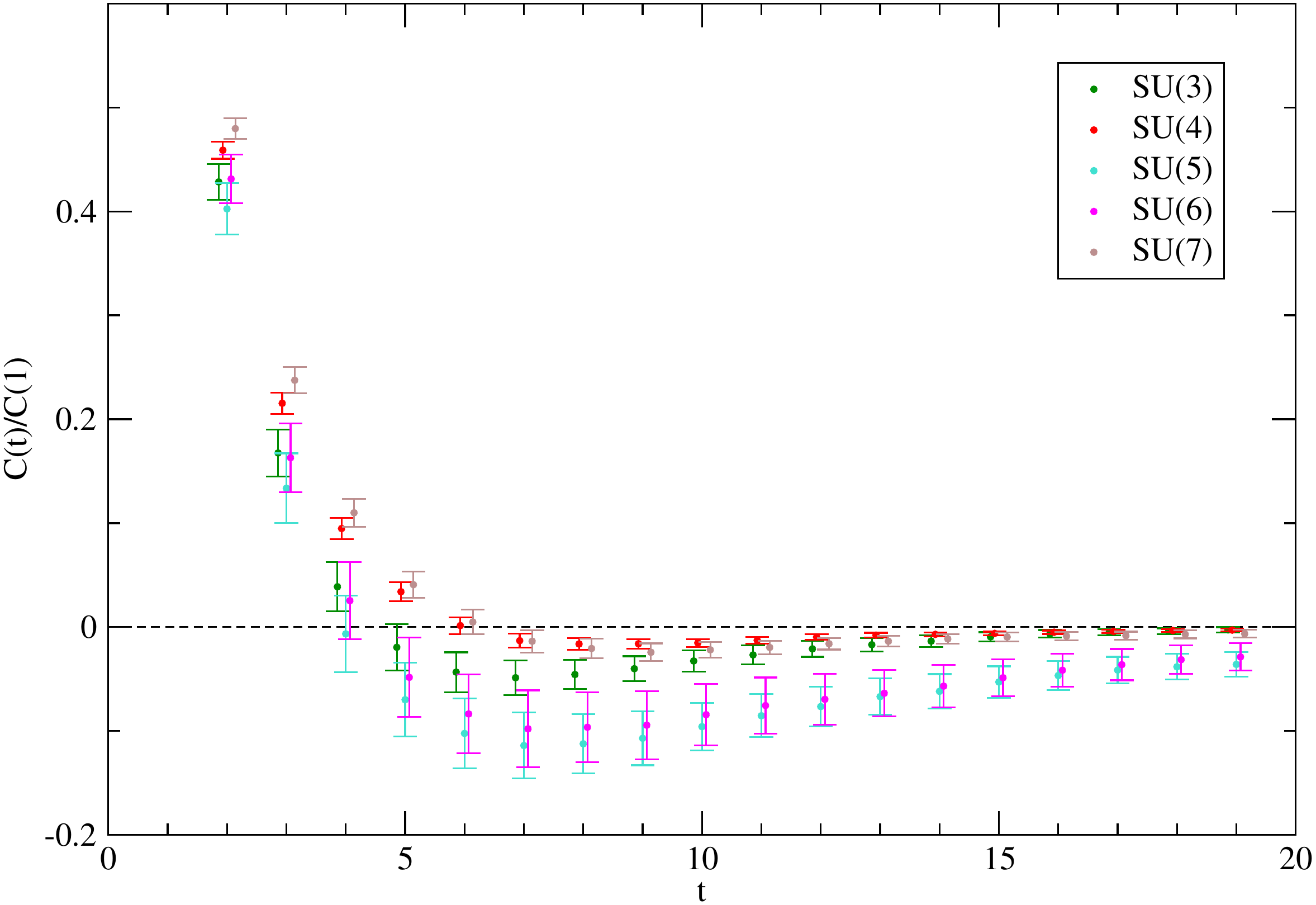} \includegraphics[width=0.48\textwidth]{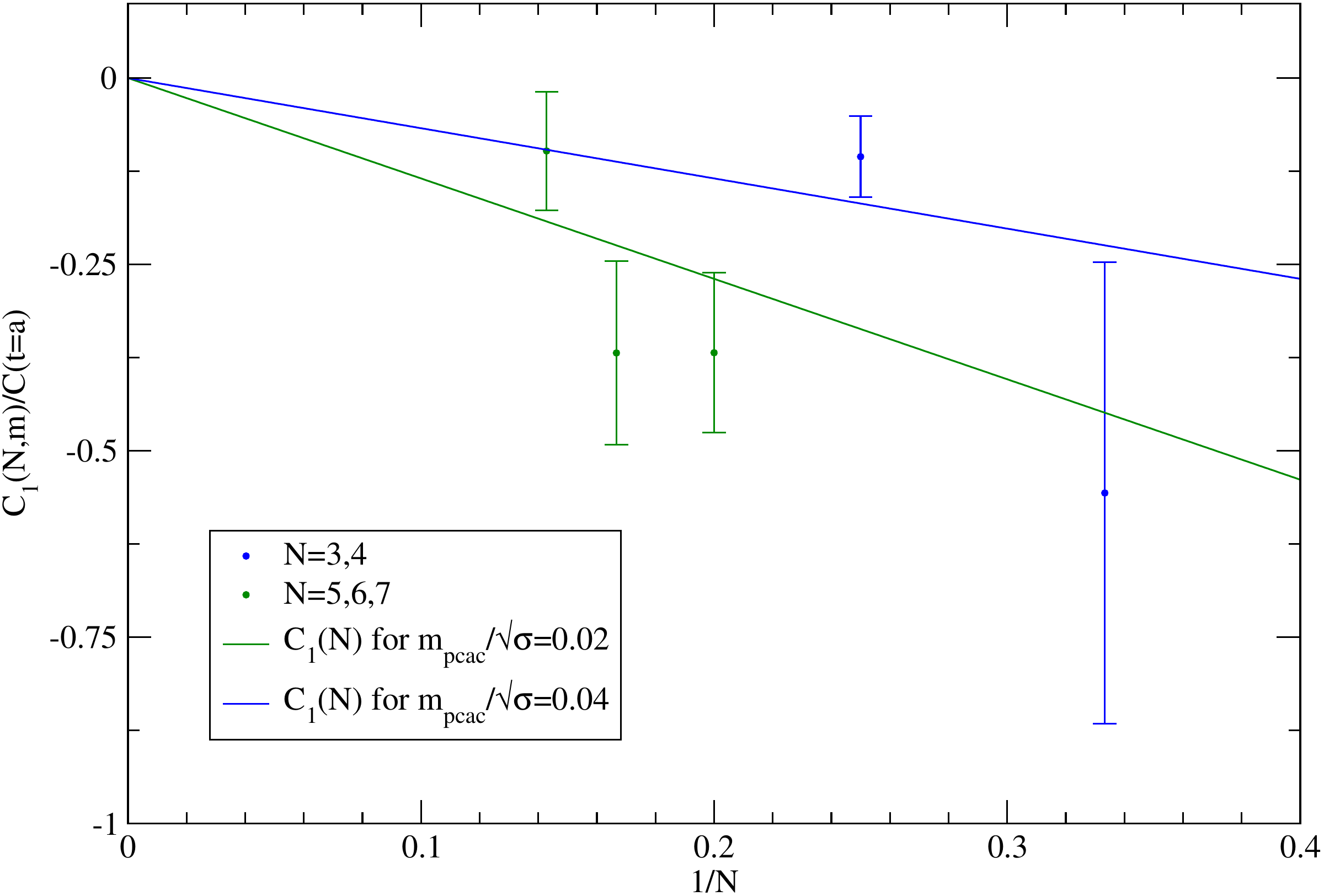}

\end{center}
\caption{Left: the $a_0$ two point functions for each $\SU(N)$ group at the lowest mass value. Notice that the smallest PCAC mass for $\SU(3,4)$ is approximately twice the one for $\SU(5,6,7)$ ($\mpcac/\sqrt{\sigma} \approx 0.04$ vs. 0.02, see tables~\ref{tab:su2table1}-\ref{tab:su17table1}). Right: the amplitudes of the negative long range contribution, $C_1(N,\mpcac)$, normalised to $C(t=1)$ for each group at the lowest quark masses. The solid curves represent their expectation according to the fit of eq.~(\ref{eq:ghostAmp}) for $\mpcac/\sqrt{\sigma}=0.02$ and for $\mpcac/\sqrt{\sigma}=0.04$.}
\label{fig:a0ghost}
\end{figure}

The analysis of the scalar mesons $a_0$ requires special attention. In the quenched theory, in which $\eta'$ also becomes a Goldstone boson violating unitarity, the scalar correlator shows a long-range negative contribution, in addition to the standard short-range exponential decay. In ref.~\cite{Bardeen:2001jm} it was shown that this effect, which is dominant and clearly visible only at the lowest quark masses, is caused by loop diagrams corresponding to an intermediate $\eta'-\pi$ state, which is light and has negative norm in the quenched approximation. Our approach consists in fitting the $a_0$ two-point functions as:
\begin{equation}\label{eq:a0corr}
C(t) = C_0 e^{-m_{a_0} t } - C_1 e^{- \mu t } ,
\end{equation}
where the ``unphysical'' quantities $C_1$ and $\mu$ are fixed using the values at large $t$ only. This approach works very well for $N \geq 5 $, where the noise in the central region of the correlator is smaller. For similar smearing/normalization we expect the amplitudes $C_i$ to be proportional to $N$, with $1/N^2$ corrections,
\begin{equation}
C_i \approx {N}\left(a_i + \frac{b_i}{N^2} + \dots \right).
\end{equation}
In particular, we expect the ratio $C_1/C_0$ to be zero at $N = \infty$, i.e. $a_1=0$ and $C_1\propto N^{-1}$. Indeed we find no evidence of negative contributions at $N=17$ for the values of the quark masses studied. Moreover, the negative contributions get smaller at higher masses, meaning that $C_1$ is at least suppressed like $\mpcac^{-1}$ (and vanishes for $m_{a_0} > 2 m_\pi$).  (Un)fortunately this behaviour is visible only at the very lowest quark mass for each of the groups studied, shown in figure~\ref{fig:a0ghost} (left), so we cannot analyse in detail the amplitude $C_1$ as a function of $\mpcac$. The best strategy to give a qualitative estimate of $C_1$ is to employ a combined $(\mpcac,N)$ fit, where we use data points from all group (figure~\ref{fig:a0ghost} - right).  This leads to the preliminary estimate 
\begin{equation}\label{eq:ghostAmp}
\frac{C_1}{C(t=a)} = -0.0269(53) \frac{ \sqrt{\sigma}}{N\cdot \mpcac },
\end{equation}
with a reduced $\chi^2$ of $1.4$. Note that for $\SU(2)$ (not shown in the figure) we were unable to obtain meaningful results.

\begin{figure}
\begin{center}
\includegraphics[width=0.48\textwidth]{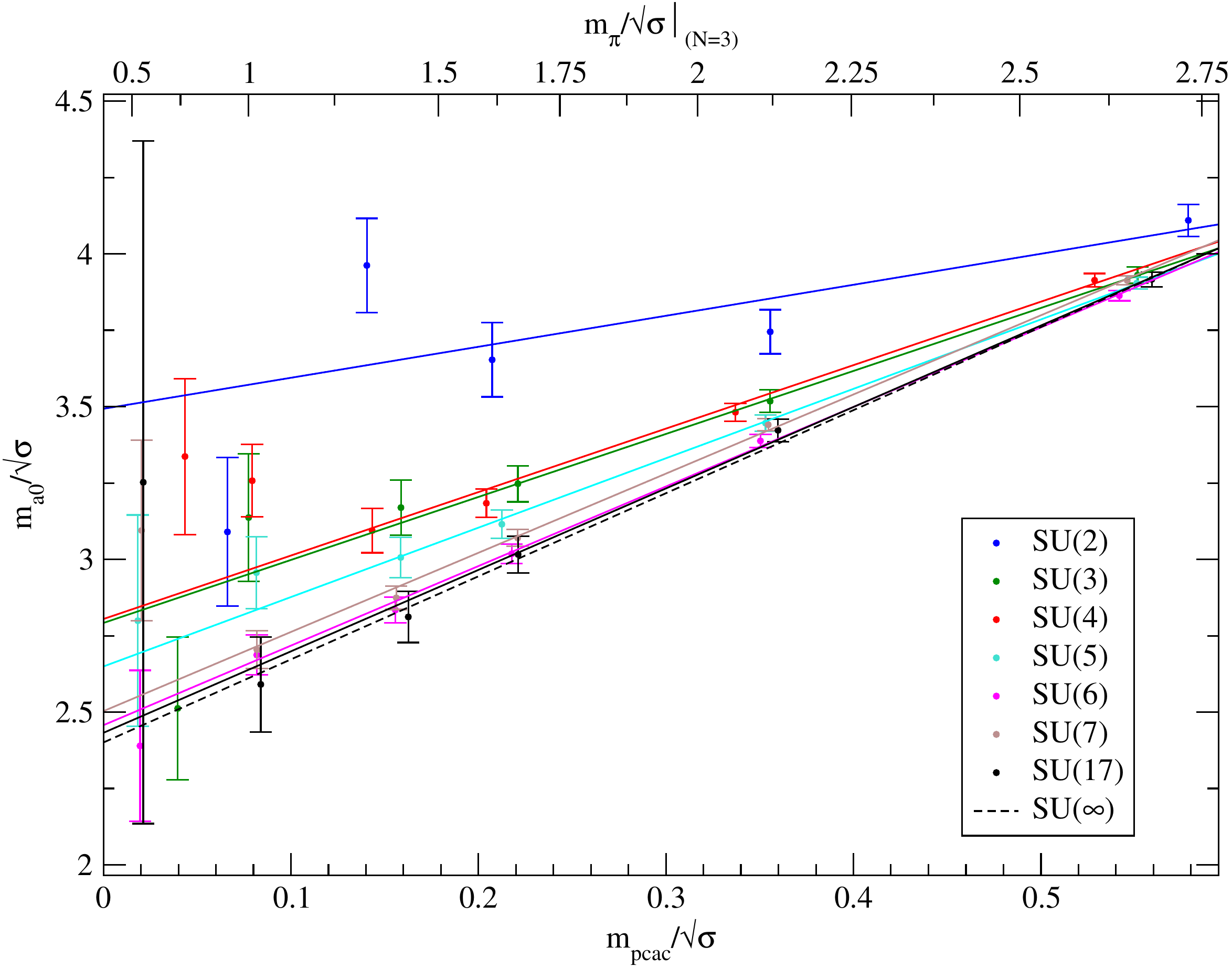} \includegraphics[width=0.48\textwidth]{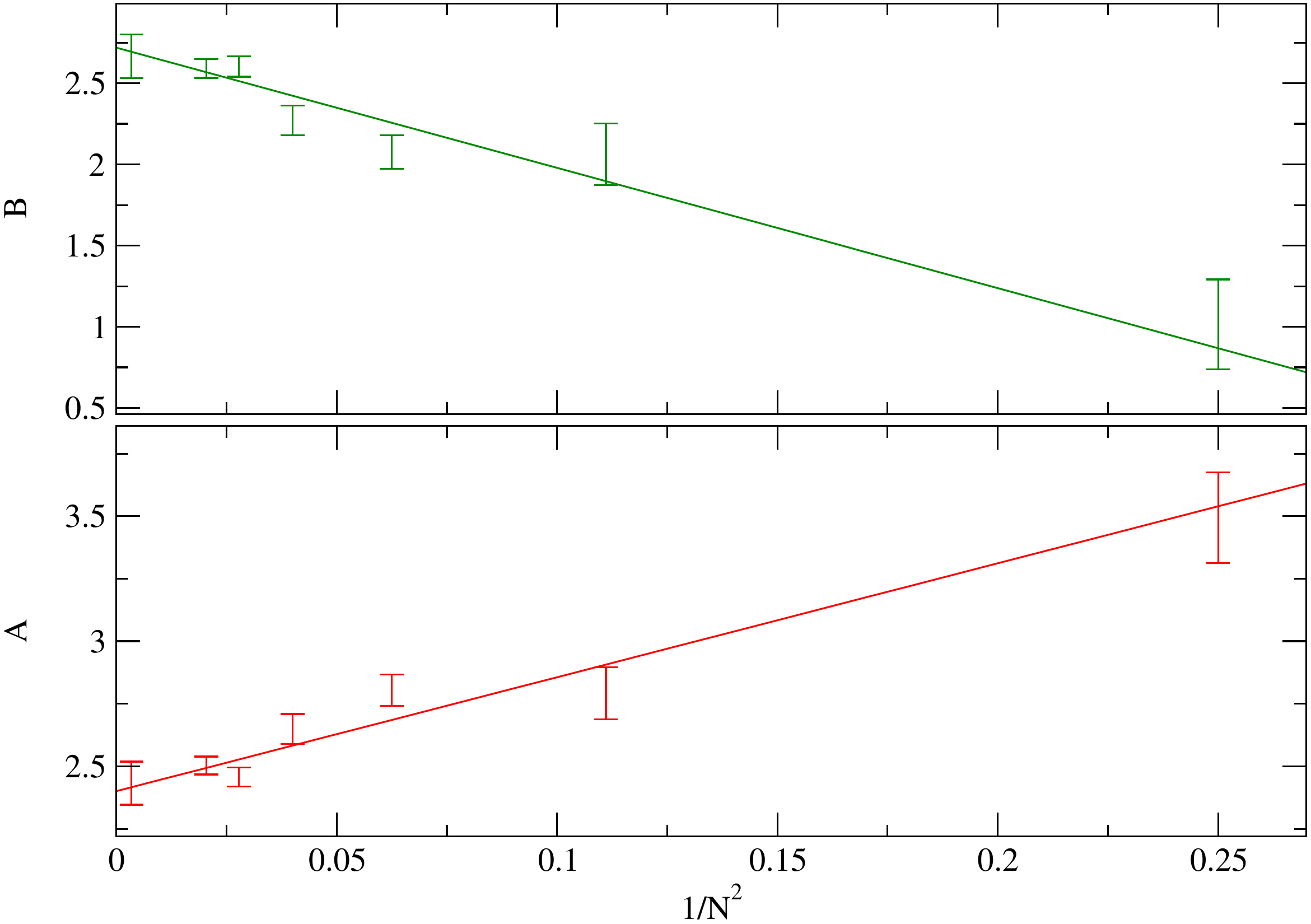}

\end{center}
\caption{Fit of the $a_0$ mass to: $m_{a_0}/\sqrt{\sigma} = A + B \cdot \mpcac /\sqrt{\sigma}$ (left) and $1/N^2$ fit of the parameters $A$ and $B$ (right).}
\label{fig:a0}
\end{figure}

The $a_0$ masses, calculated according to eq.~(\ref{eq:a0corr}), are listed in tables \ref{tab:su2table1}-\ref{tab:su17table1} and their $1/N^2$ expansion (plotted in figure~\ref{fig:a0}) reads,
\begin{equation}
\frac{m_{a_0}}{\sqrt{\sigma}} = \left(  2.402(34)  + \frac{4.25(62)}{N^2} \right) + \left(  2.721(53) - \frac{6.84(96)}{N^2} \right) \frac{\mpcac}{\sqrt{\sigma}}.
\end{equation}

\subsection{The remaining mesons}

The ground state energies for the remaining states are listed in tables \ref{tab:su2table1}--\ref{tab:su17table1}. The corresponding fits are shown in figures~\ref{fig:a1}--\ref{fig:b1} of appendix~\ref{sec:addtabfig}; in these cases, our fits do not include quadratic terms, because of the larger uncertainties of the data. We fit the remaining states to the form,
\begin{equation}\label{eq:otherParticleFit}
\frac{m_{ X }}{\sqrt{\sigma}} = \left(  A_{X,1}  +  \frac{A_{X,2}}{N^2} \right) + \left( B_{X,1}    +\frac{B_{X,2}}{N^2} \right) \frac{\mpcac}{\sqrt{\sigma}}
\end{equation}
and list the results for $A_{X,1},A_{X,2},B_{X,1}$ and $B_{X,2}$ in table~\ref{tab:otherParticlesFit}.

As discussed above, the variational method allows us to extract the first excited states in each channel, although the results for these masses are much noisier than the corresponding ground states. We list the masses in tables~\ref{tab:su2table2}--\ref{tab:su17table2} and plot in figures~\ref{fig:pistar} to \ref{fig:a0star} their quark mass dependences and the $1/N^2$ dependences of the respective fit parameters. The excited states are also fitted to eq.~(\ref{eq:otherParticleFit}) and the results are listed in table~\ref{tab:otherParticlesFit}.

\subsection{Decay constants}

We define the lattice meson decay constants\footnote{Throughout the paper we use the notation $F_\pi = f_\pi/ \sqrt{2} \approx 92$~MeV to indicate the pion decay constant and $F_\pi^{\mathrm{lat}}$ to address the same quantity calculated on the lattice up to the renormalization factor $Z_A$, i.e. $F_\pi = Z_A F_\pi^{\mathrm{lat}} $.  } $F_\pi^{\mathrm{lat}}$ and $f_\rho^{\mathrm{lat}}$ as
\begin{eqnarray}
\langle 0 | A_4| \pi \rangle &=&\sqrt{2} \, m_\pi F_\pi^{\mathrm{lat}}, \\
\langle 0 | V_k| \rho_\lambda \rangle &=& m_\rho f_\rho^{\mathrm{lat}} e_k(p,\lambda),
\end{eqnarray}
where $A_4=\bar{u}\gamma_4\gamma_5 d$ and $V_k=\bar{u}\gamma_k d$ are the non-singlet axial and vector currents, respectively, while $e_k(p,\lambda) $ denotes a polarization vector.

\begin{figure}
\begin{center}
\includegraphics[width=0.48\textwidth]{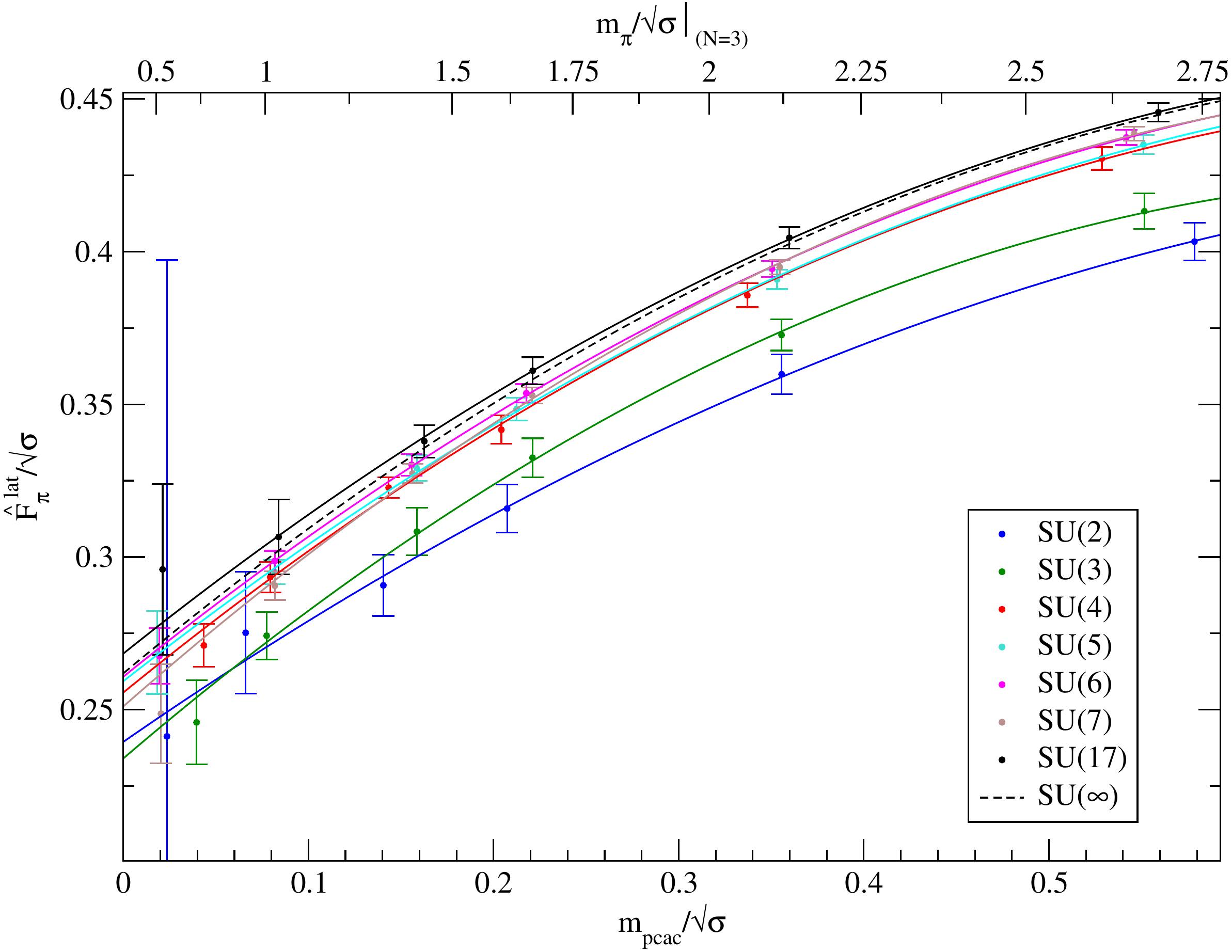} \includegraphics[width=0.48\textwidth]{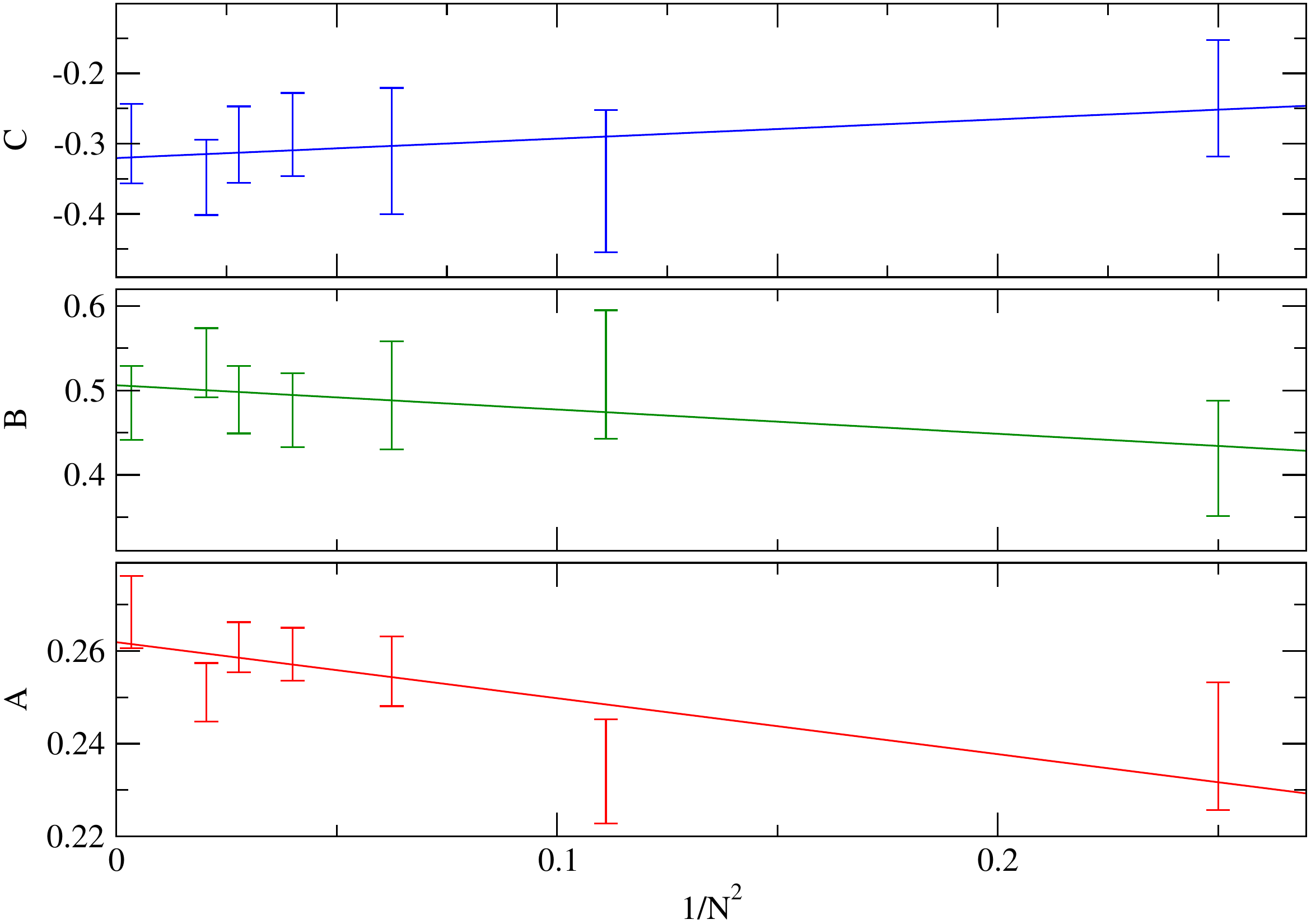}

\end{center}
\caption{Fit of the rescaled pion decay constant to: $\hat{F}_{\pi}^{\mathrm{lat}}/\sqrt{\sigma} = A + B \cdot \mpcac/\sqrt{\sigma} + C \cdot \mpcac^2/\sigma$, and $1/N^2$ fit of the resulting fit parameters $A$, $B$ and $C$.}
\label{fig:fpi}
\end{figure}

We fit  the pion and the axial correlators for large $t$ as:
\begin{align} 
\langle A_4(t) \pi_\alpha (0) \rangle &  \simeq \frac{1}{2 m_\pi} \langle 0 | A_4| \pi \rangle \langle \pi |  \pi_\alpha^\dagger | 0 \rangle e^{-t m_\pi}\equiv C_{A_4} e^{-t m_\pi}  \\
\langle  \pi_\alpha(t) \pi_\alpha (0) \rangle &  \simeq \frac{1}{2 m_\pi} \langle 0 |  \pi_\alpha| \pi \rangle \langle \pi |  \pi_\alpha^\dagger |0 \rangle e^{-t m_\pi}\equiv C_{\pi} e^{-t m_\pi}
\end{align}
(where $\pi_\alpha$ is one of the four differently smeared pion interpolators) and then compute $F_\pi^{\mathrm{lat}}$ as:
\begin{equation}
F_\pi^{\mathrm{lat}} = C_{A_4} \sqrt{\frac{1}{m_\pi  C_{\pi}}}.
\end{equation}
In order to better compare large-$N$ to $N=3$ results, we choose to normalize the decay constants as:
\begin{equation}
\hat{F}_\pi = {F}_\pi \sqrt{\frac{3}{N}}, \qquad \qquad
\hat{f}_\rho = {f}_\rho \sqrt{\frac{3}{N}}.
\end{equation} 
For the $\SU(3)$ theory, we find results which are consistent with ref.~\cite{Gockeler:1997fn}, while for larger gauge groups our measurements show that the expected $\sqrt{N}$ scaling behavior is well satisfied, see figure~\ref{fig:fpi} and tables~\ref{tab:su2table3}--\ref{tab:su17table3}. 

\begin{figure}
\begin{center}
\includegraphics[width=0.48\textwidth]{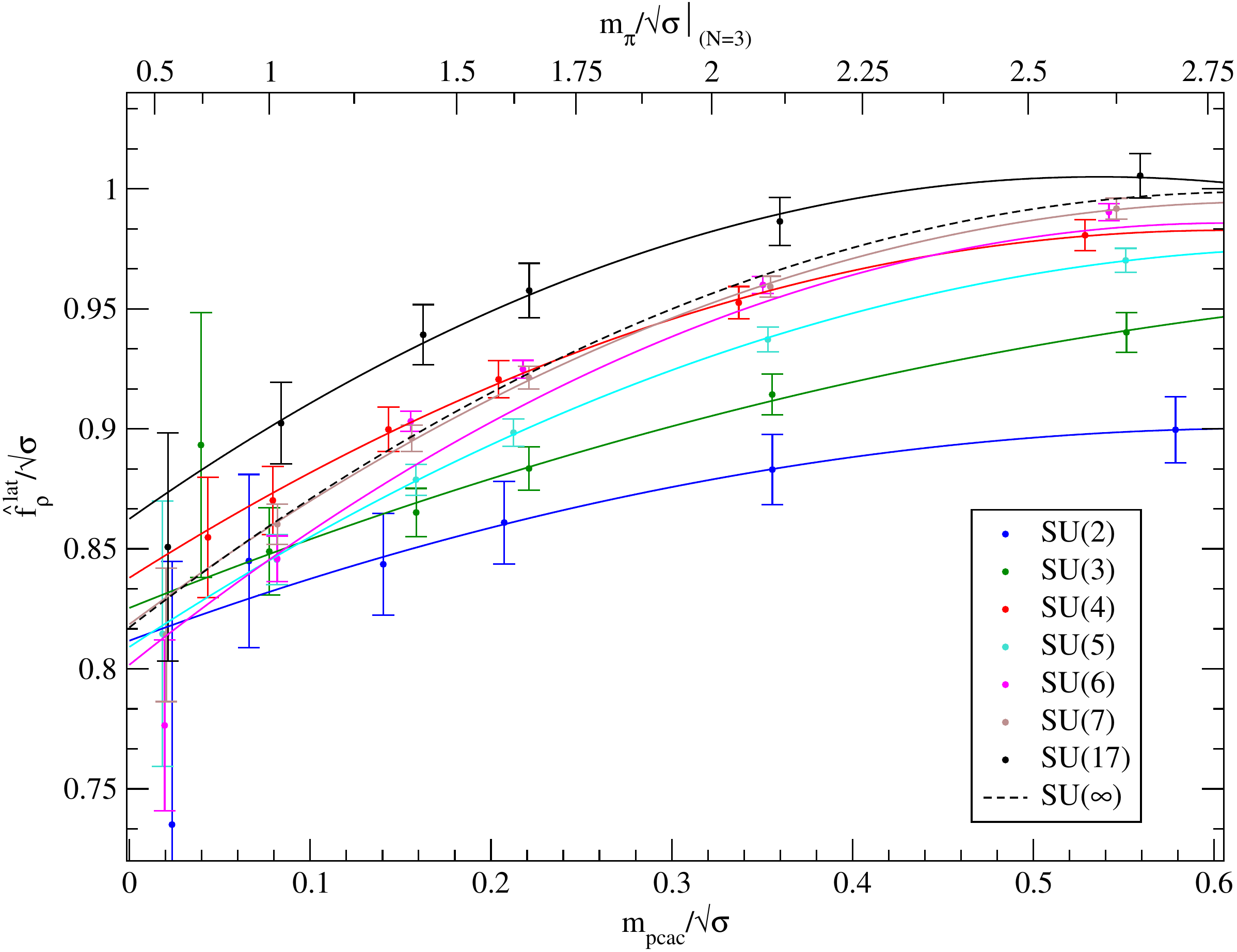} \includegraphics[width=0.48\textwidth]{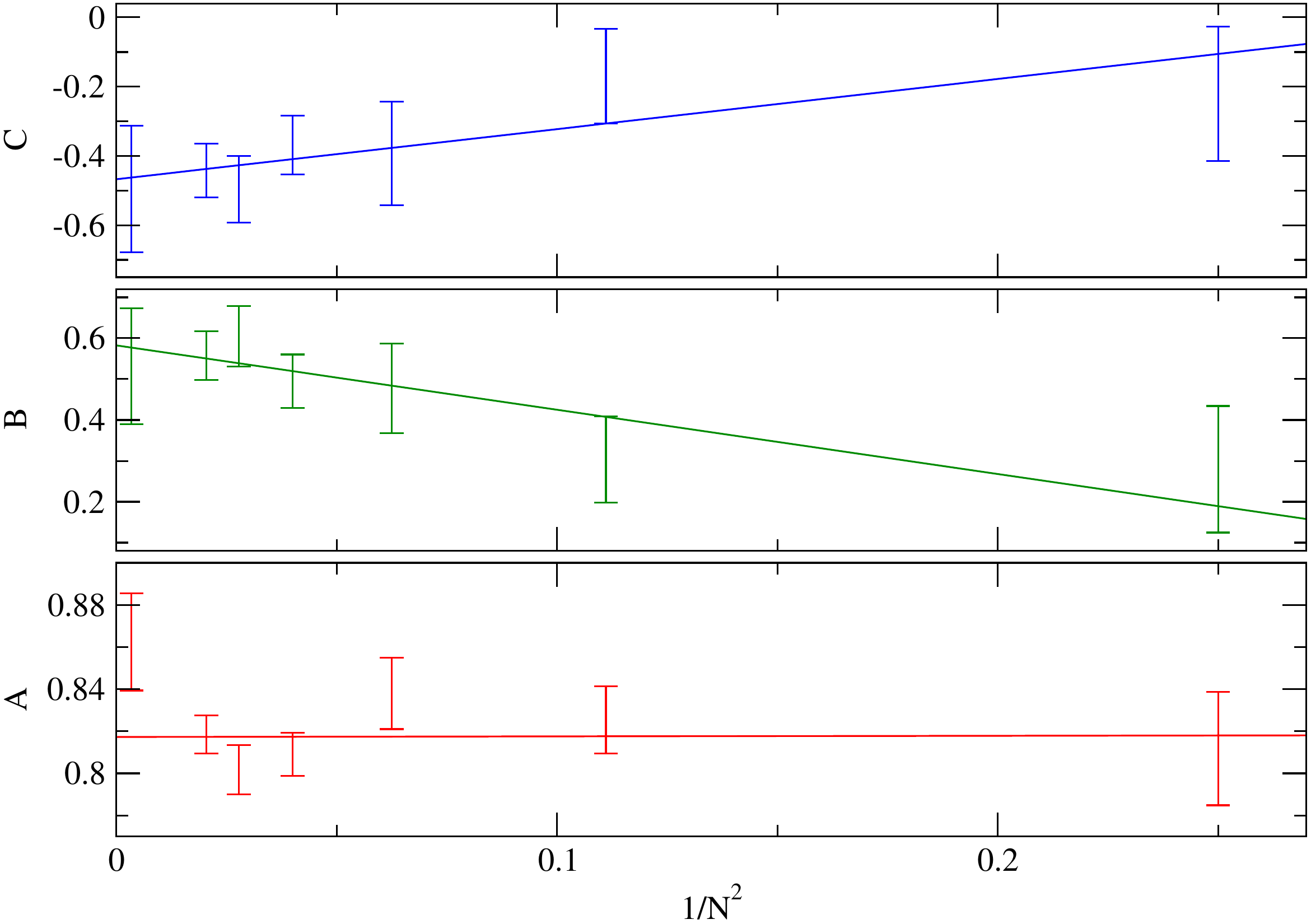}

\end{center}
\caption{Fit of the rescaled $\rho$ decay constant to: $\hat{f}_{\rho}^{\mathrm{lat}}/\sqrt{\sigma} = A + B \cdot \mpcac/\sqrt{\sigma} + C \cdot \mpcac^2/\sigma$ (left), together with the $1/N^2$ fit of the resulting parameters $A$, $B$ and $C$ (right).}
\label{fig:frho}
\end{figure}

The $1/N^2$ expansion of the (rescaled) pion decay constant reads:
\begin{eqnarray}
\frac{ \hat{F}_\pi^{\mathrm{lat}} }{\sqrt{\sigma}} &=& \left(  0.2619(37)    -\frac{0.121(56)}{N^2} \right) + \left(  0.506(24)    -\frac{0.29(30)}{N^2} \right) \frac{\mpcac}{\sqrt{\sigma}} \nonumber \\
& &+ \left(  -0.320(31)  +  \frac{0.28(37)}{N^2} \right) \frac{\mpcac^2}{\sigma}.
\end{eqnarray}
In the case of $f_\rho^{\mathrm{lat}}$, we use a similar approach---the main difference being due to presence of the polarization vector $e_k(p,\lambda)$. This satisfies the relation
\begin{equation}
\sum_\lambda e_\mu(p,\lambda) e_\nu (p,\lambda) =  g_{\mu\nu} -\frac{p_\mu p_\nu}{p^2},
\end{equation}
so that for zero momentum and for a fixed direction $\mu=\nu$ the above expression becomes one, and the computation is identical to the previous one (up to a $\sqrt{2}$ factor). To improve the statistical precision, we averaged the results over the three spatial directions. We plot the $\rho$ decay constants for different $N$ in figure~\ref{fig:frho}; the $1/N^2$ fits can be summarized as:
\begin{eqnarray}
\frac{\hat{f}_\rho^{\mathrm{lat}} }{\sqrt{\sigma}} &=& \left(  0.8173(70)    +\frac{0.00(11)}{N^2} \right) + \left(  0.582(44)    -\frac{1.57(63)}{N^2} \right) \frac{\mpcac}{\sqrt{\sigma}} \\ \nonumber 
&& + \left(  - 0.467(57)  +  \frac{1.45(80)}{N^2} \right) \frac{\mpcac^2}{\sigma}  .
\end{eqnarray}
The decay constants computed on the lattice are related to the ones in the continuum by the renormalization constants $Z_A$ and $Z_V$,
\begin{equation}
\hat{F}_\pi = Z_A \hat{F}_\pi^{\mathrm{lat}}, \qquad \qquad
\hat{f}_\rho = Z_V \hat{f}_\rho^{\mathrm{lat}}.
\end{equation}
$Z_A$ and $Z_V$ have been determined non perturbatively only for $N=3$ \cite{Gimenez:1998ue,Gockeler:1998ye} while two-loop perturbative results are known \cite{Skouroupathis:2007jd,Skouroupathis:2008mf} to converge slowly. In appendix \ref{sec:rencon} we detail our strategy to obtain estimates of $Z_A$ and $Z_V$ for $N \neq3$ and of their systematics. The results for $\hat{F}_\pi$ and $\hat{f}_\rho$ are listed in tables~\ref{tab:su2table3}--\ref{tab:su17table3}. 

\subsection{Finite volume effects}
\label{subsec:FSE}

Due to the Eguchi-Kawai volume reduction~\cite{Eguchi:1982nm}, finite size effects (FSE) are expected to be zero at infinite $N$, as long as all lattice dimensions (in physical units) are kept larger than a critical length $L_c$ \cite{Kiskis:2003rd}, so that center symmetry is not spontaneously broken. 

\begin{figure}
\begin{center}
\includegraphics[width=0.48\textwidth]{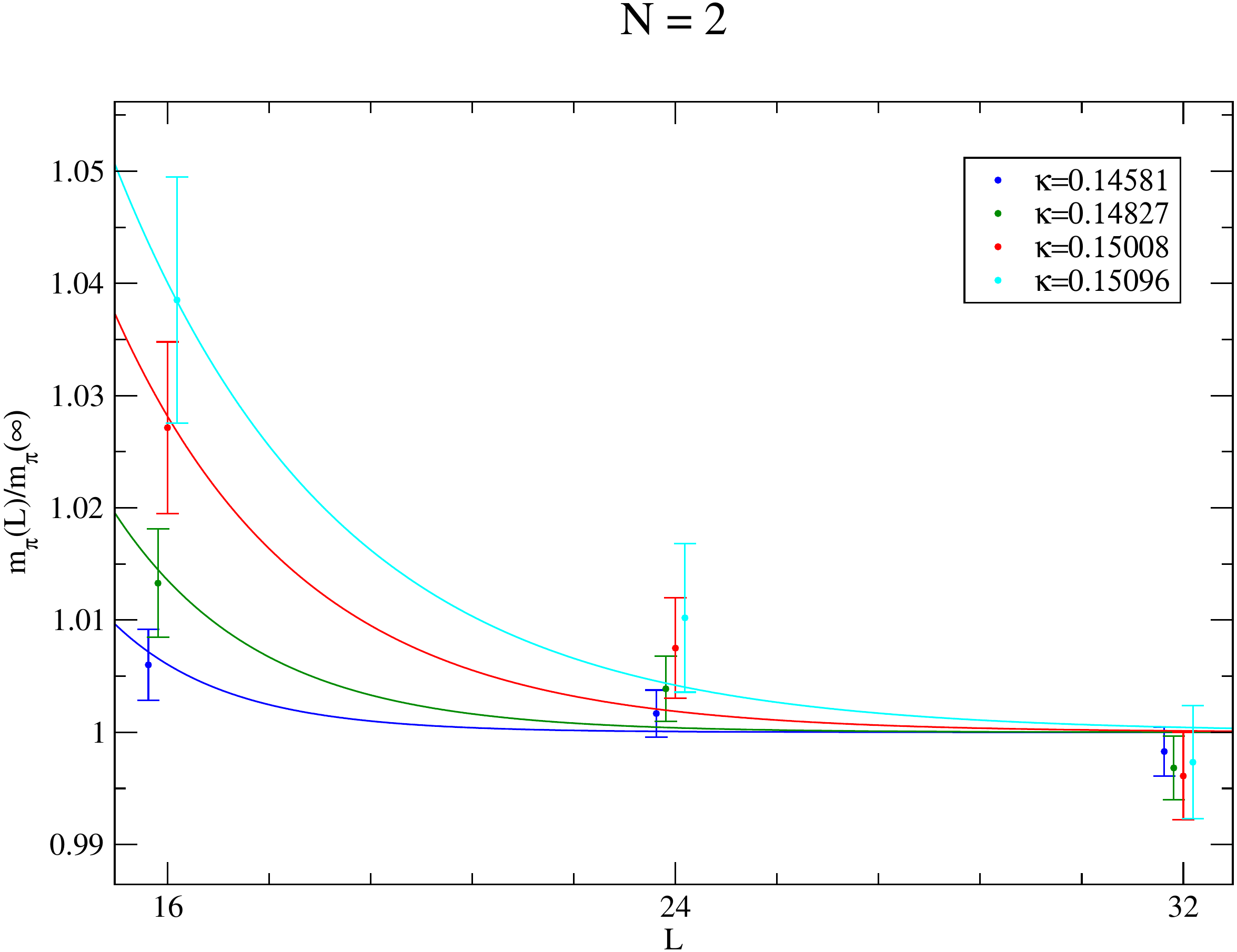} \includegraphics[width=0.48\textwidth]{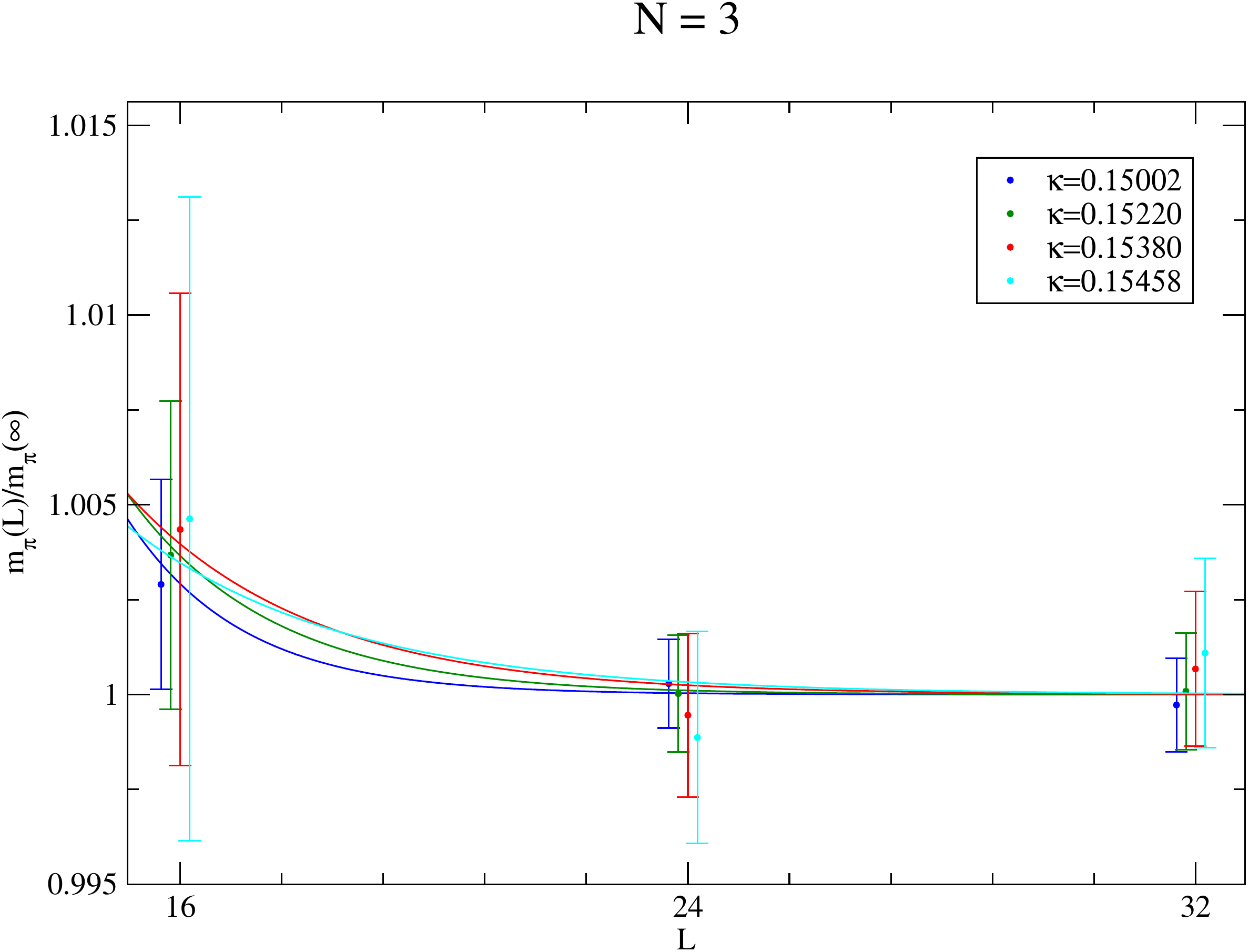}

\end{center}
\caption{Fit of $m_\pi(L) / m_\pi(\infty)$ according to eq.~(\ref{eq:fsepion})  for the $\SU(2)$ (left) and $\SU(3)$ theory (right). 
}
\label{fig:fse}
\end{figure}

At finite $N$, FSE become larger for smaller quark masses and for smaller $N$ \cite{Colangelo:2003hf,Colangelo:2005cg}:
\begin{equation}\label{eq:fsepion}
m_\pi(L) = m_\pi(\infty)  \left[1+ B  \exp (-m_\pi(\infty) L ) \right)],
\end{equation}
where the parameter $B$ vanishes in the large-$N$ limit. For $N=2$ and $3$, we carried out simulations at three volumes and fitted the pion masses to eq.~(\ref{eq:fsepion}), obtaining the results displayed in figure~\ref{fig:fse}. As one can see from these plots, FSE are drastically reduced going from 2 to 3 colors, where the data can already be fitted to a constant. For larger $N$-values we carried out simulations at two volumes only, unsurprisingly, without any evidence of FSE. While one may also carry out a similar analysis for the other particles, their finite size corrections are expected to be smaller than for the pion, and thus negligible within our statistical uncertainties. 

\subsection{Finite cut-off effects}
\label{subsec:cutoff_effects}

Before comparing our results to theoretical continuum predictions, it is important to discuss the potential impact of discretization effects on our lattice results. Strictly speaking, since our simulations were performed only at one value of the lattice spacing, it is not possible to perform a continuum extrapolation. However, following the discussion in ref.~\cite{Bali:2008an} (a study carried out at values of $a$ very close to ours), one can nevertheless get an estimate of the systematic uncertainty induced by the finite lattice cut-off, by comparing the results with those obtained at a different, coarser, lattice spacing in ref.~\cite{DelDebbio:2007wk}. In particular, in ref.~\cite{Bali:2008an} it was shown that, in spite of the $60\%$ difference between the two lattice spacings used in the two studies, the $\rho$ meson masses obtained are very close to each other (up to differences of the order of $5 \%$). Since our simulations employ unimproved Wilson fermions,
the leading lattice artefacts are expected to scale like $a$
(the same holds for refs.~\cite{DelDebbio:2007wk, Bali:2008an,Bali:2007kt} where the same action was used
and ref.~\cite{DeGrand:2012hd} where HYP-smeared clover-Wilson fermions with
tree-level coefficient were employed, resulting in $\mathcal{O}(\alpha a)$ lattice artefacts).
For the $\rho$ meson mass, the analysis carried out in ref.~\cite{Bali:2008an} showed that, by extrapolating their results and those from ref.~\cite{DelDebbio:2007wk} either linearly or quadratically in $a$, one obtains estimates of the continuum-limit value which are about $2\%$ to $5\%$ off from the result obtained at $a \simeq 0.09$~fm. In the absence of analogous data for the other states investigated in the present study, it is reasonable to take these numbers as a rough, order-of-magnitude estimate of the systematic finite cut-off errors affecting our results.

As a final remark, we observe that while taking the continuum limit can lead to a slight quantitative change in our results for physical quantities, it should not dramatically affect their (mild) dependence on $N$. In principle, interchanging the order of the continuum and large-$N$ limits of a lattice model may involve some subtleties (see ref.~\cite{Neuberger:2002bk} for a discussion). However, the results of previous studies indicate that, for lattice gauge theories and for the observables discussed here, the two limits commute~\cite{reviews,reviews-end}.

\subsection{Large-$N$ spectrum}
To conclude this section, we display in table~\ref{tab:largeNsummary} the results for the meson spectrum extrapolated to infinite $N$ at different quark masses. The results are listed in units of $\sqrt{\sigma}$ and in units of the (normalized) pion decay constant in the chiral limit,
\begin{equation}
\hat{F}_{\infty} = \left.\sqrt{\frac{3}{N}}F_\pi(m_q=0)\right|_{N\rightarrow\infty},
\end{equation}
which should be particularly useful for chiral
perturbation theory ($\chi$PT) applications~\cite{Pelaez:2006nj,Geng:2008ag,Nieves:2009ez,Nieves:2011gb} (see section~\ref{subsec:chipt}).

\begin{figure}
\begin{center}
\includegraphics[width=0.68\textwidth]{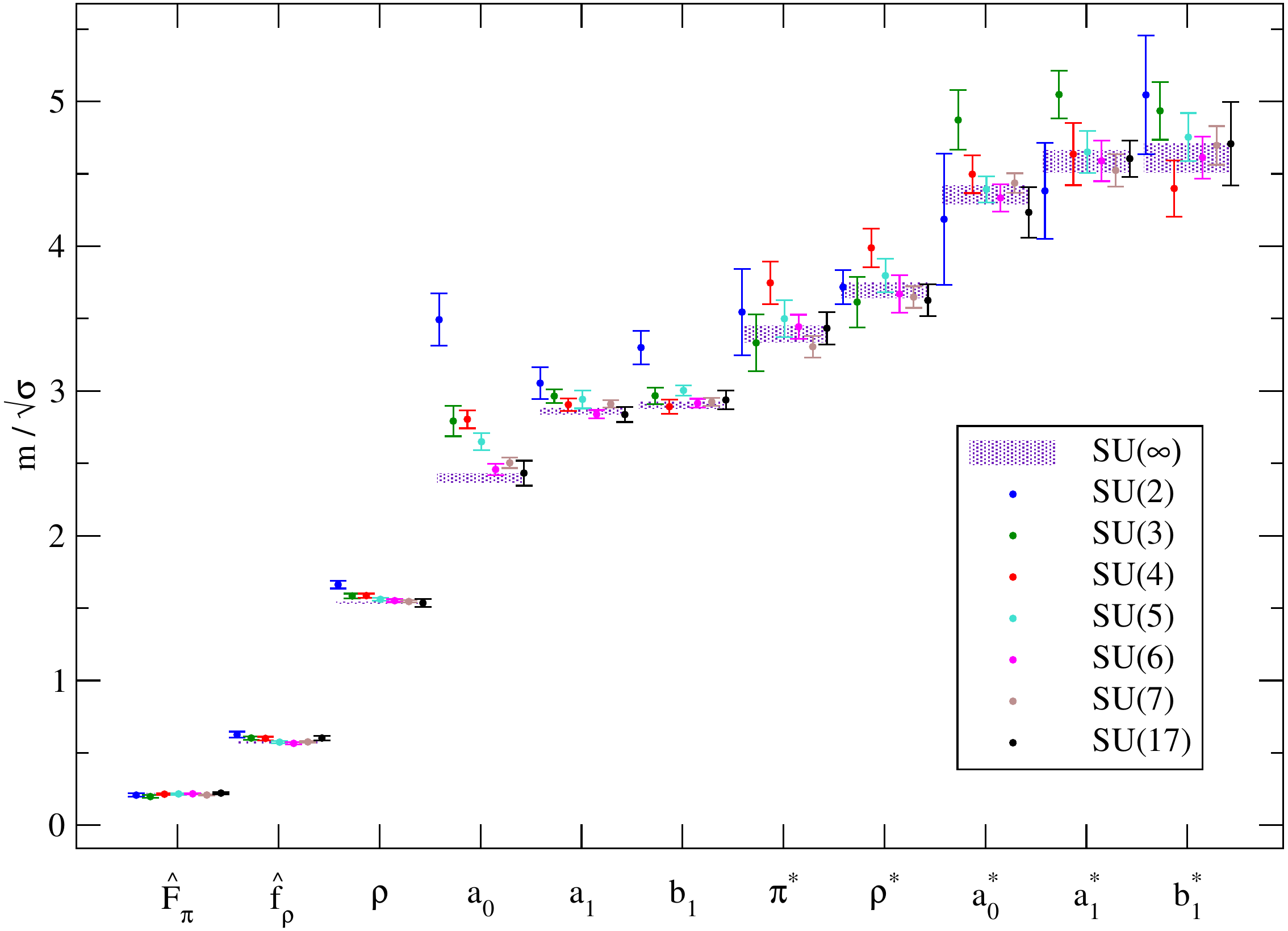} 

\end{center}
\caption{The meson spectrum for different $N$ in the chiral limit. The masses and the decay constants are given in units of the square root of the string tension for each group $\SU(N)$ and the extrapolated $N\rightarrow\infty$ values are shown as horizontal bands. }
\label{fig:global}
\end{figure}

\begin{figure}
\begin{center}
\includegraphics[width=0.48\textwidth]{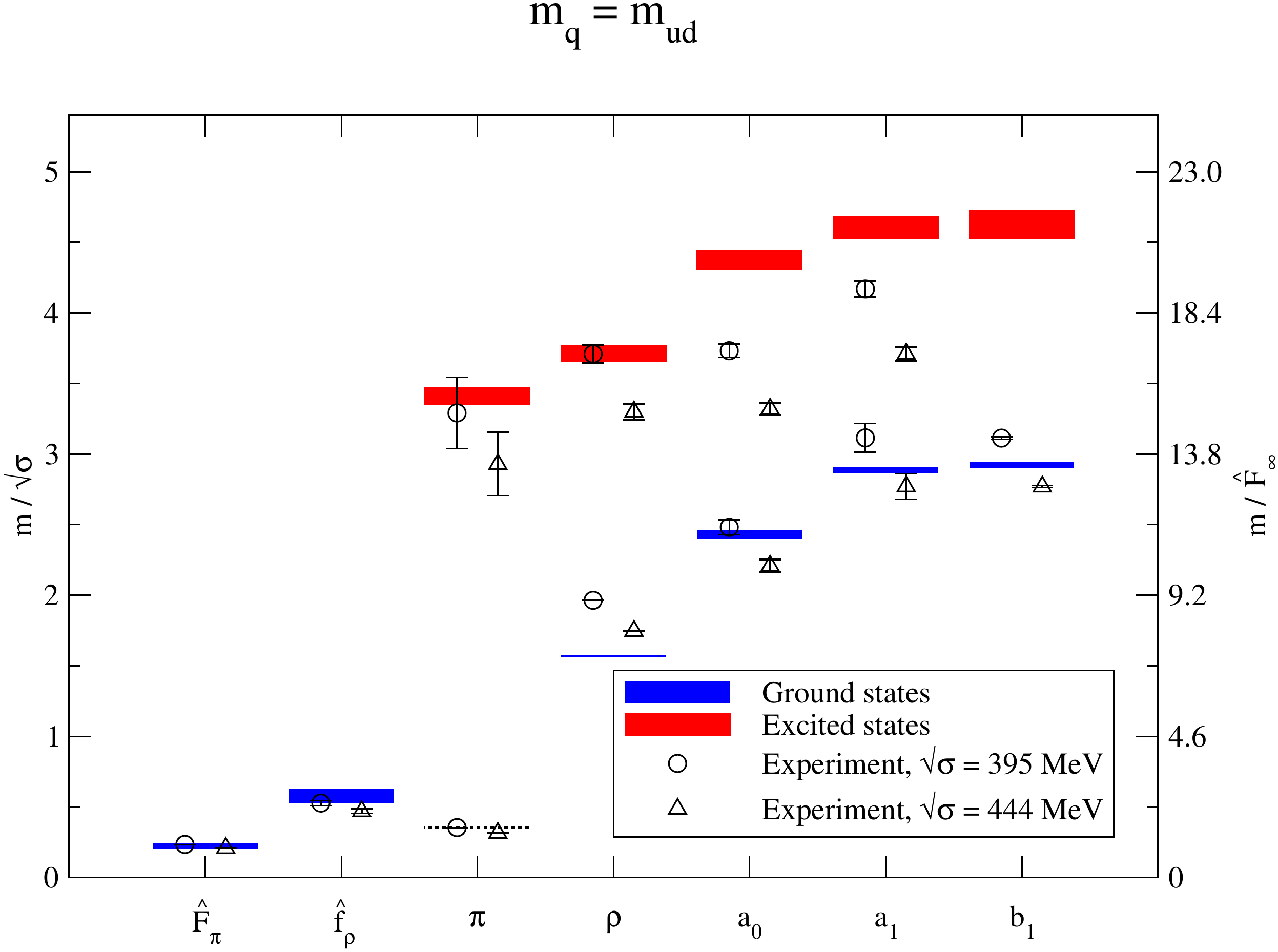} \includegraphics[width=0.48\textwidth]{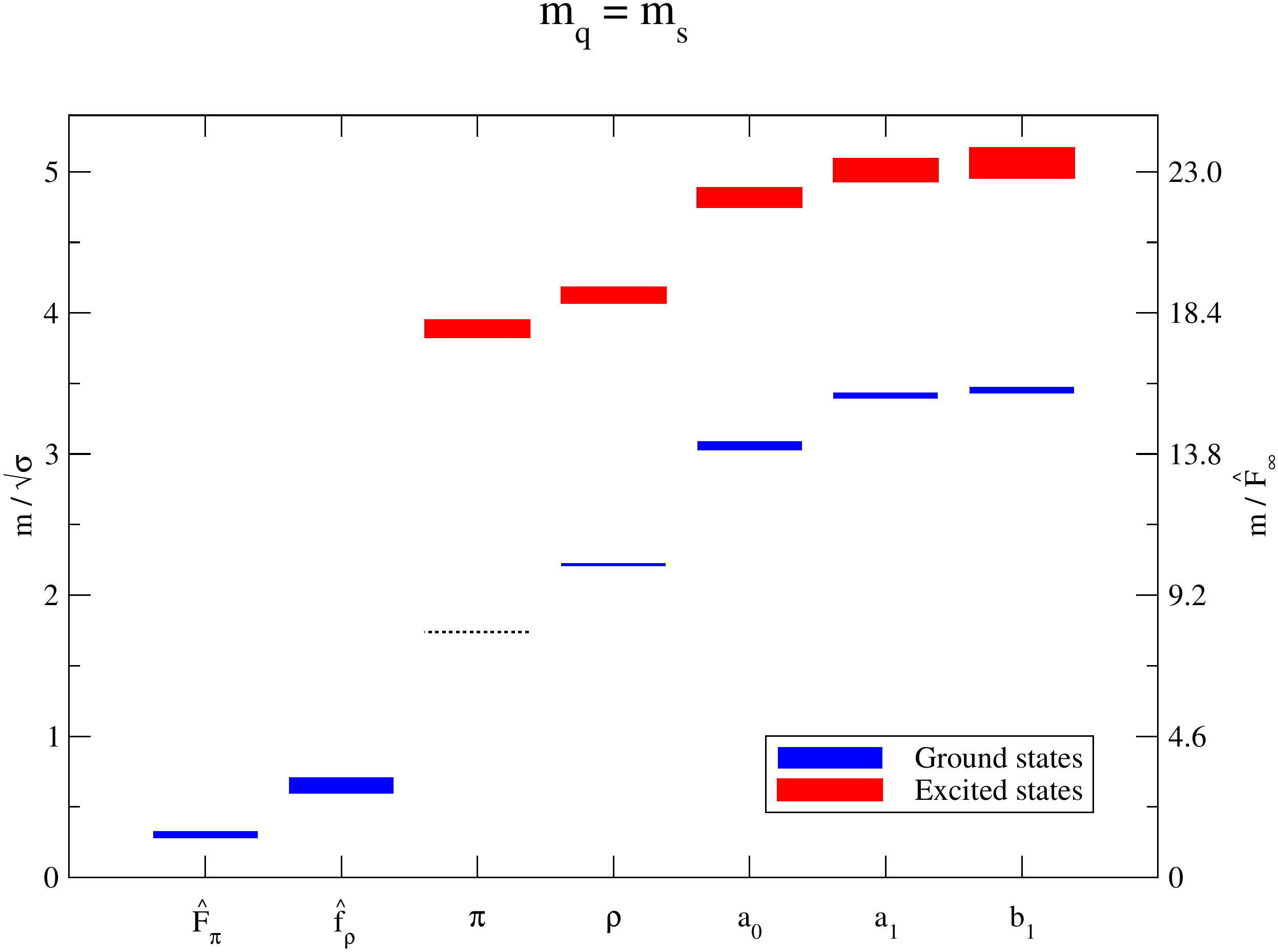}

\end{center}
\caption{Left: the large-$N$ meson spectrum at the $m_q=m_{ud}$ scale compared to the experimental  $N=3$ masses for the two values of the string tension: $\sigma = (395\mathrm{~MeV})^2$, obtained from imposing $\hat{F}_\infty = 85.9 \mathrm{~MeV}$, and the \emph{ad hoc} value $\sigma=(444\mathrm{~MeV})^2$. The pion mass is taken as input to set  $m_q=m_{ud}$.  Right: the large-$N$ meson spectrum at the $m_q=m_{s}$ scale. }
\label{fig:globalExp}
\end{figure}

Of phenomenological interest is not only the spectrum at
$m_q=0$ (figure~\ref{fig:global}) but are also the spectra at $m_q=m_{ud}$ and
at $m_q=m_s$ (figure~\ref{fig:globalExp}) where $m_{ud}$ and $m_s$ denote the
physical (isospin-averaged) light quark and strange quark
masses, respectively. We fix the former imposing at $N=\infty$
the values~\cite{Colangelo:2010et}:
\begin{eqnarray}
\hat{F}_\infty &=& 85.9 \mathrm{~MeV},\\
m_\pi(m_{ud}) &=& 138 \mathrm{~MeV}.
\end{eqnarray}
$\hat{F}_\infty$ fixes the lattice spacing to $a=0.10$~fm and the string tension to $\sigma=(395  \;\mathrm{ MeV})^2$. This value leads to a better agreement between the large-$N$ meson spectrum and the experimental masses than the previously used estimate $\sigma=(444  \;\mathrm{ MeV})^2$ (see figure~\ref{fig:globalExp}--left). We find the ratio
\begin{equation}\label{FpiRatio}
\frac{\hat{F}_{\pi}(m_{ud})}{\hat{F}_\pi(0)}=1.020(20)\,,
\end{equation}
where the renormalization constants cancel, to be compared to the value $F_\pi(m_{ud})/F_\pi(0)=1.073(15)$ for $N=3$ QCD with sea quarks~\cite{Colangelo:2010et}.
In the large-$N$ limit the mass-dependence of $F_\pi$ appears to be reduced.

The strange quark mass is obtained by fixing at $N=\infty$ the mass 
of a (fictitious) strange-antistrange pion to the value
\begin{equation}
m_\pi(m_{s}) = (m^2_{K^{\pm}}+m^2_{K^0}-m^2_{\pi^{\pm}})^{1/2}\approx 686.9\,\mathrm{MeV}.
\end{equation}
The large-$N$ meson spectrum at $m_q=m_s$ is shown in figure~\ref{fig:globalExp}~(right).

Note that our way of fixing $m_{ud}$ and $m_s$ is arbitrary and different choices of input observables of real $N=3$ QCD may give values that differ by $\mathcal{O}(1/N)$ corrections.
In fact the procedure of adjusting the quark masses described here differs from that of ref.~\cite{Bali:2013fpo}.

Setting the scale from $\hat{F}_\infty = 85.9\mathrm{~MeV}$ and fixing the quark masses as detailed above, we find the values $m_\rho = 619(9)\mathrm{~MeV}$ at $m_q=m_{ud}$ and $876(13)\mathrm{~MeV}$ at $m_q=m_{s}$, to be compared to $m_\rho = 775\mathrm{~MeV}$ and $m_\phi = 1019\mathrm{~MeV}$ in QCD. For the scalar mesons we find $m_{a_0} = 959(18)\mathrm{~MeV}$ and $1208(22)\mathrm{~MeV}$ for light and strange quarks, respectively. Note that the errors stated do not include an
overall scale uncertainty of 8\%, due to the renormalization of the pion decay constant.

\begin{table}
\begin{center}
\begin{tabular}{|c|c|c|c|c|c|c|c|} 
\cline{3-8} \multicolumn{2}{c|}{}  &\multicolumn{3}{|c|}{  ${m_\infty}/{ \sqrt{\sigma}}$}&\multicolumn{3}{|c|}{${m_\infty }/{\hat{F}_\infty} $ }\\ \hline
Particle &$J^{PC}$&$m_q = 0$ & $m_q=m_{ud}$ &$m_q=m_{s}$ &$m_q = 0$ &$m_q=m_{ud}$&$m_q=m_{s}$\\ \hline
$\pi$ &$0^{-+}$  & 0 & 0.35 & 1.74 & 0 & 1.61 & 8.00 \\ \hline
$\rho$ &$1^{--}$ & 1.5382(65) & 1.5683(65) & 2.216(11) & 7.08(10) & 7.21(10) & 10.20(15) \\ \hline
$a_0$ &$0^{++}$  & 2.401(31) & 2.428(31) & 3.059(33) & 11.04(21) & 11.17(21) & 14.07(25) \\ \hline
$a_1$ &$1^{++}$ & 2.860(21) & 2.883(21) & 3.414(23) & 13.16(21) & 13.26(21) & 15.71(24) \\ \hline
$b_1$ &$1^{+-}$  & 2.901(23) & 2.924(23) & 3.452(24) & 13.35(21) & 13.45(22) & 15.88(25) \\ \hline
$\pi^\star$ &$0^{-+}$ & 3.392(57) & 3.413(57) & 3.887(60) & 15.61(34) & 15.70(34) & 17.88(37) \\ \hline
$\rho^\star$ &$1^{--}$ & 3.696(54) & 3.714(54) & 4.127(56) & 17.00(34) & 17.08(34) & 18.99(37) \\ \hline
$a_0^\star$ & $0^{++}$ & 4.356(65) & 4.375(65) & 4.816(69) & 20.04(41) & 20.13(41) & 22.16(44) \\ \hline
$a_1^\star$ &$1^{++}$  & 4.587(75) & 4.605(75) & 5.012(81) & 21.10(46) & 21.18(46) & 23.06(49) \\ \hline
$b_1^\star$ &$1^{+-}$ & 4.609(99) & 4.628(99) & 5.06(11) & 21.20(54) & 21.29(55) & 23.29(58) \\ \hline
$\hat{F}_\pi $& -  & 0.2174(30) & 0.2216(30) & 0.3035(58) & 1 & 1.020(20) & 1.396(33) \\ \hline
$\hat{f}_\rho$ & -  & 0.5721(49) & 0.5762(49) & 0.6516(93) & 2.632(43) & 2.651(44) & 2.998(60) \\ \hline
\end{tabular}
\end{center}
\caption{The $N=\infty$ meson spectrum and decay constants in units of the square root of the
string tension $\sqrt{\sigma}$ and in units of the (normalized)
chiral pion decay constant $\hat{F}_{\infty}= F_\pi(0)\sqrt{3/N}$ in the chiral limit,
for three different values of the quark mass which are set using the pion masses as inputs. We also display
$\hat{F}_{\pi}=F_{\pi}\sqrt{3/N}$ and
$\hat{f}_{\rho}=f_{\rho}\sqrt{3/N}$ in the last rows
of the table.
The results have not been extrapolated to the continuum limit. To
account for this, based on \cite{Bali:2008an}, a systematic error
of 5\% should be associated to all values. Regarding the pion
decay constant, due to the lack of non-perturbative renormalization 
at $N=\infty$, an error of 8\%  should be associated to the values in units of $\hat{F}_\infty$ and to the decay constants in units of the string tension.  \label{tab:largeNsummary} }
\end{table}

\section{Comparison with other lattice studies}
\label{sec:rho}
The mass of the $\rho$ meson has generated some controversies in the literature: refs.~\cite{DelDebbio:2007wk, Bali:2008an,Bali:2007kt, DeGrand:2012hd} find a value that is close to the one measured in $\SU(3)$, while ref.~\cite{Hietanen:2009tu} finds a mass approximately twice as large. In this work, we confirm the results
of refs.~\cite{DelDebbio:2007wk, Bali:2008an,Bali:2007kt, DeGrand:2012hd}. Nevertheless, the discrepancy with ref.~\cite{Hietanen:2009tu} needs to be taken very seriously and properly addressed, since the results of ref.~\cite{Hietanen:2009tu} question not only the value of the $\rho$ mass, but (indirectly) the whole meson spectrum obtained in this work. The issue is not straightforward to settle, since the method used to address the problem in ref.~\cite{Hietanen:2009tu} is different from the one in refs.~\cite{DelDebbio:2007wk,  Bali:2008an,Bali:2007kt,DeGrand:2012hd}. Another related issue concerns $F_{\pi}(N=\infty)$~\cite{Narayanan:2005gh,Allton:2008pn,Baron:2010bv,Bazavov:2010hj} that was used in~\cite{Hietanen:2009tu} to set the physical scale. Our result, indicated in table~\ref{tab:largeNsummary}, is
\begin{equation}
  \label{eq:fpi:our}
  \hat{F}_{\infty}/\sqrt{\sigma} = 0.2174(30) \ ,
\end{equation}
while from~\cite{Narayanan:2005gh} one infers
\begin{equation}
  \label{eq:fpi:nn}
  \hat{F}_{\infty}/\sqrt{\sigma} = 0.277(12) \ ,
\end{equation}
where the known value~\cite{Lucini:2012wq} $T_c/\sqrt{\sigma} = 0.5949(17)$ has been used to convert the original results (provided in units of the deconfinement critical temperature $T_c$) into units of $\sqrt{\sigma}$. Note that in both estimates we have considered only the statistical error. When also the systematic errors related to the finite lattice spacing and (for the calculation described in this work) to the renormalization constant are taken into account, the results agree within two standard deviations, showing an acceptable level of consistency. Instead, the discrepancy of the results for the mass of the $\rho$ meson demands a thorough inspection of the numerical techniques and of the calculations that lead to the conflicting numerical values.

Refs.~\cite{DelDebbio:2007wk, Bali:2008an,Bali:2007kt,DeGrand:2012hd} use the conventional techniques we described in section \ref{sec:latticesetup}
 to extract the mass of the $\rho$ from the long-distance behavior of correlators of fermion bilinears carrying the quantum numbers of the state of interest,  at relatively low $N$ (up to 7); this method has been well tested in standard QCD calculations. On the other hand, the authors of~\cite{Hietanen:2009tu} extract the mass at larger $N$ ($N = 17$ and $N=19$), but using smaller lattice sizes, from momentum space correlators which are obtained injecting allowed discrete momenta in the lattice configurations. Although the quenched momentum technique used in~\cite{Hietanen:2009tu} is less well tested, one cannot exclude the possibility that it gives better results at large $N$. Also note that the fermion discretizations differ: Wilson fermions were employed in~\cite{DelDebbio:2007wk, Bali:2008an,Bali:2007kt} and in the present work, clover Wilson fermions in~\cite{DeGrand:2012hd} and overlap fermions in~\cite{Hietanen:2009tu}.  

Taking the technical differences into account, in principle the discrepancy may have several interpretations: 
\begin{itemize}
\item it may be due to large artifacts in either of the lattice computations (which are performed at finite quark masses and lattice spacings, and use different discretizations for the Dirac operator---characterized, in particular, by different chiral properties);
\item the quenched momentum computation in refs.~\cite{Hietanen:2009tu} is based on an expected cancellation of finite-volume effects in the large-$N$ limit. If such cancellation is not complete, the numerical results could have been contaminated by the associated systematic effect.\footnote{Although, as we have already discussed, we do find some level of consistency for $\hat{F}_{\infty}$, it is possible that the technique has different systematics depending on the quantum numbers of the meson.} A (possibly related) issue could be the mixing with some excited state(s), given that ref.~\cite{Hietanen:2009tu} extracts the meson spectrum by computing quark propagators at a small number of low momenta; 
\item large-$N$ gauge theories have a complicated phase structure~\cite{Kiskis:2003rd}, which might create metastabilities. It is hence possible that the two calculations are in different phases, with (possibly) only one being in the phase connected with the infinite volume limit;
\item the discrepancy may actually indicate a ``physical'' effect: the results reported in ref.~\cite{Hietanen:2009tu} are obtained at much larger values of $N$ than those investigated in refs.~\cite{DelDebbio:2007wk, Bali:2008an,Bali:2007kt,DeGrand:2012hd}, and it may happen that, for a moderate number of colors, the finite-$N$ correction coefficients conspire to hide the deviations from the large-$N$ limit, which might become visible only for much larger values of $N$. This interpretation, however, is at odds with the very smooth $N$-dependence of all other observables that have been investigated so far. 
\end{itemize}
Since it is of paramount importance to resolve the root causes of the discrepancy, we performed a dedicated $\SU(17)$ calculation at values of the parameters that are very similar to those of ref.~\cite{Hietanen:2009tu} ($\beta=208.45$ vs. $\beta=208.08$), to enable a direct comparison. Interestingly, setting the scale by the string tension, it turns out that refs.~\cite{Bali:2008an,DeGrand:2012hd} and the present study use a similar lattice spacing as~\cite{Hietanen:2009tu}. From this data set, we obtained values of the $\rho$ mass that are compatible with our results at lower $N$.  Finally, we have monitored the average local value of the Polyakov loop and of the plaquette in the four directions, to check that the whole lattice is in the phase relevant for the continuum limit of the large-$N$ theory. One might object that since fluctuations are reduced at large $N$, extracting masses with connected correlators might not be the correct strategy. This issue has been addressed in~\cite{Bali:2008an}, where it has been shown that for mesons the relative precision of the mass in fact improves as $N$ increases.

As a consequence of this analysis, where we find $N=17$ results to be in good agreement with $N\leq7$, the possible sources of the discrepancy can be narrowed down to the different fermion discretization, the different method for extracting the mass and the influence of the bulk phase on the numerical results. Concerning this last point, we remark that while in our calculation we have kept fixed a physical scale (provided in our case by the string tension), in~\cite{Narayanan:2005gh,Hietanen:2009tu} $\beta$ is varied across the gauge groups using the large-$N$ perturbative $\beta$-function. This might underestimate the value of $\beta$ at which the bulk is present at large-$N$ (one can compare the $\beta$ used in~\cite{Narayanan:2005gh} with the investigation of the bulk phase reported in~\cite{Lucini:2005vg} to realise that there could be an influence of this unwanted phase on some of the results of ~\cite{Narayanan:2005gh}). This could be the source of the small difference in $\hat{F}_{\infty}$. In fact, it was noticed already in~\cite{Hietanen:2009tu} that $b = \beta/(2 N^2) = 0.35$, which was crucial for obtaining the value of $\hat{F}_{\infty}$ in~\cite{Narayanan:2005gh}, gives a variation of the mass of the $\rho$ of around 30\% with respect to the value obtained at $b = 0.36$~$(\beta=208.08)$, and this variation was ascribed to bulk effects. It is then reasonable to assume that the same effects might have influenced the extraction of $\hat{F}_{\infty}$ in~\cite{Narayanan:2005gh}. 

To sum up, while the techniques described in this work are well established, with known systematic effects (which we have discussed in detail), those of~\cite{Narayanan:2005gh,Hietanen:2009tu} are new and further work is needed before the reliability of the corresponding results can be fully assessed. Should future investigations confirm them, clearing the doubts raised in this paper, the implications will be instructive and a new intriguing picture of the relationship between QCD and the large-$N$ limit might emerge. In the absence of such an analysis and supported by the numerical results for $N = 17$ obtained  in this work, 
the numerical evidence that we have presented shows a large-$N$ meson spectrum and decay constants that are close to the SU(3) ones, with the $N = 3$ values described by modest $1/N^2$ corrections to the observables at $N = \infty$.

\section{Comparison with analytical predictions}
\label{sec:postdictions}

In this section, we relate our numerical results to analytical predictions. We mainly focus on calculations derived in the context of gauge/gravity models (subsection~\ref{subsec:holography}), and briefly mention some of those in chiral perturbation theory (subsection~\ref{subsec:chipt}). In the literature there exist also analytical large-$N$ studies of the spectrum of mesons with different radial and/or spin quantum numbers (e.g. ref.~\cite{Masjuan:2012gc}). However, our numerical results are limited to ground states and first radial excitations.

\subsection{Holographic models}
\label{subsec:holography}

Part of the motivation for studying the meson spectrum in the large-$N$ limit comes from the Maldacena conjecture~\cite{Maldacena:1997re,Gubser:1998bc,Witten:1998qj}, namely from the expectation that gauge theories admit a dual description in terms of string theories, defined in a higher-dimensional spacetime. In particular, the ``holographic dictionary'' relating quantities between the two types of theories states that, when the number of color charges in the gauge theory tends to infinity, the string coupling $g_s$ in the dual string theory tends to zero. Under these conditions, loop effects on the string side of the correspondence can be neglected, i.e. the theory reduces to its ``classical string'' limit. (In addition, the limit in which the gauge theory is strongly coupled corresponds to the limit in which the string length tends to zero. Hence, if the gauge theory has a large number of colors \emph{and} is strongly coupled, its string dual reduces to a classical gravity theory, in an appropriate, curved, higher-dimensional spacetime, opening up the possibility of analytical treatment).

The first explicit version of this conjecture was formulated by studying a system of $N$ D3 branes (i.e. $3+1$-dimensional hyperplanes) in type IIB string theory, defined in a ten-dimensional spacetime given by the direct product of a five dimensional anti-de~Sitter spacetime with a five-dimensional sphere, AdS$_5 \times S^5$: it was realized that the low-energy dynamics of such a system can be equivalently described in terms of supersymmetric Yang-Mills (SYM) theory, with $\U(N)$ gauge group and $\mathcal{N}=4$ supercharges, and that the two theories share the same global symmetries. The parameters of the gauge theory (the number of colors $N$ and the 't~Hooft coupling $\lambda$) are mapped to those of the string theory (the string coupling $g_s$, and the ratio of the string length $l_s$ to the radius of the space $R$) as:
\begin{equation}
\lambda/N = 4\pi g_s ,\;\;\; \lambda = (l_s/R)^{-4}.
\end{equation}

For the $\mathcal{N}=4$ SYM theory, the validity of the gauge/string correspondence is supported by many pieces of mathematical evidence, and no counter-examples to it have ever been found (although a general proof has not been formulated yet). \emph{If} the correspondence is indeed true, and generic (i.e., if \emph{any} gauge theory admits a holographic description in terms of a string model), then it would be possible to study the strong-coupling regime of gauge theories (at least of those with a large number of colors) via simple, ``classical'' calculations on the string theory side. This possibility has motivated a huge theoretical effort, aimed at extending the holographic correspondence to theories which are more ``QCD-like''---including, in particular, theories with matter fields in the fundamental representation of the gauge group, exhibiting a non-trivial running coupling, with a spectrum of confined states and spontaneous chiral symmetry breaking at low energies. This topic is extensively discussed in ref.~\cite{Erdmenger:2007cm}; in the following, we summarize the main issues relevant for our present discussion, and for a comparison with lattice results.

Before reviewing the holographic approach to meson spectra in the large-$N$ limit, however, we would like to warn the reader that, while conceptually extremely interesting, these types of computations are not yet at a level of accuracy which enables precise comparison with lattice results (or with experimental data). Although the gauge/string correspondence has led to dramatic theoretical progress in several aspects of strongly coupled field theories, its application to \emph{quantitatively} address questions of a direct, phenomenological relevance is still limited. For these reasons, it should be understood that the theoretical predictions discussed in this context have to be taken \emph{cum grano salis}.

First of all, it is worth emphasizing that the $\mathcal{N}=4$ SYM theory, at least at zero or low temperatures, is \emph{qualitatively} very different from QCD: in particular, it is non-confining, it does not have a chiral condensate nor a discrete spectrum, it is maximally supersymmetric, conformally invariant, and it does not have elementary fields in the fundamental representation of the gauge group. One possible way to modify the  AdS/CFT correspondence, in order to obtain a gauge theory with features closer to those of QCD is based on the so-called ``top-down'' approach, i.e., on a deformation of the original AdS/CFT setup with the addition of extra ingredients. In particular, in order to have a gauge theory with fields in the fundamental representation of the gauge group, one can modify the dual string model by adding a set of $n_f$ ``flavor'' D7 branes~\cite{Karch:2002sh} (see also, e.g., ref.~\cite{Babington:2003vm} for a discussion): open strings stretching from the original D3 branes to one of these D7 branes correspond to fundamental matter fields in the gauge theory, with a mass proportional to the separation between the D7 branes and the D3 branes. This reduces the supersymmetry of the corresponding gauge theory down to $\mathcal{N}=2$, and leads to a global $\U(n_f)$ ``flavor'' symmetry (which is the remnant of the gauge symmetry on the D7-branes). On the other hand, open strings that connect two D7 branes are interpreted as the holographic duals of ``mesons''. In this context, ``mesons'' denote tightly
bound states of a quark and its antiquark.

Assuming that the number $n_f$ of D7 branes remains finite, and hence much smaller than the number $N$ of D3 branes (which is taken to infinity), one can work in the so-called ``probe approximation''~\cite{Karch:2002sh}, in which the D7 branes do not affect the geometry of the spacetime in which the string theory is defined. It is interesting to note that, in the dual gauge theory, this corresponds to a ``quenched approximation'' (which, historically, has been used for a long time also in numerical simulations of lattice QCD~\cite{Aoki:1999yr}), in which quarks are propagating in a background generated by the gauge fields only. This approximation becomes exact in the 't~Hooft limit, in which the $1/N$ suppression of virtual quark loops leads to a unitary, quenched theory, whose dynamics is purely determined by planar gluon loops.

Within the probe approximation in holography, one can easily derive the structure of the meson spectrum: for quarks of non-vanishing mass $m_q$ (that is, for a non-vanishing separation between the D3 and the D7 branes), one finds the spectrum to be~\cite{Kruczenski:2003be}:
\begin{equation}
\label{meson_excitation_spectrum_probe_approximation}
M^{(n)} = 4 \pi m_q \sqrt{\frac{(n+1)(n+2)}{\lambda}},
\end{equation}
where $n$ denotes the quantum number describing the radial excitation of the meson.
Some observations are in order:
\begin{enumerate}
\item eq.~(\ref{meson_excitation_spectrum_probe_approximation}) shows that, in the limit of light quarks, the meson mass vanishes---and it does so linearly in $m_q$. This is in contrast to the situation in QCD, where, in general, in the chiral limit mesons remain massive (due to the existence of a mass gap) or, in the case of pseudoscalar mesons which are the Nambu-Goldstone bosons associated with the spontaneous breakdown of chiral symmetry, with masses that vanish like $\sqrt{m_q}$; 
\item the dependence of $M^{(n)}$ on $n$ described by eq.~(\ref{meson_excitation_spectrum_probe_approximation}) is different from the one characteristic of Regge trajectories ($M^{(n)} \propto \sqrt{n}$):  in principle, the correct functional dependence characterizing the spectrum in the large-$N$ limit could unambiguously be identified through lattice calculations. In practice however it turns out that the lattice determination of even the first few excited states is a technically very demanding task (see ref.~\cite{Morningstar:2008mc} and references therein for a discussion), hence in this paper we do not attempt a systematic computation in this direction, and limit our investigation to the first excited state in some channels only;
\item at finite quark mass and finite but large 't~Hooft coupling $\lambda$, eq.~(\ref{meson_excitation_spectrum_probe_approximation}) implies that the mass of the lightest meson is finite, but parametrically suppressed with respect to $m_q$:
\begin{equation}
\label{meson_ground_state_probe_approximation}
m_\pi^{(0)} = 4 \pi m_q \sqrt{\frac{2}{\lambda}},
\end{equation}
and, hence, that the constituent quark and antiquark form a tightly bound state.
\end{enumerate}

With the caveats mentioned above, it is nevertheless possible to consider a quantitative comparison of our results with holographic predictions in the probe approximation. As an example, we consider the dependence of the $\rho$ vector meson mass on the mass of the pion, which can be compared with the prediction derived in ref.~\cite{Babington:2003vm} in a setup involving a background with a non-constant dilaton~\cite{Constable:1999ch}:
\begin{equation}
\label{slope_prediction}
\frac{m_\rho(m_{\pi})}{m_{\rho}(0)} \simeq 1+0.307 \left[ \frac{m_\pi}{m_{\rho}(0)}
\right]^2.
\end{equation}
Our numerical result for the same quantity in the large-$N$ limit, reported in eq.~(\ref{eq:slopeRho1}), is remarkably close to the above expectation.

The inclusion of backreaction effects of the D7 branes on the spacetime geometry makes the meson spectrum computation more complicated, but it is still possible to study the meson mass dependence for a probe brane in a background which includes curvature effects due to the flavor branes. The result reads:
\begin{equation}
\label{meson_excitation_spectrum_backreaction}
M^2 \propto (n+1)(n+2) m_q^2 \ln \left( m_q^2 \right).
\end{equation}

Various authors have proposed different constructions, characterized by less super\-symmetry---see, e.g., refs.~\cite{Klebanov:1998hh, Polchinski:2000uf, Maldacena:2000yy, Sakai:2003wu, Sakai:2004cn}. In particular, ref.~\cite{Sakai:2004cn} discusses a setup involving a set of D4, D8 and $\overline{\mbox{D8}}$ branes, with a compactified direction of radius inversely proportional to a characteristic Kaluza-Klein scale, and derives a dual description featuring massless pions and interesting predictions for other meson masses. According to this construction, the ratio between the squared masses of the states with the quantum numbers corresponding to the $a_1(1260)$ and the $\rho$ mesons turns out to be:
\begin{equation}
\label{Sakai_Sugimoto_a_1_over_rho}
\frac{m^2_{a_1(1260)}}{m^2_\rho} \simeq 2.4, 
\end{equation}
which is compatible with the experimental unquenched $N=3$ value $2.5(1)$~\cite{Nakamura:2010zzi}. This value can also be compared to our result for this quantity, extrapolated to the chiral and large-$N$ limits, which is $3.6(3)$. 

An even less favorable comparison holds for the ratio of the fundamental and first excited states in the vector channel:
\begin{equation}
\label{Sakai_Sugimoto_rhostar_over_rho}
\frac{m^2_{\rho(1450)}}{m^2_\rho} \simeq 4.3. 
\end{equation}
While the experimentally observed value for this quantity is $3.57(12)$~\cite{Nakamura:2010zzi}, our $N=\infty$ result reads  $6.0(4)$. Regarding the (isovector) scalar channel, the prediction of this model for the squared ratio of the lightest state mass to the lightest vector reads:
\begin{equation}
\label{Sakai_Sugimoto_a_0_over_rho}
\frac{m^2_{a_0(1450)}}{m^2_\rho} \simeq 4.9, 
\end{equation}
to be compared with the real-world value\footnote{Strictly speaking, the lightest experimentally measured state with quantum numbers $I^G(J^{PC})=1^-(0^{++})$ is the $a_0(980)$, rather than the $a_0(1450)$. However, the correct identification of scalar mesons is a particularly challenging problem, and the identification of the $a_0(980)$ as a genuine meson is somewhat controversial: most likely, the $a_0(980)$ wave function has a large $K\overline{K}$ component~\cite{Nakamura:2010zzi}, and hence this state could be interpreted as a two-meson resonance or a tetraquark.} $3.61(9)$~\cite{Nakamura:2010zzi}. Our result, extrapolated to vanishing quark mass and for $N \to \infty$, is $8.4(6)$.

Finally, we mention that a different type of strategy for holographic studies was pioneered in refs.~\cite{hepph0501128, DaRold_Pomarol}: it goes under the name of ``AdS/QCD'', and consists of constructing a gravity dual in a curved higher-dimensional (typically: five-dimensional) space, reproducing the known features of QCD (a related approach has been discussed in refs.~\cite{IHQCD,IHQCD_2,IHQCD_3,IHQCD_4,IHQCD_5,IHQCD_6}). Contrary to the constructions mentioned above, here one follows a ``bottom-up'' approach, which would be related to a non-critical string theory setup~\cite{Polyakov:1998ju, Klebanov:2004ya}. An important caveat in such constructions, however, is that string corrections may be quantitatively non-negligible at finite values of the 't~Hooft coupling, hence in this case the gravity approximation could be unjustified. However, these effective models appear to capture salient features of QCD, including, in particular, confinement and chiral symmetry breaking, and typically yield quantitatively rather accurate predictions for certain physical quantities (or ratios thereof). The quantitative predictions of the model proposed in ref.~\cite{hepph0501128} were worked out using two different methods. One possibility is to fix the values of the free parameters of the model, by setting the masses of $\pi$ and $\rho$, and the pion decay constant to their physical values. This led to the predictions:
\begin{equation}
\label{hepph0501128_predictions_I}
m_{a_1(1260)} = 1363~\mbox{MeV}, \;\; 
\tilde F_\rho= (329~\mbox{MeV})^2 \;\;
\end{equation}
(note that, upon conversion to our notations, this corresponds to $f_\rho=198~\mbox{MeV}$). Alternatively, one can perform the best fit for all of the seven parameters simultaneously, which results in:
\begin{eqnarray}
\label{hepph0501128_predictions_II}
m_\pi = 141~\mbox{MeV}, \;\; 
m_\rho = 832~\mbox{MeV}, \;\; 
m_{a_1(1260)} =& 1220~\mbox{MeV}, \nonumber \\
F_\pi= 84~\mbox{MeV},  \;\;  
\tilde F_\rho= (353~\mbox{MeV})^2& \;\; 
\end{eqnarray}
(which, in our conventions, would correspond to $f_\rho=212~\mbox{MeV}$).

In principle, these values can also be compared to our results extrapolated to the large-$N$ limit. Note, however, that, given that the parameters of this model involve input from experimental data, the values obtained for $N \to \infty$ are not necessarily expected to be in better agreement than those for $N=3$. However, our lattice computations at different values of $N$ can provide helpful insights into the consistency of the model: since the holographic construction is a \emph{gravitational} one, i.e. it is based on the approximation of an infinite number of colors, the lattice results can reveal the quantitative impact of corrections due to the finiteness of $N$, and therefore provide a non-trivial test of the validity of the model. Our results reveal that, in most cases, the finite-$N$ corrections evaluated at $N=3$ amount for relative corrections well below $10\%$.

Another AdS/QCD model was discussed in refs.~\cite{Iatrakis:2010zf, Iatrakis:2010jb}: its effective action includes an open-string tachyon, which is responsible for chiral symmetry breaking in the dual gauge theory. The authors of these works were able to reproduce the experimentally observed masses of several low-spin mesons to precisions around $10\%$ to $15\%$.

\subsection{Chiral perturbation theory}
\label{subsec:chipt}

As is well-known, chiral perturbation theory~\cite{chipt_general,chipt_general_1,chipt_general_2,chipt_general_3,chipt_general_4,chipt_general_5} is an effective low-energy theory describing the dynamics of the lightest mesons in QCD. It relies on the parametric separation of the chiral symmetry breaking scale $\sim4\pi F_\pi$ and the (nearly) zero mass of the (pseudo-) Goldstone bosons. In the case of a QCD-like theory with $n_f$ flavors of light quarks, $\chi$PT describes the fields associated with the light mesons, in terms of the components of a field taking values in $\U(n_f)$. Its dynamics is governed by an effective Lagrangian (constrained by the symmetries of the theory), which can be organized in a systematic expansion, according to the number of derivatives and of factors involving a possible explicit mass term, in which the coefficients of the different terms are low-energy constants (LECs), whose numerical values can be fixed using phenomenological input, and compared with the expectations from large-$N$ counting rules. In particular, inspection of the terms contributing to the lowest order shows that the effective Lagrangian is proportional to the square of the pion decay constant, i.e. to $N$. Essentially, this implies that, for $N \to \infty$, the effective theory for light mesons becomes exact at tree level. However, we should point out that the construction of $\chi$PT for large-$N$ QCD (and the constraints that can be derived from it) is a topic which involves some subtleties: the interested reader can find a clear exposition of this subject in ref.~\cite{Kaiser:2000gs}. In addition, there exist studies of large-$N$ $\chi$PT for the baryonic sector~\cite{Jenkins:1995gc}, too.

While a systematic comparison of large-$N$ $\chi$PT predictions with lattice results (including, in particular, the study of the $N$-scaling of various LEC's) is a task that would go way beyond the scope of the present work, for our present purposes it is worthwhile mentioning refs.~\cite{Pelaez:2006nj,Geng:2008ag,Nieves:2009ez,Nieves:2011gb}, in which the $N$-dependence of the masses for some of the mesons presented here was investigated in $\chi$PT. These studies address the full theory with sea quarks. In this case meson masses tended to increase with the number of colors relative to $\hat{F} = \sqrt{3/N} F_{\pi}(0)$, as opposed to our quenched results. 
The present work and $\chi$PT results can be compared by studying ratios of quantities defined at $N=\infty$, where both theories are quenched. Comparing our results shown in table~\ref{tab:largeNsummary} with those in refs.~\cite{Pelaez:2006nj,Geng:2008ag,Nieves:2009ez,Nieves:2011gb} we find a discrepancy of order 30\%, however the systematics of their approach are hard to estimate, in particular because experimental data are only available for $N=3$. In the future our $N=\infty$ results could be used to constrain effective field theory predictions.

\section{Conclusions}
\label{sec:conclusions}

We have computed decay constants as well as the ground and first excited
state masses of mesons in the large-$N$ limit of QCD by lattice simulations
of the $N=2, 3, 4, 5, 6, 7$ and 17 quenched theories. In all channels but the scalar the dependence on the number of
colors for $N\geq 3$ is mild. The corrections are well parameterised by an expansion in $1/N^2$. 

We detect statistically significant quenched chiral logarithms
of the pion mass for $N\leq 4$ which can be described by a
$\delta$-parameter~\cite{Chen:2003im}: $m_{\pi}^2\propto m_q^{1/(1+\delta)}$
as $m_q\rightarrow 0$. The observed rapid decay of $\delta$ towards
large $N$-values suggests in this case the subleading $1/N^3$
contribution to dominate over the leading $1/N$ contribution at $N=3$
where we find $\delta(N=3)=0.069(14)$.

Extrapolating our quenched results to the large-$N$ limit~(which is unitary and the same as that of
QCD with sea quarks), we determine the meson spectrum in the chiral limit as well as at physical light
and strange quark masses. We find the scalar to be 
about 1.5 times heavier than the vector particle at quark masses
smaller than that of the strange quark. This is of particular relevance
to the phenomenology of scalar mesons \cite{Pelaez:2006nj,Geng:2008ag,Nieves:2009ez,Nieves:2011gb}.

We also clarified a discrepancy among previous studies: our results for the $\rho$ meson mass for $\SU(17)$ are compatible with the large-$N$ extrapolation of studies carried out for smaller values of $N$~\cite{DelDebbio:2007wk, Bali:2008an,Bali:2007kt, DeGrand:2012hd}, and are in contrast to the findings obtained in ref.~\cite{Hietanen:2009tu} at $N=17,19$ using different techniques. This suggests that the disagreement with the rest of the literature may be due to some technical aspect of ref.~\cite{Hietanen:2009tu}.

Finally, we compared our numerical results with analytical predictions, including, in particular, some of those derived from holographic models, finding  qualitative (and semi-quantitative) agreement.  Our results for the masses of various states
 (expressed in units of the $\rho$ meson mass) exhibit a systematic tendency towards values which are larger than those obtained from holographic computations.
Expressed in units of the string tension (or of the pion decay constant normalized by $\sqrt{N}$), we find meson masses to decrease
with an increasing number of colors. This may be different in QCD with sea quarks. 

\acknowledgments
This work is partially supported by the EU ITN STRONGnet (grant 238353), by the German DFG (SFB/TR 55), by the British STFC (grants ST/J000329/1 and ST/G000506/1) and by the Academy of Finland (project 1134018). L.D.D. is supported by an STFC Consolidated Grant and B.L. by a Royal Society University Research Fellowship. The simulations were performed on the Athene and iDataCool clusters in Regensburg, at LRZ Munich, at the Finnish IT Center for Science (CSC) in Espoo, and on the Swansea BlueGene/P system (part of the DiRAC Facility, jointly funded by STFC, the Large Facilities Capital Fund of BIS and Swansea University). We thank Nilmani Mathur, Rajamani Narayanan and
Carlos N{\'u}\~{n}ez for discussions and Stefan~Solbrig for assistance in the simulations.

\appendix
\section{Renormalization constants}
\label{sec:rencon}

The perturbative expansions for the non-singlet renormalization constants $Z_A$ and $Z_V$ are known to order $\lambda^2$. As discussed above, in this work we considered two definitions of the quark mass, namely:
\begin{equation}
m_q=\frac{1}{2a}\left(\frac{1}{\kappa}-\frac{1}{\kappa_c}\right)=
\frac{1}{2a\kappa}-m_c,
\end{equation}
and the definition through the axial Ward identity:
\begin{equation}
\mpcac=\frac{\left\langle0|\partial_{\mu}A_{\mu}(x)|\pi\right\rangle}{2
\left\langle 0| P(x)|\pi\right\rangle},
\end{equation}
where $A_{\mu}=\bar{q}\gamma_{\mu}\gamma_{5}q$ and $P=\bar{q}\gamma_5q$ are the non-singlet axial and pseudoscalar local currents for a quark of mass $m_q$.

Converting the lattice results to the $\overline{\mathrm{MS}}$ scheme
at a scale $\mu=\pi/a$ amounts to:
\begin{equation}
m_{\overline{\mathrm{MS}}}(\pi/a)=\frac{1}{Z_S}m_q=\frac{Z_A}{Z_P}\mpcac ,
\end{equation}
with the renormalization constants~\cite{Skouroupathis:2007jd,Skouroupathis:2008mf}:
\begin{align}
\label{eq:zs}
Z_S&=1-0.515360027(4)\left(1-\frac{1}{N^2}\right)\frac{\lambda}{4\pi}-
\left(0.6155(2)-\frac{0.7325(2)}{N^2}\right)\left(1-\frac{1}{N^2}\right)\left(\frac{\lambda}{4\pi}\right)^2,\\
\label{eq:zp}
Z_P&=1-0.580734161(4)\left(1-\frac{1}{N^2}\right)\frac{\lambda}{4\pi}-
\left(0.8420(2)-\frac{0.9284(2)}{N^2}\right)\left(1-\frac{1}{N^2}\right)\left(\frac{\lambda}{4\pi}\right)^2,\\
\label{eq:zv}
Z_V&=1-0.8203561429(3)\left(1-\frac{1}{N^2}\right)\frac{\lambda}{4\pi}-
\left(1.01790(5)-\frac{0.85455(2)}{N^2}\right)\left(1-\frac{1}{N^2}\right)\left(\frac{\lambda}{4\pi}\right)^2,\\
\label{eq:za}
Z_A&=1-0.4693595879(2)\left(1-\frac{1}{N^2}\right)\frac{\lambda}{4\pi}-
\left(0.09173(5)-\frac{0.15084(2)}{N^2}\right)\left(1-\frac{1}{N^2}\right)\left(\frac{\lambda}{4\pi}\right)^2,
\end{align}
where $\lambda=2N^2/\beta$ is the 't Hooft coupling in the lattice scheme. $Z_S$ and $Z_P$ above are defined at the lattice cut-off scale $\mu=\pi/a$, while $Z_V$ and $Z_A$ are scale-independent. The scale dependence also cancels from the ratio $Z_S/Z_P$. For the sake of completeness, we also quote the expansion of
$m_c=1/(2a\kappa_c)$~\cite{Follana:2000mn, Caracciolo:2001ki}:
\begin{eqnarray}
m_c&=&4-2.046522156925001893\left(1-\frac{1}{N^2}\right)\frac{\lambda}{4\pi} \nonumber \\
&& \qquad -\left(2.7691775(3)-\frac{2.6160500(3)}{N^2}\right)
\left(1-\frac{1}{N^2}\right)\left(\frac{\lambda}{4\pi}\right)^2.
\end{eqnarray}

As is well-known, lattice perturbation theory converges slowly, but the convergence can usually be boosted, using an improved definition of the expansion parameter, e.g., $\lambda_{\overline{\mathrm{MS}}}(\pi/a)$, taking the latter value from a lattice computation. Alternatively, one can replace the bare lattice coupling with a value
computed from a lattice observable~\cite{Makeenko:1981bb,Samuel:1984ub}.

Expanding the average plaquette
\begin{equation}
U_\square = 1 - c_1 \lambda - c_2 \lambda^2  +\dots
\end{equation}

leads to one- and two-loop improved couplings of the form
\begin{align}
\lambda_E^{(1)}&=-d_1\ln U_{\Box},\\
\lambda_E^{(2)}&=\lambda_E^{(1)}-d_1d_2 \left( \ln U_{\Box}\right)^2,
\end{align}
where~\cite{DiGiacomo:1981wt,Athenodorou:2007hi}:
\begin{align}
d_1&=\frac{1}{c_1}=8\left(1-\frac{1}{N^2}\right),\\
d_2&=\frac12+\frac{c_2}{c_1^2}= \frac{0.826844\, N^2 -1}{N^2 -1}.
\end{align}
The resulting values are displayed in table~\ref{tab:thooft}. To first and second order accuracy in perturbation theory, $\lambda^{(1)}_E$ and $\lambda^{(2)}_E$ agree with $\lambda$, respectively.

In order to obtain improved $\mathcal{O}(\lambda)$ results, we replace
$\lambda$ by $\lambda_E^{(1)}$ in eqs.~(\ref{eq:zs})--(\ref{eq:za}) above,
while for improved $\mathcal{O}(\lambda^2)$ renormalization constants
we replace $\lambda$ by $\lambda_E^{(2)}$
and $\lambda^2$ by ${\lambda_E^{(1)}}^2$,
for consistency to this order in perturbation theory.

In table~\ref{tab:Zfact}, we display the resulting perturbative renormalization constants and critical hopping parameters. Our non-perturbative results on
$\kappa_c$ can be found in table~\ref{tab:pcacParams}.
The one-loop improved $Z_A$- and $Z_V$-factors seem to be closest to the corresponding
non-perturbative results~\cite{Gimenez:1998ue,Gockeler:1998ye} (see tables~\ref{tab:Zfact},\ref{tab:ZfactNonPert}) obtained for $N=3$ at a similar coupling ($\lambda=3.0$ vs. our $\lambda=2.991$). 
In the absence of non-perturbative results for $N \neq 3$, we use these
``1-loop E"  $Z_A$- and $Z_V$-factors also for the other $\SU(N)$ groups, opting for a conservative $8\%$ estimate of their relative uncertainties.
Note that the differences at $N=3$ to the non-perturbative results are 7.5~\% and 6.5~\%, respectively.

The quantity  $Z_P/(Z_A Z_S)$, needed to relate the quark to the PCAC mass, is listed as well in table~\ref{tab:Zfact}. We find a 
huge difference between our perturbative expectations and
the non-perturbative results for $N=3$ of refs.~\cite{Gimenez:1998ue,Gockeler:1998ye}, obtained at $\lambda=3.0$ and listed in table~\ref{tab:ZfactNonPert}.
Therefore, we fix the latter ratio non-perturbatively from the fit eq.~(\ref{eq:pcacParamEq}).

\begin{table}
\begin{center}
\begin{tabular}{|c|c|c|c|c|c|}\hline
$N$&$\beta$&$U_{\Box}$&$\lambda$&$\lambda_E^{(1)}$&$\lambda_E^{(2)}$\\\hline
 2&2.4645&0.644718(3)&3.246&4.682&3.101\\
 3&6.0175&0.595595(3)&2.991&4.664&2.718\\
 4&11.0280&0.578794(2)&2.902&4.666&2.586\\
 5&17.5350&0.573069(2)&2.851&4.640&2.522\\
 6&25.4520&0.568682(2)&2.829&4.644&2.490\\
 7&34.8343&0.566504(2)&2.813&4.641&2.471\\
17&208.45&0.562729(4)&2.773&4.616&2.423\\ \hline
\end{tabular}
\end{center}
\caption{Plaquette values, and 't~Hooft couplings: one- and two-loop improved ($E$).\label{tab:thooft}}
\end{table}

\begin{table}[!h]
\begin{center}
\begin{tabular}{|c|c|c|c|c|c|}\hline
 & $N$&1-loop&2-loop&1-loop $E$&2-loop $E$ \\ \hline
\multirow{2}{*}{$Z_S$ }& 2 & 0.9 & 0.879 & 0.856 & 0.86 \\
& 3 & 0.891 & 0.864 & 0.83 & 0.836 \\
& 4 & 0.888 & 0.86 & 0.821 & 0.827 \\
& 5 & 0.888 & 0.859 & 0.817 & 0.824 \\
& 6 & 0.887 & 0.858 & 0.815 & 0.822 \\
& 7 & 0.887 & 0.857 & 0.814 & 0.821 \\
& 17 & 0.887 & 0.857 & 0.811 & 0.819 \\\hline
\multirow{2}{*}{$Z_P$ }& 2 & 0.887 & 0.857 & 0.838 & 0.829 \\
& 3 & 0.877 & 0.84 & 0.808 & 0.798 \\ 
& 4 & 0.874 & 0.835 & 0.798 & 0.787 \\ 
& 5 & 0.873 & 0.834 & 0.794 & 0.783 \\
& 6 & 0.873 & 0.833 & 0.791 & 0.78 \\ 
& 7 & 0.873 & 0.832 & 0.79 & 0.778 \\ 
& 17 & 0.872 & 0.832 & 0.787 & 0.776  \\\hline
\multirow{2}{*}{$Z_V$} & 2 & 0.841 & 0.801 & 0.771 & 0.764 \\
& 3 & 0.826 & 0.78 & 0.729 & 0.729 \\ 
& 4 & 0.822 & 0.774 & 0.714 & 0.717 \\
& 5 & 0.821 & 0.773 & 0.709 & 0.713 \\
& 6 & 0.82 & 0.771 & 0.705 & 0.71 \\ 
& 7 & 0.82 & 0.771 & 0.703 & 0.708 \\
& 17 & 0.82 & 0.77 & 0.7 & 0.706 \\\hline
 \multirow{2}{*}{$Z_A$ }&2 & 0.909 & 0.906 & 0.869 & 0.907 \\
 & 3 & 0.901 & 0.897 & 0.845 & 0.901 \\
 & 4 & 0.898 & 0.894 & 0.837 & 0.899 \\
 &  5 & 0.898 & 0.894 & 0.834 & 0.898 \\
 &  6 & 0.897 & 0.893 & 0.831 & 0.898 \\
  & 7 & 0.897 & 0.893 & 0.83 & 0.898 \\
  & 17 & 0.897 & 0.892 & 0.828 & 0.898 \\\hline 
  \multirow{2}{*}{ $\kappa_c$ } & 2 & 0.13875 & 0.14295 & 0.14585 & 0.14702 \\
  & 3 &0.14018 & 0.14526 & 0.15038 & 0.15137 \\
  & 4 & 0.14057 & 0.145912 & 0.15209 & 0.15298 \\
   & 5 & 0.14068 & 0.14609 & 0.15269 & 0.15359 \\
& 6  & 0.14076 & 0.14623 & 0.15316 & 0.15396 \\
& 7 &  0.14080 & 0.14629 & 0.15339 & 0.15417 \\
& 17 & 0.14085 & 0.14637 & 0.15380 & 0.15453 \\\hline
\multirow{2}{*}{{\Large$ \frac{ Z_P}{Z_A\cdot Z_S } $ }} & 2 &  1.085 & 1.076 & 1.126 & 1.063  \\
& 3 & 1.093 & 1.084 & 1.152 & 1.06   \\
 & 4 & 1.095 & 1.086 & 1.162 & 1.058   \\
  & 5 & 1.096 & 1.087 & 1.166 & 1.057  \\
  & 6 &  1.097 & 1.087 & 1.168 & 1.057 \\
   & 7 &  1.097 & 1.087 & 1.17 & 1.056  \\
    & 17 & 1.097 & 1.087 & 1.172 & 1.056  \\\hline
\end{tabular}
\end{center}
\caption{Perturbative $Z$-factors and $\kappa_c^{-1}$: one-loop, two-loop,
bare and improved ($E$).\label{tab:Zfact}}
\end{table}
\clearpage

 \begin{table}[!h]
\begin{center}
\begin{tabular}{|c|c|c|c|c|c|}\hline
Ref. & $Z_S$& $Z_P$ &$Z_A$& $Z_V$ & $Z_P/(Z_A Z_S)$ \\ \hline
 \cite{Gimenez:1998ue}&0.68(1)& 0.45(6)&0.81(1) &0.71(2) & 0.82(11) \\ \hline
\cite{Gockeler:1998ye} &0.7718(16)&0.4934(24)&0.7821(9)&0.6833(8)&0.8174(44)\\ \hline
\end{tabular}
\end{center}
\caption{Non-perturbative $Z$-factors from refs.~\cite{Gimenez:1998ue,Gockeler:1998ye} obtained for $N=3$ at $\lambda=3.0$ (close to our $\lambda=2.991$).\label{tab:ZfactNonPert}  }
\end{table}

\section{Additional tables and figures}
\label{sec:addtabfig}

This appendix includes additional tables and figures.

\begin{figure}[!h]
\begin{center}
\includegraphics[width=0.48\textwidth]{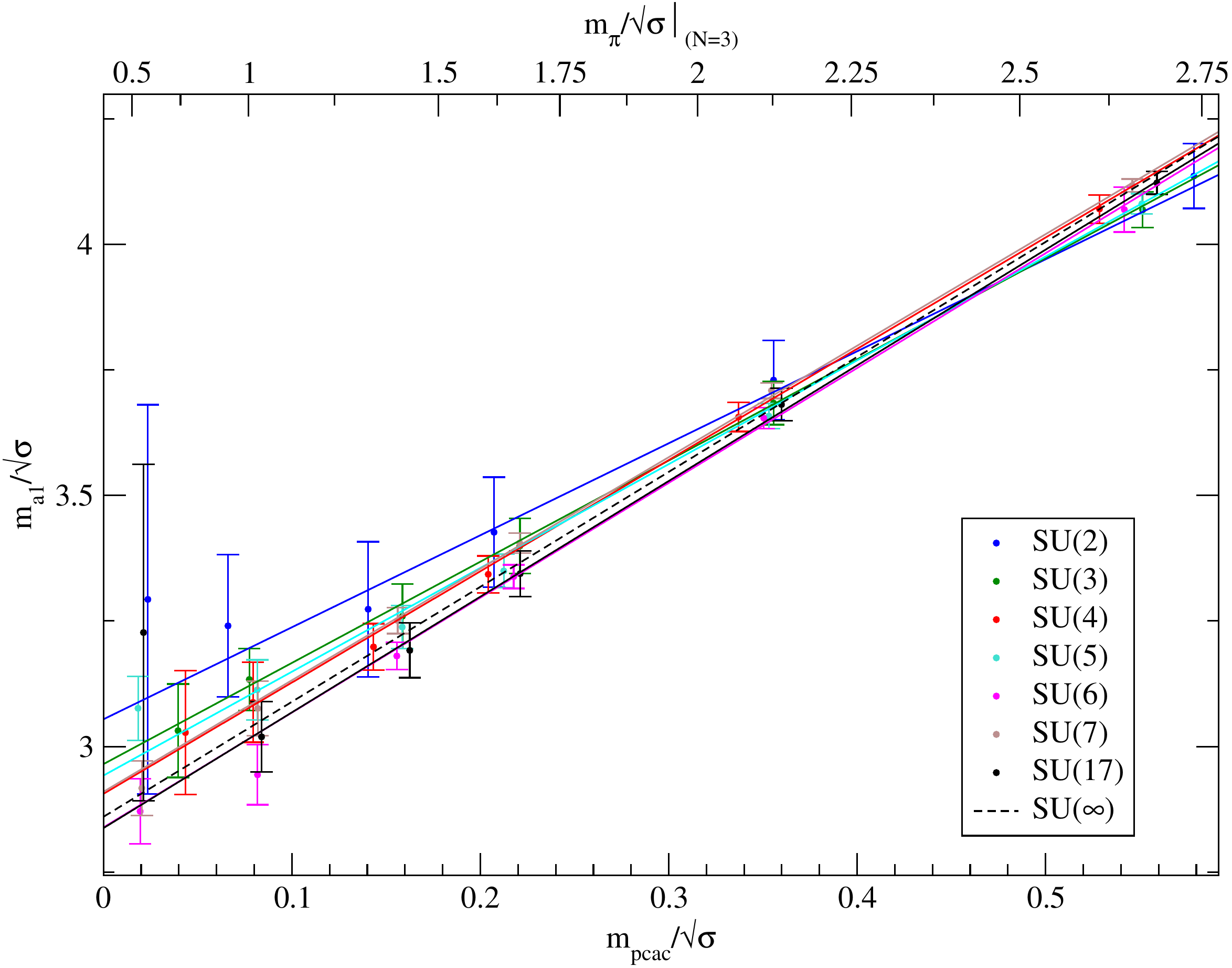} \includegraphics[width=0.48\textwidth]{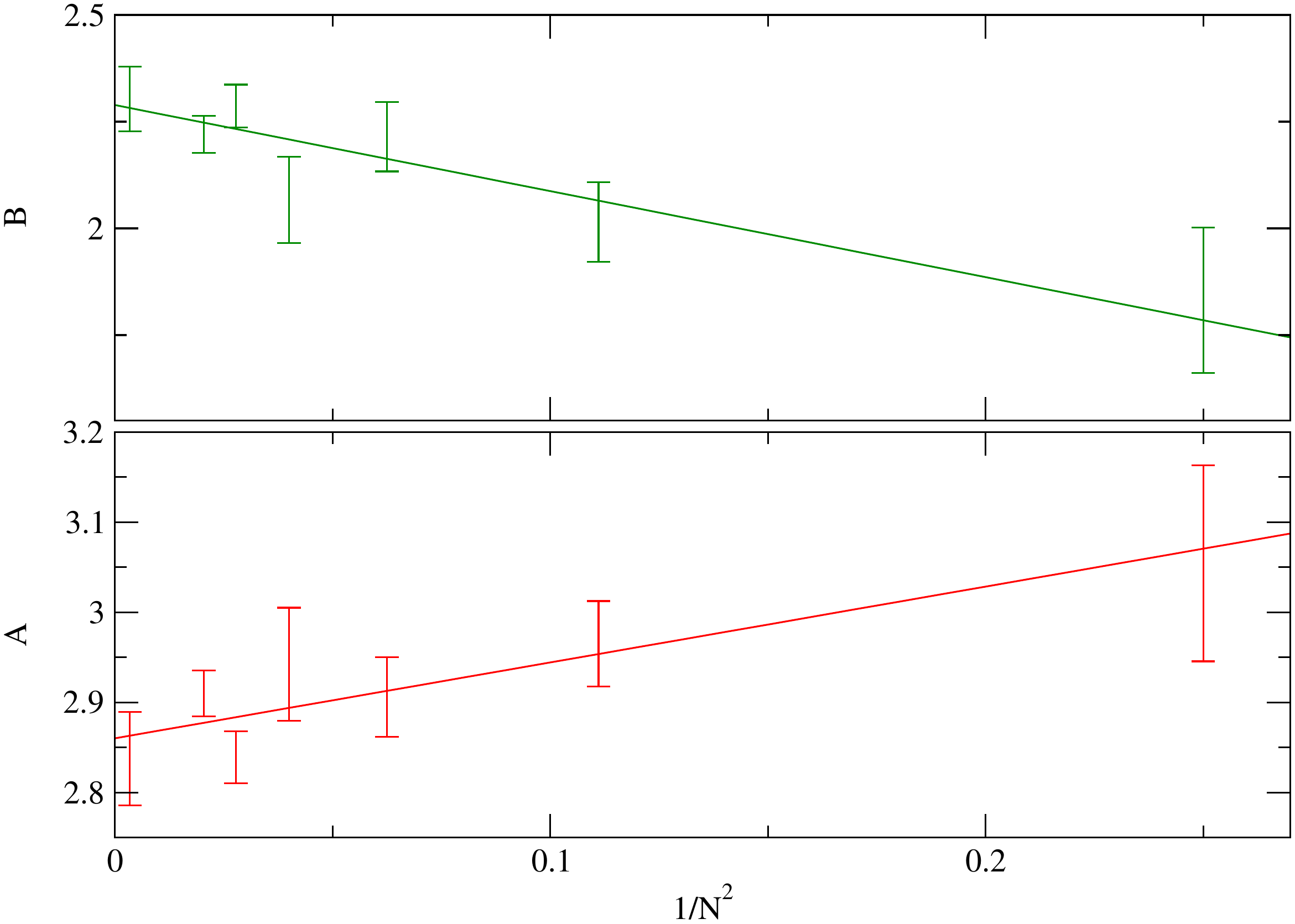}

\end{center}
\caption{Fit of the $a_1$ mass to: $m_{a_1}/\sqrt{\sigma} = A + B \cdot \mpcac /\sqrt{\sigma}$ (left) and $1/N^2$ fit of the parameters $A$ and $B$ (right).}
\label{fig:a1}
\end{figure}

\begin{figure}[!h]
\begin{center}
\includegraphics[width=0.48\textwidth]{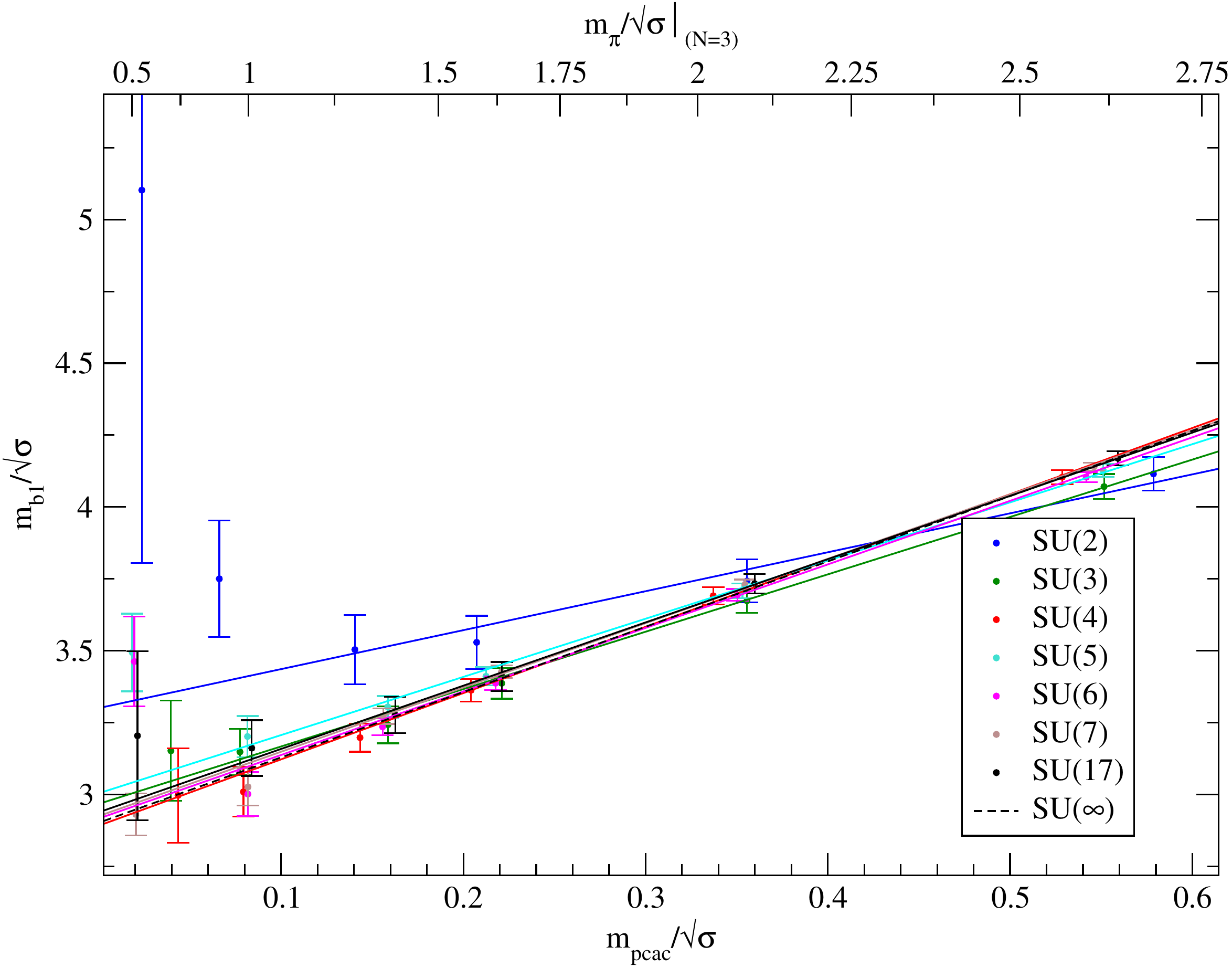} \includegraphics[width=0.48\textwidth]{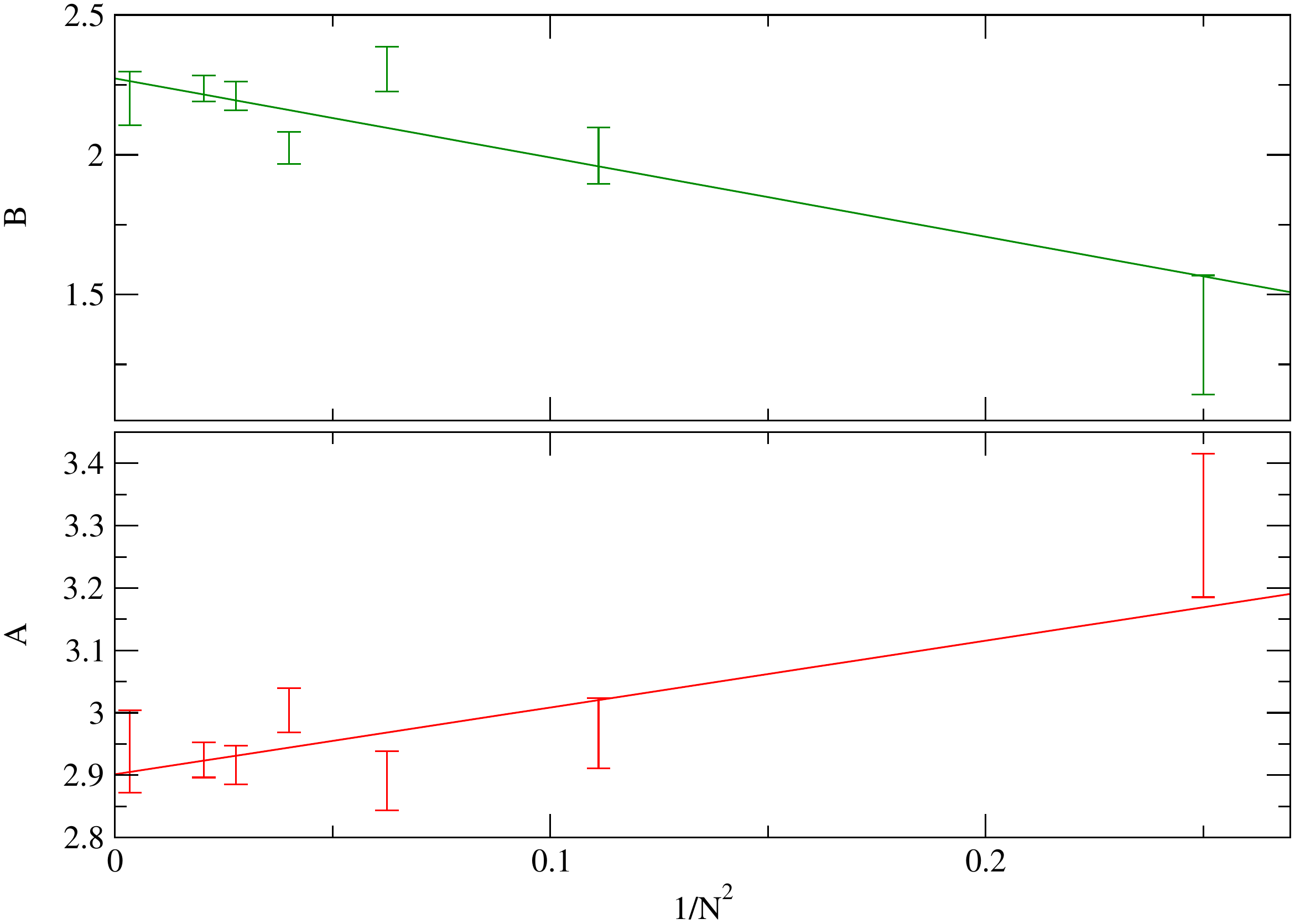}

\end{center}
\caption{Same as figure~\ref{fig:a1}, but for the mass of the $b_1$ state. }
\label{fig:b1}
\end{figure}

\begin{figure}[!h]
\begin{center}
\includegraphics[width=0.48\textwidth]{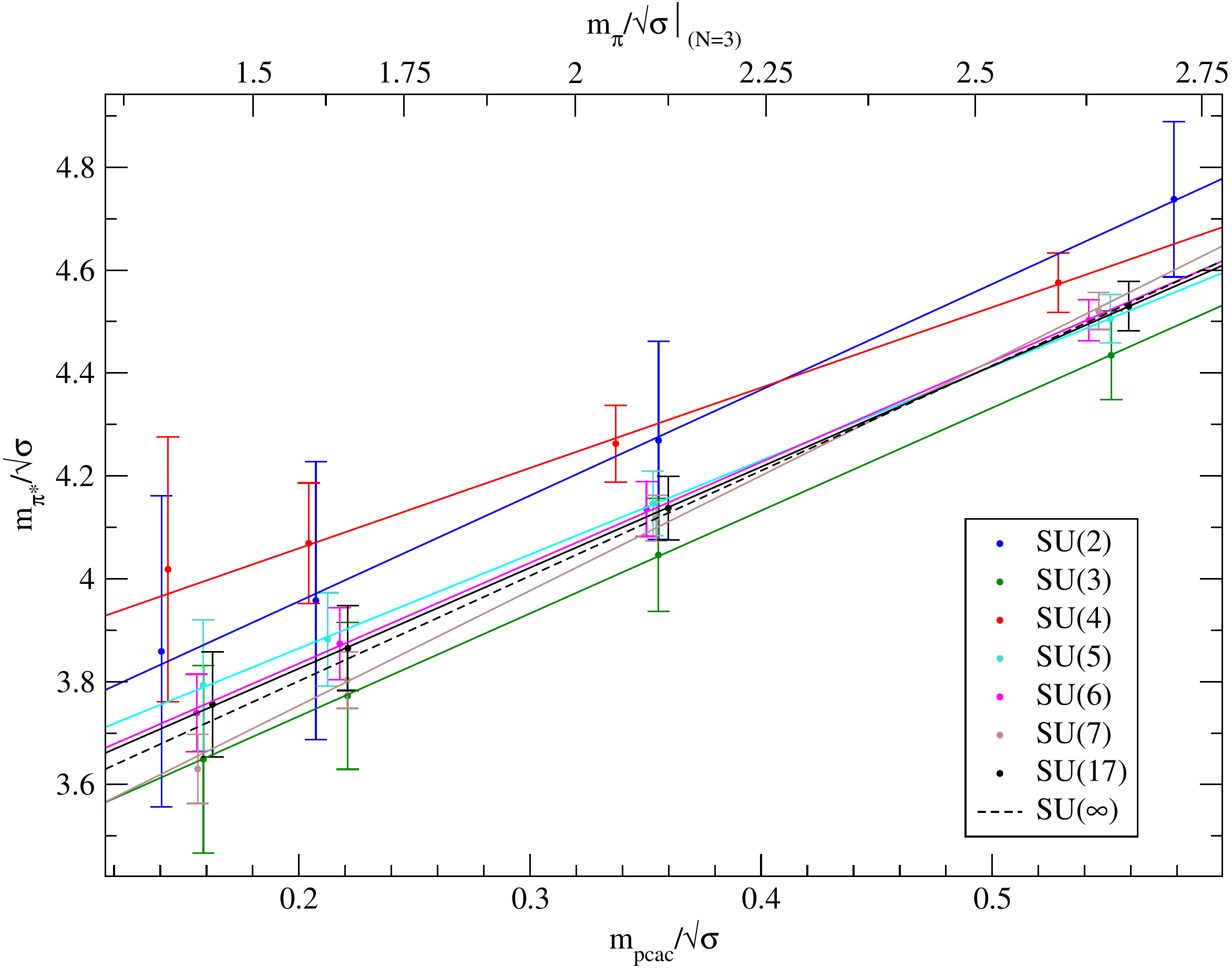} \includegraphics[width=0.48\textwidth]{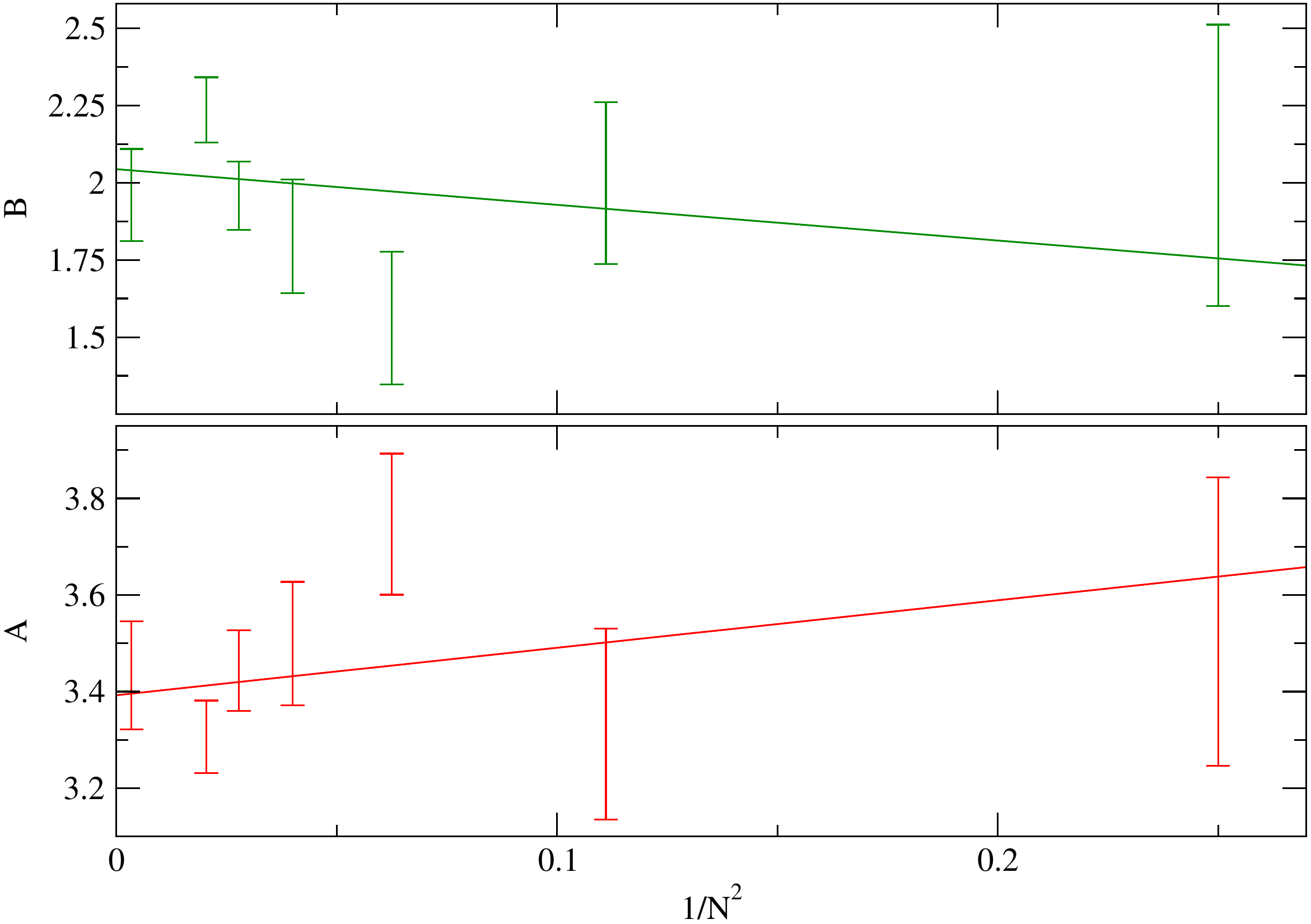}
\end{center}
\caption{Same as figure~\ref{fig:a1}, but for the mass of the $\pi^\star$ state. }
\label{fig:pistar}
\end{figure}

\begin{figure}[!h]
\begin{center}
\includegraphics[width=0.48\textwidth]{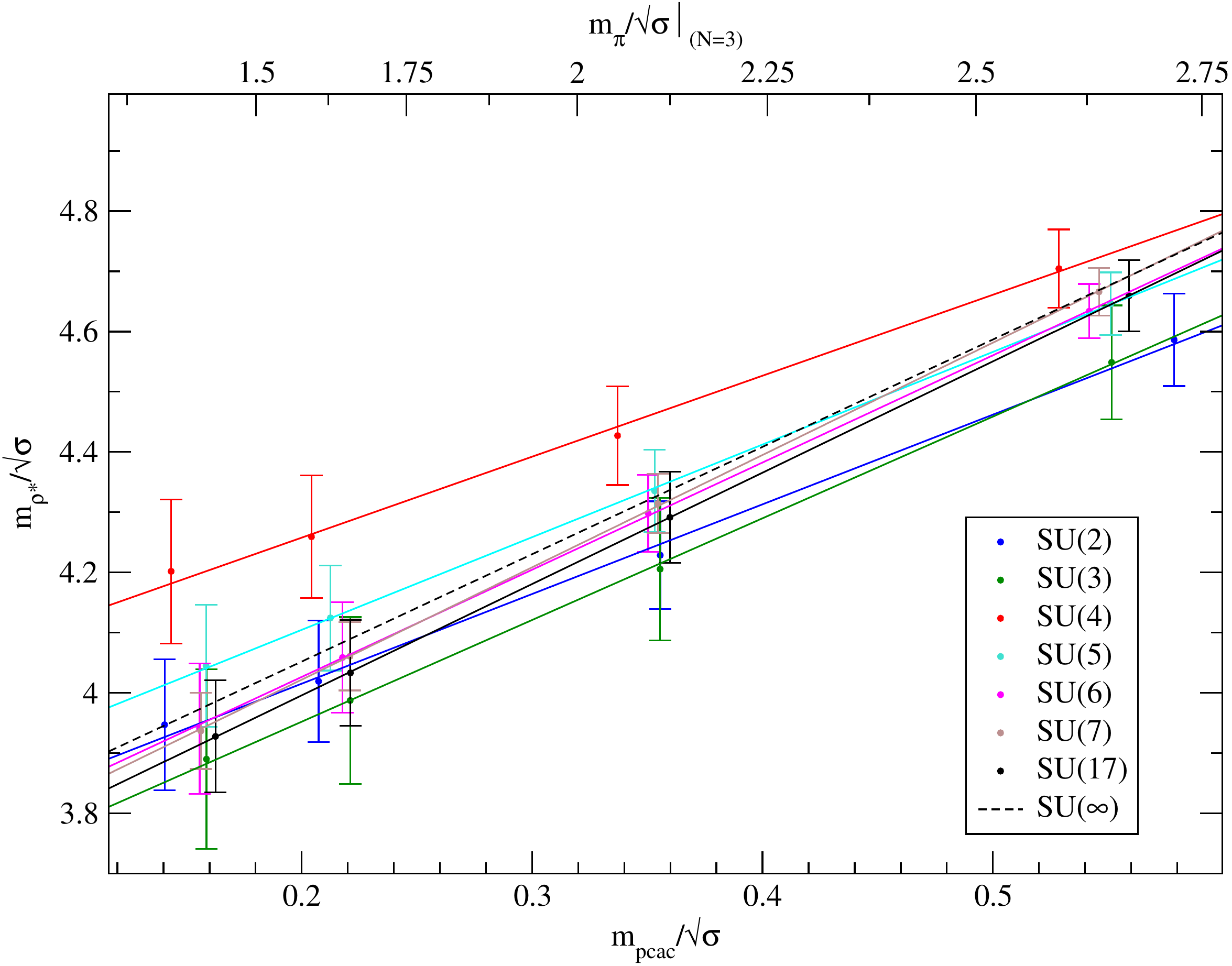} \includegraphics[width=0.48\textwidth]{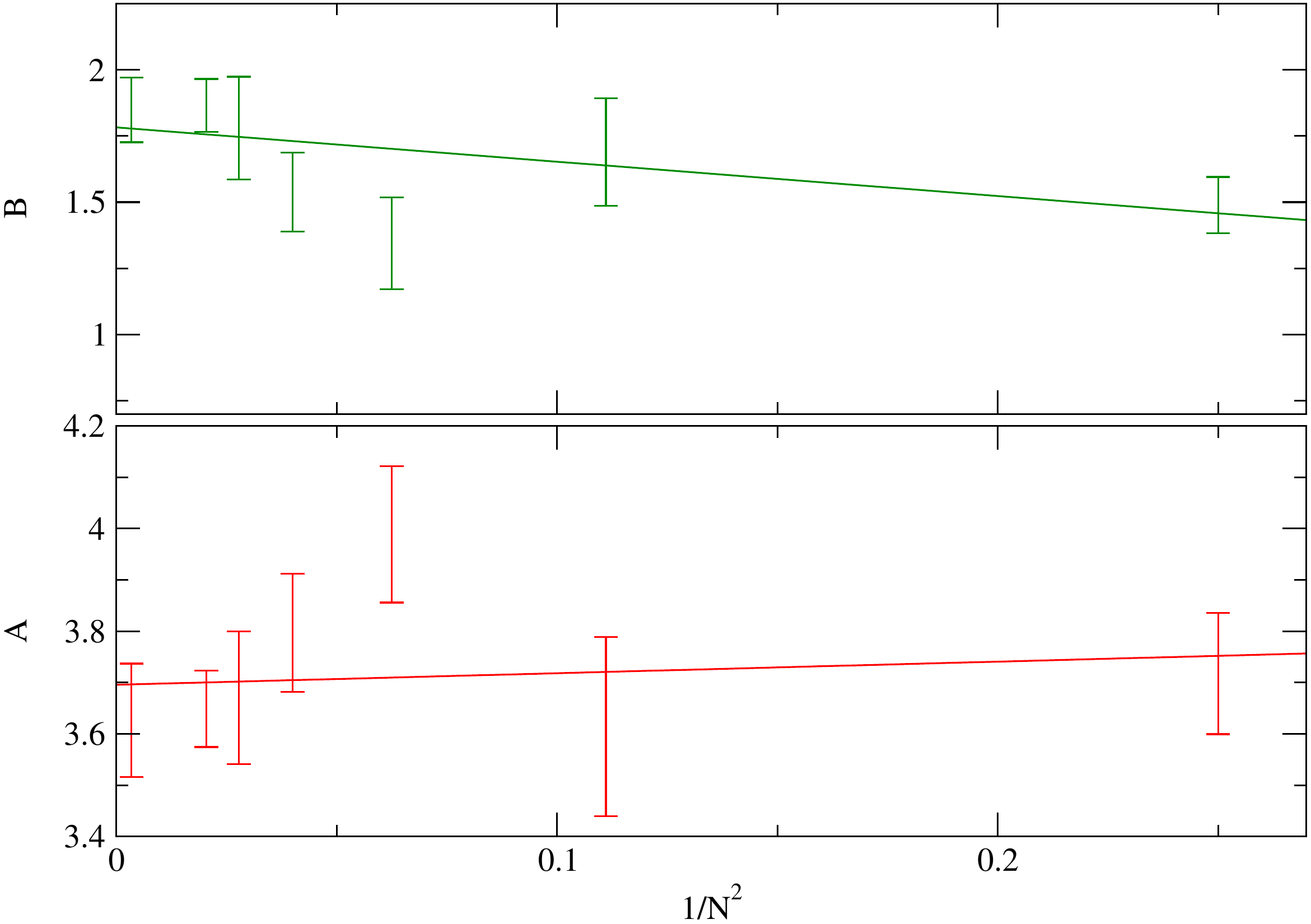}
\end{center}
\caption{Same as figure~\ref{fig:a1}, but for the mass of the $\rho^\star$ state.}
\label{fig:rhostar}
\end{figure}

\begin{figure}[!h]
\begin{center}
\includegraphics[width=0.48\textwidth]{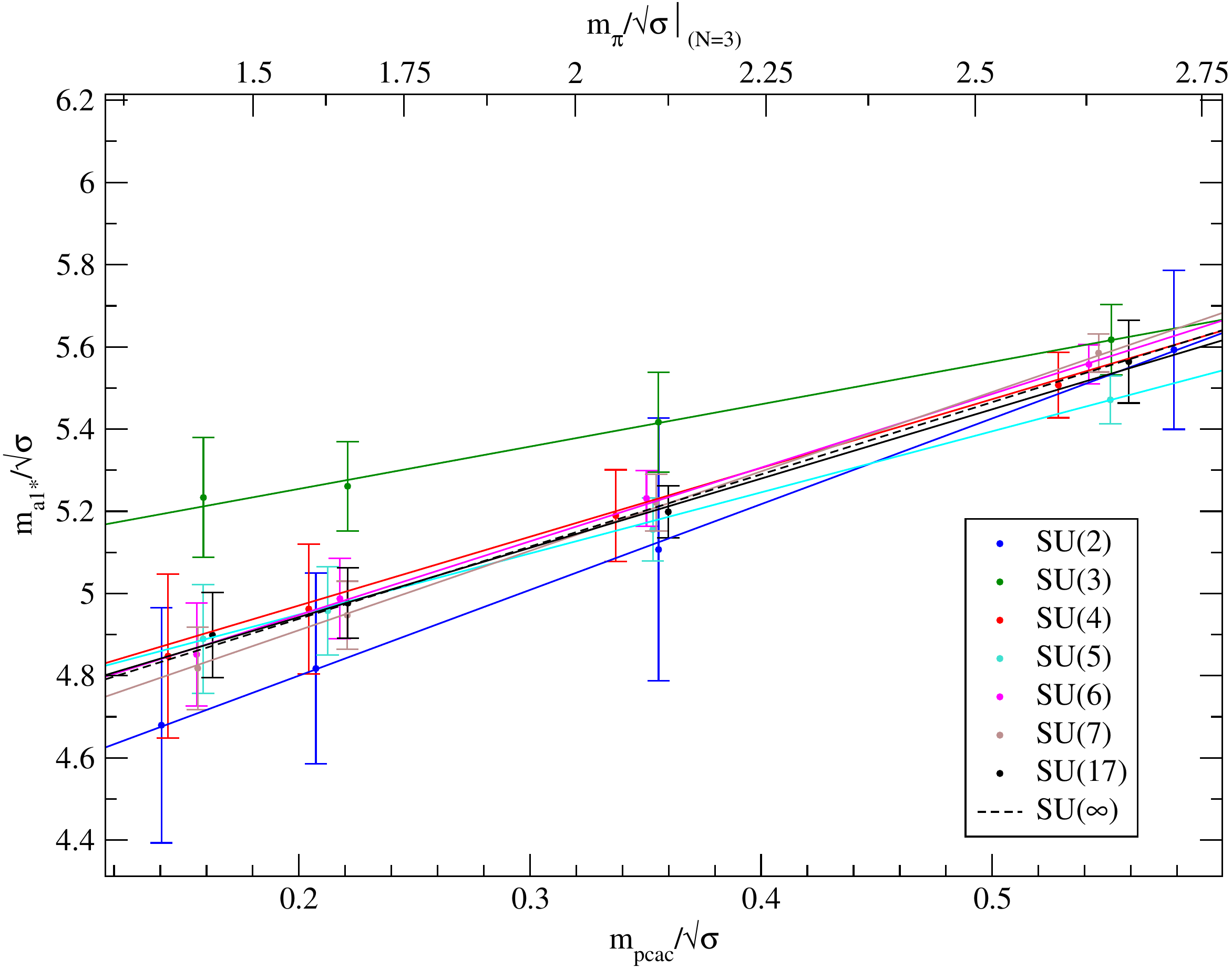} \includegraphics[width=0.48\textwidth]{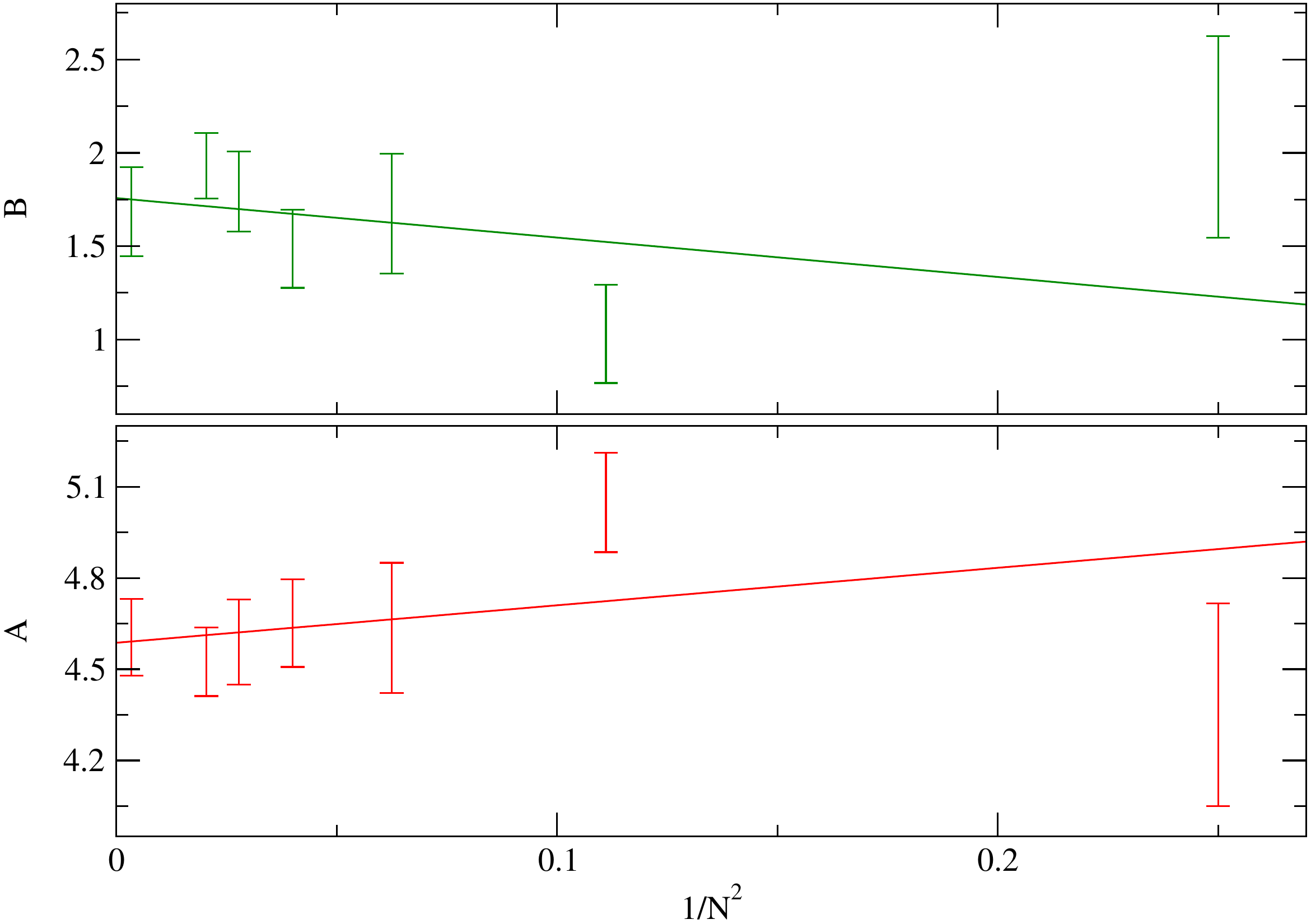}
\end{center}
\caption{Same as figure~\ref{fig:a1}, but for the mass of the $a_1^\star$ state.}
\label{fig:a1star}
\end{figure}

\begin{figure}[!h]
\begin{center}
\includegraphics[width=0.48\textwidth]{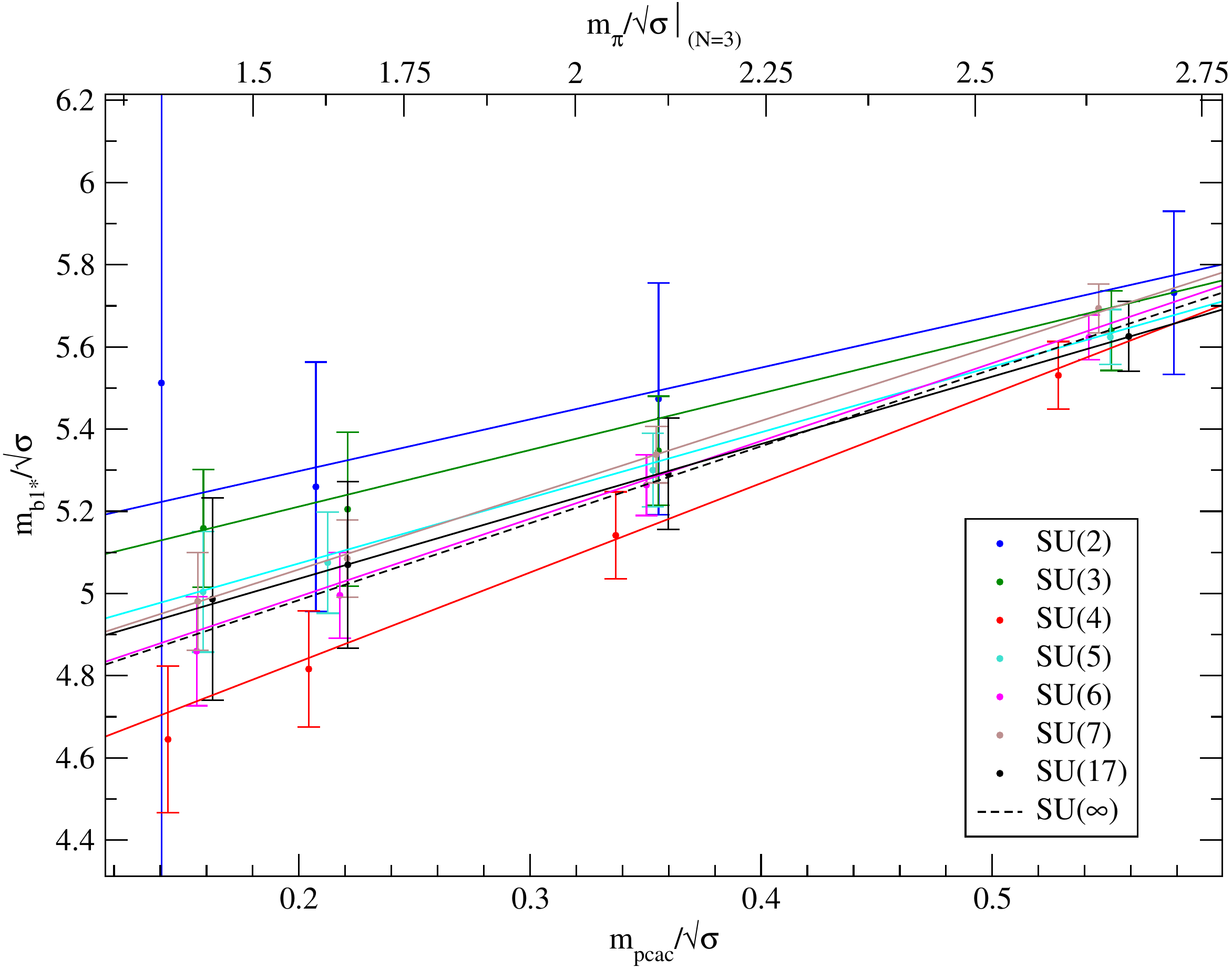} \includegraphics[width=0.48\textwidth]{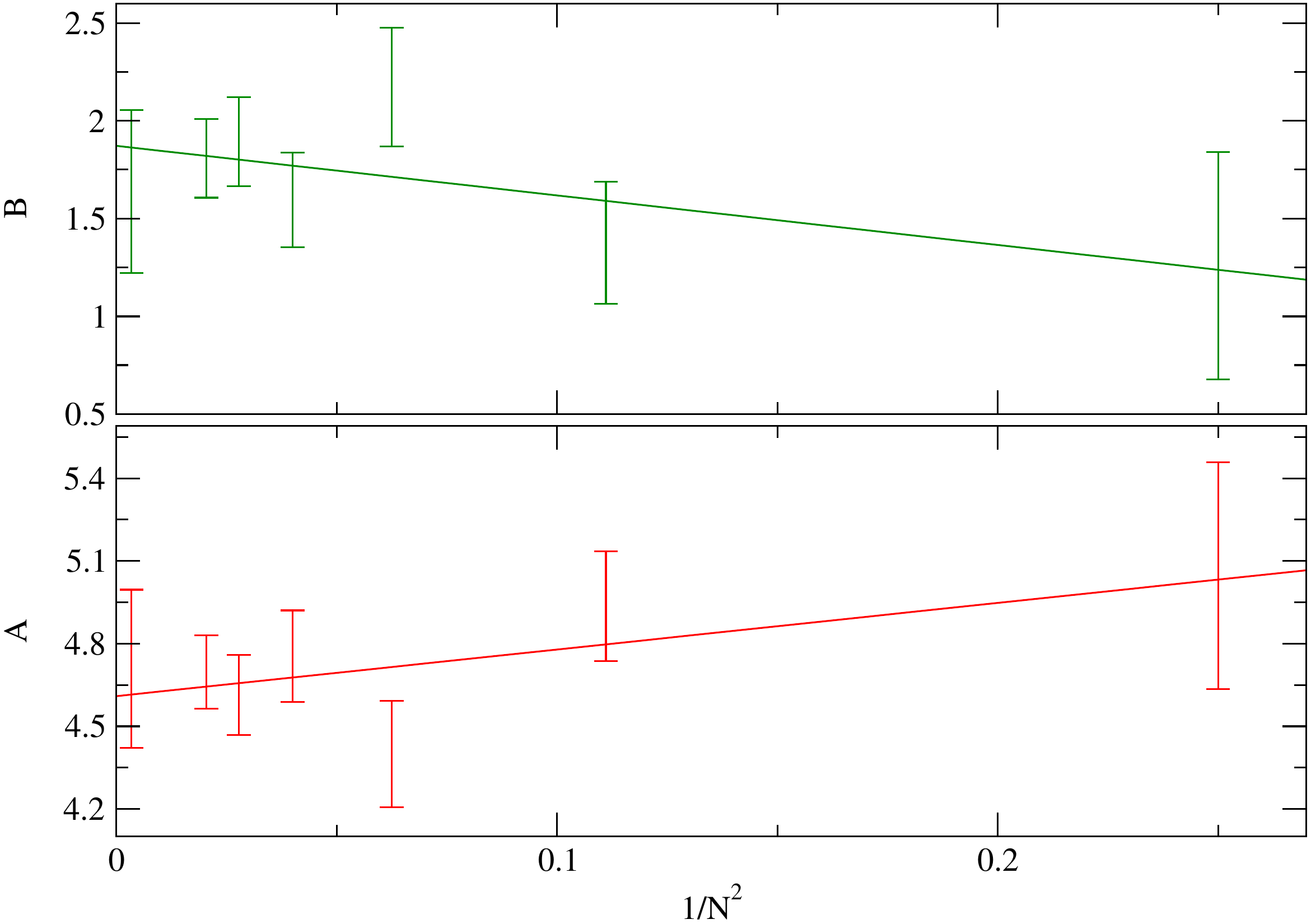}
\end{center}
\caption{Same as figure~\ref{fig:a1}, but for the mass of the $b_1^\star$ state. }
\label{fig:b1star}
\end{figure}

\begin{figure}[!h]
\begin{center}
\includegraphics[width=0.48\textwidth]{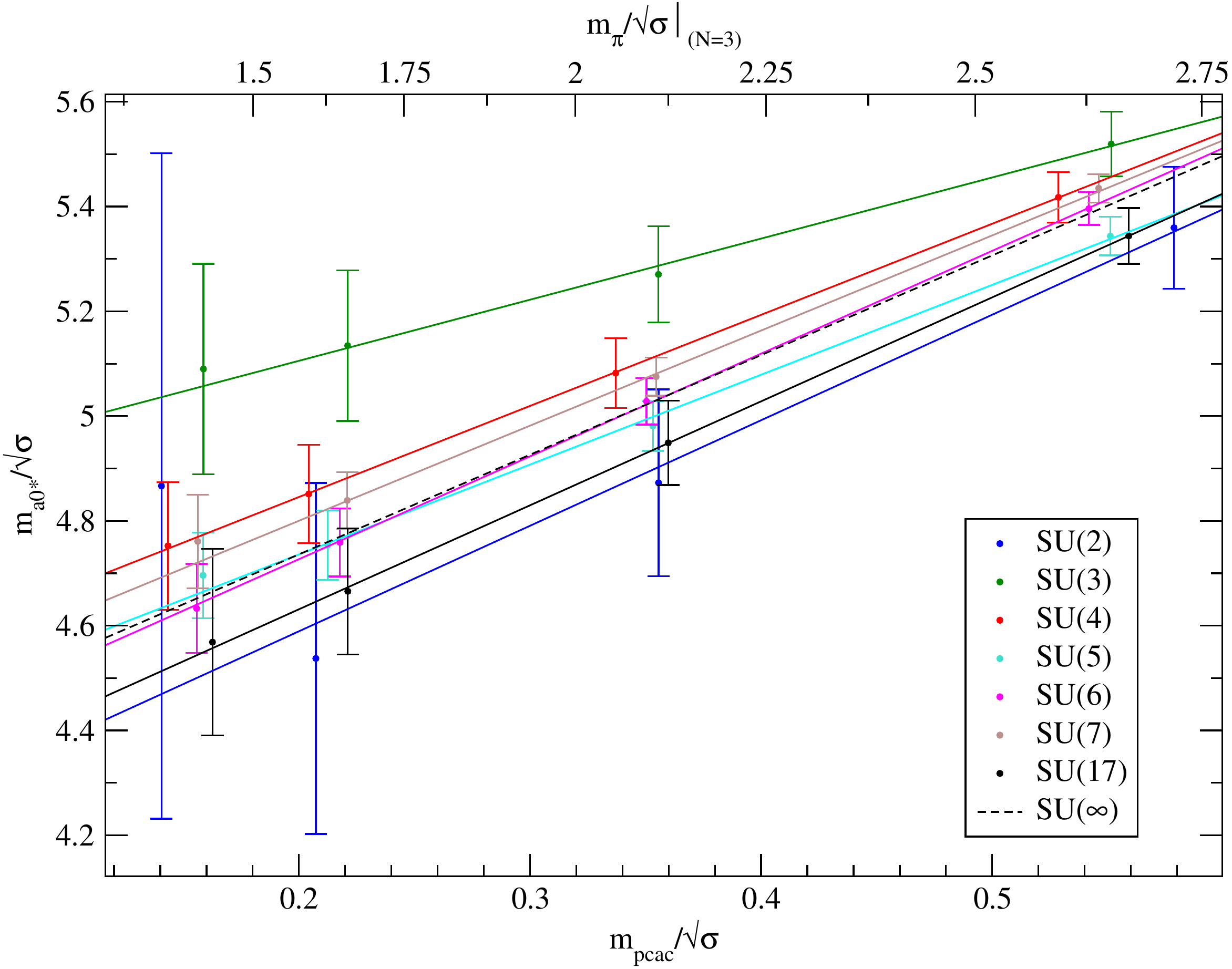} \includegraphics[width=0.48\textwidth]{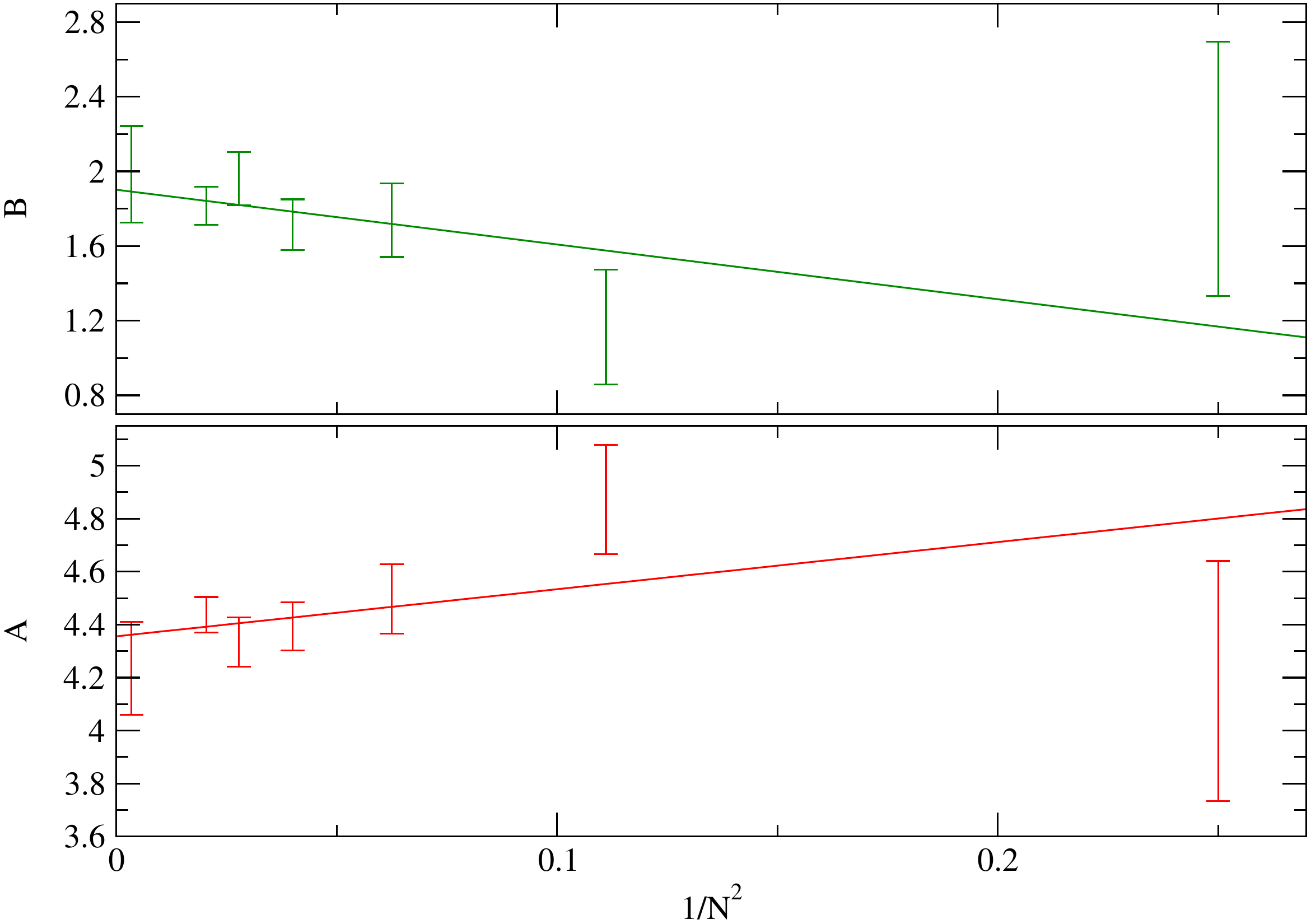}
\end{center}
\caption{Same as figure~\ref{fig:a1}, but for the mass of the $a_0^\star$ state. }
\label{fig:a0star}
\end{figure}

\clearpage

\begin{table}[!h]
\begin{center}
\begin{tabular}{|c|c|c|c|}\hline
$N$&$Z_P/(Z_A Z_S)$&$A$&$\kappa_c$ \\ \hline
2& 0.6831(75)  & 0.972(74)    & 0.152880(22)(81) \\ \hline
3& 0.7519(42)&   0.739(38)    & 0.156670(8)(86)   \\ \hline
4& 0.7812(36) &  0.607(32)    & 0.158218(7)(87)    \\ \hline
5& 0.8050(25)&   0.483(21)    & 0.158596(5)(87)   \\ \hline
6& 0.8098(22)&   0.462(18)    & 0.159103(4)(87)    \\ \hline
7& 0.8133(19)&   0.447(16)    & 0.159326(4)(87)    \\ \hline
17& 0.8286(55)&  0.370(42)   & 0.159590(12)(87)  \\ \hline
\end{tabular}
\end{center}
\caption{Fit parameters of the PCAC mass.\label{tab:pcacParams}}
\end{table}

\begin{table}[!h]
\begin{center}
\begin{tabular}{|c|c|c|c|c|}\hline
$N$&$\Delta\kappa_c/\Delta\beta$&$\Delta\beta/\Delta(a\sqrt{\sigma})$& $\delta(a\sqrt{\sigma})$&$\delta\kappa_c$ \\ \hline
2& 0.0588172 & 1.37097 &0.001&0.0000806 \\ \hline
3& 0.033977 &2.54391 &0.001&0.0000864   \\ \hline
4& 0.019335 & 4.4773&  0.001 & 0.0000866 \\ \hline
6& 0.00891258 &9.78578& 0.001& 0.0000872\\ \hline
\end{tabular}
\end{center}
\caption{Systematic errors of $\kappa_c$ and its dependencies, according to eq.~(\ref{eq:systemerr}).\label{tab:kappacsyserr}}
\end{table}

\begin{table}[!h]
\begin{center}
\begin{tabular}{|c|c|c|c|c|c|c|c|}\hline
$\kappa$&$N_s^3 \times N_t$&$\mpcac/\sqrt{\sigma}$ &$m_{\pi}/\sqrt{\sigma}$&$m_{\rho}/\sqrt{\sigma} $&$m_{a_1}/\sqrt{\sigma} $&$m_{b_1}/\sqrt{\sigma} $&$m_{a_0}/\sqrt{\sigma} $\\ \hline
0.14581 & $ 16 ^3\times 32 $ & 0.5786(11) &   2.7042(85) & 2.931(13) & 4.038(50) & 4.043(49) & 3.952(59) \\
 & $ 24 ^3\times 48 $ & 0.57866(90) & 2.6926(56) & 2.9110(82) & 4.136(64) & 4.115(58) & 4.110(52) \\ 
 & $ 32 ^3\times 64 $ & 0.5776(12)  & 2.6835(58) & 2.9122(93) & 4.133(76) & 4.125(70) & 4.085(76) \\ \hline
0.14827 & $ 16 ^3\times 32 $ & 0.3551(12) &   2.131(10) & 2.464(18) & 3.622(72) & 3.602(69) & 3.565(75) \\
 & $ 24 ^3\times 48 $ & 0.35564(84) & 2.1116(61) & 2.438(10) & 3.730(79) & 3.743(75) & 3.745(72) \\ 
 & $ 32 ^3\times 64 $ &0.3548(11)  & 2.0968(59) & 2.431(11) & 3.707(92) & 3.679(78) & 3.71(11) \\\hline
0.15008 & $ 16 ^3\times 32 $ & 0.2054(14) &   1.664(12) & 2.134(24) & 3.33(11) & 3.29(10) & 3.49(11) \\
 & $ 24 ^3\times 48 $ & 0.20732(83) & 1.6322(72) & 2.104(16) & 3.43(11) & 3.529(92) & 3.65(12) \\ 
 & $ 32 ^3\times 64 $ &0.2065(10)  & 1.6138(63) & 2.086(18) & 3.40(11) & 3.353(94) & 3.57(20) \\\hline
0.15096 & $ 16 ^3\times 32 $ & 0.1363(17) &   1.400(15) & 1.977(29) & 3.18(16) & 3.14(15) & 3.97(22) \\
 & $ 24 ^3\times 48 $ & 0.14047(88) & 1.3617(89) & 1.947(23) & 3.27(13) & 3.50(12) & 3.96(15) \\ 
 & $ 32 ^3\times 64 $ & 0.13964(99) & 1.3444(68) & 1.933(26) & 3.25(14) & 3.22(12) & 3.68(39) \\\hline
0.151959 & $ 24 ^3\times 48 $ & 0.0661(15) & 0.960(15) & 1.818(32) & 3.24(14) & 3.75(20) & 1.97(36) \\ 
 & $ 32 ^3\times 64 $ & 0.0660(11) & 0.9558(100) & 1.813(40) & 3.17(16) & 3.38(20) & 3.67(16) \\ \hline
0.152496 & $ 24 ^3\times 48 $ & 0.0236(23) & 0.639(39) & 1.812(71) & 3.29(39) & 5.1(1.3) & \\ 
 & $ 32 ^3\times 64 $ & 0.0186(18) & 0.538(30) & 1.730(60) & 3.18(15) & 3.64(26) & 2.88(32) \\ \hline
\end{tabular}
\end{center}
\caption{Ground state meson masses of $\SU(2)$ gauge theory.\label{tab:su2table1}}
\end{table}
 
 \clearpage

\begin{table}[!h]
\begin{center}
\begin{tabular}{|c|c|c|c|c|c|c|c|}\hline
$\kappa$&$N_s^3 \times N_t$&$\mpcac/\sqrt{\sigma}$ &$m_{\pi}/\sqrt{\sigma}$&$m_{\rho}/\sqrt{\sigma} $&$m_{a_1}/\sqrt{\sigma} $&$m_{b_1}/\sqrt{\sigma} $&$m_{a_0}/\sqrt{\sigma} $\\ \hline
0.15002 & $ 16 ^3\times 32 $ & 0.5514(11) &   2.6597(73) & 2.915(11) & 4.081(35) & 4.090(33) & 3.938(37) \\
 & $ 24 ^3\times 48 $ & 0.55154(55) & 2.6528(31) & 2.9066(47) & 4.068(36) & 4.071(44) & 3.930(27) \\ 
 & $ 32 ^3\times 64 $ &  0.55138(58) & 2.6513(33) & 2.9033(58) & 4.117(57) & 4.129(54) & 3.982(43) \\\hline
0.1522 & $ 16 ^3\times 32 $ & 0.3555(11) &   2.1186(86) & 2.476(13) & 3.678(42) & 3.684(42) & 3.512(50) \\
 & $ 24 ^3\times 48 $ & 0.35558(52) & 2.1109(32) & 2.4642(57) & 3.684(43) & 3.674(42) & 3.518(37) \\
 & $ 32 ^3\times 64 $ & 0.35558(51) & 2.1111(32) & 2.4631(70) & 3.739(68) & 3.769(65) & 3.577(67) \\\hline
0.1538 & $ 16 ^3\times 32 $ & 0.2209(12) &   1.669(10) & 2.156(16) & 3.390(55) & 3.415(56) & 3.266(83) \\
 & $ 24 ^3\times 48 $ & 0.22105(49) & 1.6606(36) & 2.1362(76) & 3.400(55) & 3.386(53) & 3.248(59) \\
 & $ 32 ^3\times 64 $ &0.22132(47)   & 1.6626(34) & 2.1399(92) & 3.464(86) & 3.529(83) & 3.299(91) \\\hline
0.15458 & $ 16 ^3\times 32 $ & 0.1582(13) &   1.418(12) & 2.006(20) & 3.260(68) & 3.323(73) & 3.25(12) \\
 & $ 24 ^3\times 48 $ & 0.15864(48) & 1.4099(40) & 1.9781(98) & 3.259(64) & 3.242(64) & 3.169(90) \\
 & $ 32 ^3\times 64 $ &  0.15907(44) & 1.4130(35) & 1.979(13) & 3.32(10) & 3.422(98) & 3.163(98) \\\hline
0.155638 & $ 24 ^3\times 48 $ & 0.07741(50) & 0.9993(56) & 1.788(19) & 3.134(62) & 3.148(80) & 3.51(21) \\
 & $ 32 ^3\times 64 $ & 0.07818(42) & 1.0030(39) & 1.796(24) & 3.064(87) & 3.26(10) & 3.20(26) \\ \hline
0.15613 & $ 24 ^3\times 48 $ & 0.03958(69) & 0.7306(86) & 1.710(25) & 3.032(93) & 3.15(18) & 2.51(23) \\  
 & $ 32 ^3\times 64 $ & 0.04128(43) & 0.7411(54) & 1.749(26) & 2.88(11) & 3.23(17) & 3.07(61) \\ \hline
\end{tabular}
\end{center}
\caption{$\SU(3)$ ground state meson masses.\label{tab:su3table1}}
\end{table}

\begin{table}[!h]
\begin{center}
\begin{tabular}{|c|c|c|c|c|c|c|c|}\hline
$\kappa$&$N_s^3 \times N_t$&$\mpcac/\sqrt{\sigma}$ &$m_{\pi}/\sqrt{\sigma}$&$m_{\rho}/\sqrt{\sigma} $&$m_{a_1}/\sqrt{\sigma} $&$m_{b_1}/\sqrt{\sigma} $&$m_{a_0}/\sqrt{\sigma} $\\ \hline
0.15184 & $ 16 ^3\times 32 $ & 0.52889(72) &  2.6147(71) & 2.882(11) & 3.999(41) & 4.050(44) & 3.855(39) \\
 & $ 24 ^3\times 48 $ & 0.52863(38) & 2.6168(24) & 2.8921(38) & 4.070(28) & 4.104(24) & 3.914(22) \\\hline
0.154 & $ 16 ^3\times 32 $ & 0.33716(70) &  2.0641(83) & 2.437(16) & 3.575(56) & 3.639(60) & 3.406(56) \\
 & $ 24 ^3\times 48 $ & 0.33710(36) & 2.0665(24) & 2.4513(46) & 3.656(29) & 3.691(30) & 3.482(29) \\\hline
0.15559 & $ 16 ^3\times 32 $ & 0.20413(73) &  1.5965(97) & 2.106(25) & 3.235(80) & 3.331(91) & 3.137(65) \\
 & $ 24 ^3\times 48 $ & 0.20423(35) & 1.5993(25) & 2.1227(64) & 3.343(37) & 3.363(40) & 3.184(47) \\\hline
0.15635 & $ 16 ^3\times 32 $ & 0.14303(77) &  1.337(11) & 1.954(31) & 3.04(10) & 3.19(13) & 3.03(11) \\
 & $ 24 ^3\times 48 $ & 0.14332(35) & 1.3390(26) & 1.9651(85) & 3.198(46) & 3.198(49) & 3.095(73) \\\hline
0.157173 & $ 24 ^3\times 48 $ & 0.07935(36) & 1.0001(32) & 1.802(13) & 3.088(80) & 3.009(87) & 3.26(12) \\\hline
0.15764 & $ 24 ^3\times 48 $ & 0.04350(39) & 0.7498(40) & 1.709(20) & 3.03(12) & 3.00(16) & 3.34(26)\\  \hline
\end{tabular}
\end{center}
\caption{$\SU(4)$ ground state meson masses.\label{tab:su4table1}}
\end{table}

\begin{table}[!h]
\begin{center}
\begin{tabular}{|c|c|c|c|c|c|c|c|}\hline
$\kappa$&$N_s^3 \times N_t$&$\mpcac/\sqrt{\sigma}$ &$m_{\pi}/\sqrt{\sigma}$&$m_{\rho}/\sqrt{\sigma} $&$m_{a_1}/\sqrt{\sigma} $&$m_{b_1}/\sqrt{\sigma} $&$m_{a_0}/\sqrt{\sigma} $\\ \hline
0.15205 & $ 16 ^3\times 32 $ & 0.55192(54) & 2.6736(37) & 2.9321(55) & 4.070(28) & 4.114(31) & 3.872(30) \\
 & $ 24 ^3\times 48 $ & 0.55112(33) & 2.6680(18) & 2.9331(28) & 4.080(20) & 4.127(23) & 3.904(20) \\\hline
0.15426 & $ 16 ^3\times 32 $ & 0.35403(54) & 2.1156(43) & 2.4774(72) & 3.671(36) & 3.711(43) & 3.420(47) \\
 & $ 24 ^3\times 48 $ & 0.35314(31) & 2.1065(20) & 2.4769(36) & 3.658(25) & 3.708(25) & 3.446(27) \\\hline
0.15592 & $ 16 ^3\times 32 $ & 0.21349(56) & 1.6322(54) & 2.1292(100) & 3.374(54) & 3.402(67) & 3.095(61) \\
 & $ 24 ^3\times 48 $ & 0.21247(29) & 1.6192(22) & 2.1263(54) & 3.350(33) & 3.411(32) & 3.115(46) \\\hline
0.15658 & $ 16 ^3\times 32 $ & 0.15960(58) & 1.4102(61) & 1.989(12) & 3.257(69) & 3.276(90) & 3.021(98) \\
 & $ 24 ^3\times 48 $ & 0.15852(29) & 1.3954(24) & 1.9851(69) & 3.238(43) & 3.304(37) & 3.006(66) \\\hline
0.157548 & $ 24 ^3\times 48 $ & 0.08158(29) & 1.0017(29) & 1.779(13) & 3.113(60) & 3.201(72) & 2.96(12) \\\hline
0.158355 & $ 24 ^3\times 48 $ & 0.01833(38) & 0.4869(68) & 1.655(32) & 3.076(64) & 3.49(14) & 2.80(35) \\ \hline
\end{tabular}
\end{center}
\caption{$\SU(5)$ ground state meson masses.\label{tab:su5table1}}
\end{table}

\begin{table}[!h]
\begin{center}
\begin{tabular}{|c|c|c|c|c|c|c|c|}\hline
$\kappa$&$N_s^3 \times N_t$&$\mpcac/\sqrt{\sigma}$ &$m_{\pi}/\sqrt{\sigma}$&$m_{\rho}/\sqrt{\sigma} $&$m_{a_1}/\sqrt{\sigma} $&$m_{b_1}/\sqrt{\sigma} $&$m_{a_0}/\sqrt{\sigma} $\\ \hline
0.15264 & $ 16 ^3\times 32 $ & 0.54200(48) & 2.6579(35) & 2.9299(51) & 4.064(29) & 4.091(27) & 3.910(29) \\
 & $ 24 ^3\times 48 $ & 0.54185(26) & 2.6574(16) & 2.9279(25) & 4.069(44) & 4.104(17) & 3.863(17) \\ \hline
0.15479 & $ 16 ^3\times 32 $ & 0.35056(49) & 2.1079(42) & 2.4868(67) & 3.630(40) & 3.664(38) & 3.456(42) \\
 & $ 24 ^3\times 48 $ & 0.35042(24) & 2.1086(17) & 2.4832(30) & 3.654(21) & 3.694(20) & 3.388(22) \\ \hline
0.15636 & $ 16 ^3\times 32 $ & 0.21769(51) & 1.6436(50) & 2.1549(90) & 3.355(39) & 3.392(35) & 3.127(72) \\
 & $ 24 ^3\times 48 $ & 0.21774(23) & 1.6464(18) & 2.1504(42) & 3.338(24) & 3.387(24) & 3.018(31) \\ \hline
0.15712 & $ 16 ^3\times 32 $ & 0.15545(53) & 1.3820(57) & 1.990(12) & 3.206(51) & 3.232(47) & 3.00(11) \\
 & $ 24 ^3\times 48 $ & 0.15570(22) & 1.3867(20) & 1.9864(54) & 3.180(27) & 3.235(29) & 2.834(42) \\ \hline
0.158051 & $ 24 ^3\times 48 $ & 0.08177(22) & 1.0013(25) & 1.7819(100) & 2.944(60) & 3.001(76) & 2.687(65) \\ \hline
0.158845 & $ 24 ^3\times 48 $ & 0.01948(50) & 0.4989(55) & 1.569(36) & 2.871(65) & 3.46(16) & 2.39(25) \\ \hline
\end{tabular}
\end{center}
\caption{$\SU(6)$ ground state meson masses.\label{tab:su6table1}}
\end{table}

\begin{table}[!h]
\begin{center}
\begin{tabular}{|c|c|c|c|c|c|c|c|}\hline
$\kappa$&$N_s^3 \times N_t$&$\mpcac/\sqrt{\sigma}$ &$m_{\pi}/\sqrt{\sigma}$&$m_{\rho}/\sqrt{\sigma} $&$m_{a_1}/\sqrt{\sigma} $&$m_{b_1}/\sqrt{\sigma} $&$m_{a_0}/\sqrt{\sigma} $\\ \hline
0.152816 & $ 16 ^3\times 32 $ & 0.54657(39) & 2.6767(31) & 2.9489(47) & 4.105(26) & 4.117(27) & 3.913(36) \\
 & $ 24 ^3\times 48 $ & 0.54604(22) & 2.6727(13) & 2.9461(20) & 4.117(13) & 4.140(14) & 3.913(15) \\ \hline
0.154967 & $ 16 ^3\times 32 $ & 0.35511(39) & 2.1322(34) & 2.5100(55) & 3.689(32) & 3.712(33) & 3.464(25) \\
 & $ 24 ^3\times 48 $ & 0.35455(21) & 2.1242(14) & 2.5009(25) & 3.709(16) & 3.730(17) & 3.441(19) \\ \hline
0.156547 & $ 16 ^3\times 32 $ & 0.22147(41) & 1.6689(40) & 2.1781(75) & 3.356(46) & 3.412(51) & 3.147(30) \\
 & $ 24 ^3\times 48 $ & 0.22093(20) & 1.6599(16) & 2.1631(36) & 3.405(20) & 3.427(22) & 3.070(28) \\ \hline
0.157339 & $ 16 ^3\times 32 $ & 0.15669(42) & 1.3983(45) & 2.0096(98) & 3.171(61) & 3.262(72) & 3.013(47) \\
 & $ 24 ^3\times 48 $ & 0.15619(20) & 1.3892(17) & 1.9895(48) & 3.251(26) & 3.272(27) & 2.874(38) \\ \hline
0.158273 & $ 24 ^3\times 48 $ & 0.08183(20) & 1.0007(19) & 1.7785(81) & 3.076(54) & 3.026(65) & 2.704(62) \\ \hline
0.159062 & $ 24 ^3\times 48 $ & 0.02034(24) & 0.4960(40) & 1.594(28) & 2.917(54) & 2.930(73) & 3.09(30) \\ \hline
\end{tabular}
\end{center}
\caption{$\SU(7)$ ground state meson masses.\label{tab:su7table1}}
\end{table}

\begin{table}[!h]
\begin{center}
\begin{tabular}{|c|c|c|c|c|c|c|c|}\hline
$\kappa$&$N_s^3 \times N_t$&$\mpcac/\sqrt{\sigma}$ &$m_{\pi}/\sqrt{\sigma}$&$m_{\rho}/\sqrt{\sigma} $&$m_{a_1}/\sqrt{\sigma} $&$m_{b_1}/\sqrt{\sigma} $&$m_{a_0}/\sqrt{\sigma} $\\ \hline
0.15298 & $ 12 ^3\times 24 $ & 0.55913(42) & 2.7030(56) & 2.9633(77) & 4.122(23) & 4.170(24) & 3.917(24) \\ \hline 
0.15521 & $ 12 ^3\times 24 $ & 0.35983(44) & 2.1356(69) & 2.499(10) & 3.681(33) & 3.733(34) & 3.423(37) \\ \hline
0.15684 & $ 12 ^3\times 24 $ & 0.22114(46) & 1.6545(86) & 2.148(14) & 3.344(46) & 3.410(50) & 3.016(60) \\\hline
0.15755 & $ 12 ^3\times 24 $ & 0.16257(47) & 1.4104(99) & 1.992(17) & 3.191(54) & 3.277(63) & 2.812(84) \\\hline
0.158531 & $ 12 ^3\times 24 $ & 0.08390(68) & 0.998(12) & 1.776(22) & 3.019(70) & 3.161(97) & 2.59(16) \\\hline
0.15931 & $ 12 ^3\times 24 $ & 0.0212(11) & 0.474(38) & 1.580(66) & 3.23(34) & 3.20(29) & 3.3(11) \\ \hline
\end{tabular}
\end{center}
\caption{$\SU(17)$ ground state meson masses.\label{tab:su17table1}}
\end{table}

\begin{table}[!h]
\begin{center}
\begin{tabular}{|c|c|c|c|c|c|c|}\hline
$\kappa$&$N_s^3 \times N_t$&$m_{\pi^\star}/\sqrt{\sigma}$&$m_{\rho^\star}/\sqrt{\sigma} $&$m_{a_1^\star}/\sqrt{\sigma} $&$m_{b_1^\star}/\sqrt{\sigma} $&$m_{a_0^\star}/\sqrt{\sigma} $\\ \hline
0.14581 & $ 16 ^3\times 32 $ & 4.56(11) & 4.67(13) & 5.56(12) & 5.61(11) & 5.51(13) \\
 & $ 24 ^3\times 48 $ & 4.74(15) & 4.586(77) & 5.59(19) & 5.73(20) & 5.36(12) \\
 & $ 32 ^3\times 64 $ & 4.69(13) & 4.86(14) & 5.62(12) & 5.84(14) & 5.34(14) \\ \hline
0.14827 & $ 16 ^3\times 32 $ & 4.16(15) & 4.35(16) & 5.24(18) & 5.32(14) & 5.28(21) \\
 & $ 24 ^3\times 48 $ & 4.27(19) & 4.229(90) & 5.11(32) & 5.47(28) & 4.87(18) \\
 & $ 32 ^3\times 64 $ & 4.28(16) & 4.52(17) & 5.26(15) & 5.61(18) & 4.87(18) \\ \hline
0.15008 & $ 16 ^3\times 32 $ & 3.85(20) & 4.20(19) & 4.96(27) & 5.17(19) & 4.92(36) \\
 & $ 24 ^3\times 48 $ & 3.96(27) & 4.02(10) & 4.82(23) & 5.26(30) & 4.54(34) \\
 & $ 32 ^3\times 64 $ & 3.95(19) & 4.27(18) & 5.06(21) & 5.31(22) & 4.59(25) \\ \hline
0.15096 & $ 16 ^3\times 32 $ & 3.71(27) & 4.16(21) & 4.73(31) & 5.08(24) & 5.09(67) \\
 & $ 24 ^3\times 48 $ & 3.86(30) & 3.95(11) & 4.68(29) & 5.5(1.4) & 4.87(64) \\
 & $ 32 ^3\times 64 $ & 3.75(23) & 4.14(20) & 5.00(27) & 5.50(34) & 4.51(35) \\ \hline
\end{tabular}
\end{center}
\caption{First excited state meson masses of $\SU(2)$ gauge theory.\label{tab:su2table2}}
\end{table}

\begin{table}[!h]
\begin{center}
\begin{tabular}{|c|c|c|c|c|c|c|}\hline
$\kappa$&$N_s^3 \times N_t$&$m_{\pi^\star}/\sqrt{\sigma}$&$m_{\rho^\star}/\sqrt{\sigma} $&$m_{a_1^\star}/\sqrt{\sigma} $&$m_{b_1^\star}/\sqrt{\sigma} $&$m_{a_0^\star}/\sqrt{\sigma} $\\ \hline
0.15002 & $ 16 ^3\times 32 $ & 4.599(78) & 4.73(13) & 5.666(80) & 5.821(82) & 5.535(95) \\
 & $ 24 ^3\times 48 $ & 4.435(86) & 4.549(95) & 5.617(85) & 5.639(97) & 5.519(62) \\
 & $ 32 ^3\times 64 $ & 4.417(82) & 4.504(79) & 5.39(11) & 5.51(13) & 5.134(96) \\ \hline
0.1522 & $ 16 ^3\times 32 $ & 4.239(98) & 4.39(16) & 5.43(11) & 5.66(11) & 5.26(14) \\
 & $ 24 ^3\times 48 $ & 4.05(11) & 4.21(12) & 5.42(12) & 5.35(13) & 5.270(92) \\
 & $ 32 ^3\times 64 $ & 4.07(11) & 4.195(100) & 5.11(16) & 5.21(16) & 4.70(14) \\ \hline
0.1538 & $ 16 ^3\times 32 $ & 3.99(13) & 4.18(18) & 5.33(16) & 5.70(14) & 5.12(24) \\
 & $ 24 ^3\times 48 $ & 3.77(14) & 3.99(14) & 5.26(11) & 5.20(19) & 5.13(14) \\
 & $ 32 ^3\times 64 $ & 3.81(17) & 4.01(12) & 4.96(16) & 5.02(21) & 4.34(20) \\ \hline
0.15458 & $ 16 ^3\times 32 $ & 3.88(15) & 4.11(20) & 5.35(20) & 5.82(15) & 5.22(37) \\
 & $ 24 ^3\times 48 $ & 3.65(18) & 3.89(15) & 5.23(15) & 5.16(14) & 5.09(20) \\
 & $ 32 ^3\times 64 $ & 3.68(23) & 3.95(14) & 4.86(20) & 4.96(26) & 4.44(18) \\ \hline
\end{tabular}
\end{center}
\caption{$\SU(3)$ first excited state meson masses.\label{tab:su3table2}}
\end{table}

\clearpage

\begin{table}[!h]
\begin{center}
\begin{tabular}{|c|c|c|c|c|c|c|}\hline
$\kappa$&$N_s^3 \times N_t$&$m_{\pi^\star}/\sqrt{\sigma}$&$m_{\rho^\star}/\sqrt{\sigma} $&$m_{a_1^\star}/\sqrt{\sigma} $&$m_{b_1^\star}/\sqrt{\sigma} $&$m_{a_0^\star}/\sqrt{\sigma} $\\ \hline
0.15184 & $ 16 ^3\times 32 $ & 4.84(11) & 4.67(17) & 5.65(17) & 5.72(21) & 5.51(17) \\
 & $ 24 ^3\times 48 $ & 4.576(58) & 4.705(65) & 5.507(80) & 5.531(82) & 5.418(48) \\ \hline
0.154 & $ 16 ^3\times 32 $ & 4.57(13) & 4.39(21) & 5.35(22) & 5.45(27) & 5.20(22) \\
 & $ 24 ^3\times 48 $ & 4.263(74) & 4.427(82) & 5.19(11) & 5.14(11) & 5.082(66) \\ \hline
0.15559 & $ 16 ^3\times 32 $ & 4.37(16) & 4.23(26) & 5.08(27) & 5.16(31) & 4.92(32) \\
 & $ 24 ^3\times 48 $ & 4.07(12) & 4.26(10) & 4.96(16) & 4.82(14) & 4.851(94) \\ \hline
0.15635 & $ 16 ^3\times 32 $ & 4.26(20) & 4.17(30) & 4.93(32) & 5.00(36) & 4.71(43) \\
 & $ 24 ^3\times 48 $ & 4.02(26) & 4.20(12) & 4.85(20) & 4.64(18) & 4.75(12) \\ \hline
\end{tabular}
\end{center}
\caption{$\SU(4)$ first excited state meson masses.\label{tab:su4table2}}
\end{table}

\begin{table}[!h]
\begin{center}
\begin{tabular}{|c|c|c|c|c|c|c|}\hline
$\kappa$&$N_s^3 \times N_t$&$m_{\pi^\star}/\sqrt{\sigma}$&$m_{\rho^\star}/\sqrt{\sigma} $&$m_{a_1^\star}/\sqrt{\sigma} $&$m_{b_1^\star}/\sqrt{\sigma} $&$m_{a_0^\star}/\sqrt{\sigma} $\\ \hline
0.15205 & $ 16 ^3\times 32 $ & 4.646(62) & 4.795(71) & 5.586(89) & 5.671(88) & 5.36(10) \\
 & $ 24 ^3\times 48 $ & 4.505(47) & 4.646(52) & 5.471(58) & 5.624(67) & 5.344(37) \\ \hline
0.15426 & $ 16 ^3\times 32 $ & 4.282(79) & 4.484(88) & 5.32(17) & 5.38(13) & 4.91(17) \\
 & $ 24 ^3\times 48 $ & 4.146(63) & 4.336(68) & 5.155(77) & 5.300(90) & 4.981(48) \\ \hline
0.15592 & $ 16 ^3\times 32 $ & 4.00(10) & 4.28(10) & 5.04(12) & 5.18(12) & 4.41(25) \\
 & $ 24 ^3\times 48 $ & 3.882(90) & 4.125(87) & 4.96(11) & 5.08(12) & 4.754(66) \\ \hline
0.15658 & $ 16 ^3\times 32 $ & 3.88(12) & 4.20(11) & 4.93(15) & 5.09(15) & 4.10(30) \\
 & $ 24 ^3\times 48 $ & 3.79(13) & 4.05(10) & 4.89(13) & 5.00(15) & 4.696(82) \\ \hline
\end{tabular}
\end{center}
\caption{$\SU(5)$ first excited state meson masses.\label{tab:su5table2}}
\end{table}

\begin{table}[!h]
\begin{center}
\begin{tabular}{|c|c|c|c|c|c|c|}\hline
$\kappa$&$N_s^3 \times N_t$&$m_{\pi^\star}/\sqrt{\sigma}$&$m_{\rho^\star}/\sqrt{\sigma} $&$m_{a_1^\star}/\sqrt{\sigma} $&$m_{b_1^\star}/\sqrt{\sigma} $&$m_{a_0^\star}/\sqrt{\sigma} $\\ \hline
0.15264 & $ 16 ^3\times 32 $ & 4.494(62) & 4.644(72) & 5.519(52) & 5.674(58) & 5.22(16) \\
 & $ 24 ^3\times 48 $ & 4.503(40) & 4.634(45) & 5.557(48) & 5.623(54) & 5.396(31) \\ \hline
0.15479 & $ 16 ^3\times 32 $ & 4.131(78) & 4.329(91) & 5.130(73) & 5.291(80) & 4.83(24) \\
 & $ 24 ^3\times 48 $ & 4.136(53) & 4.298(64) & 5.231(68) & 5.264(74) & 5.028(44) \\ \hline
0.15636 & $ 16 ^3\times 32 $ & 3.90(10) & 4.14(12) & 4.846(99) & 5.00(11) & 4.59(12) \\
 & $ 24 ^3\times 48 $ & 3.874(70) & 4.059(92) & 4.987(98) & 5.00(10) & 4.759(65) \\ \hline
0.15712 & $ 16 ^3\times 32 $ & 3.82(13) & 4.08(14) & 4.71(12) & 4.85(13) & 4.17(28) \\
 & $ 24 ^3\times 48 $ & 3.739(75) & 3.94(11) & 4.85(13) & 4.86(13) & 4.633(85) \\ \hline
\end{tabular} 
\end{center}
\caption{$\SU(6)$ first excited state meson masses.\label{tab:su6table2}}
\end{table}

\begin{table}[!h]
\begin{center}
\begin{tabular}{|c|c|c|c|c|c|c|}\hline
$\kappa$&$N_s^3 \times N_t$&$m_{\pi^\star}/\sqrt{\sigma}$&$m_{\rho^\star}/\sqrt{\sigma} $&$m_{a_1^\star}/\sqrt{\sigma} $&$m_{b_1^\star}/\sqrt{\sigma} $&$m_{a_0^\star}/\sqrt{\sigma} $\\ \hline
0.152816 & $ 16 ^3\times 32 $ & 4.588(38) & 4.743(40) & 5.630(43) & 5.701(52) & 5.386(50) \\
 & $ 24 ^3\times 48 $ & 4.520(36) & 4.666(39) & 5.585(46) & 5.694(59) & 5.435(27) \\ \hline
0.154967 & $ 16 ^3\times 32 $ & 4.181(49) & 4.356(53) & 5.297(57) & 5.431(57) & 5.024(63) \\ 
 & $ 24 ^3\times 48 $ & 4.118(44) & 4.315(49) & 5.221(69) & 5.338(69) & 5.075(36) \\ \hline
0.156547 & $ 16 ^3\times 32 $ & 3.928(62) & 4.143(64) & 5.075(79) & 5.190(81) & 4.753(92) \\
 & $ 24 ^3\times 48 $ & 3.803(55) & 4.061(57) & 4.947(83) & 5.085(94) & 4.839(54) \\ \hline
0.157339 & $ 16 ^3\times 32 $ & 3.814(75) & 4.054(72) & 4.972(100) & 5.07(10) & 4.63(12) \\
 & $ 24 ^3\times 48 $ & 3.630(67) & 3.937(63) & 4.82(10) & 4.98(12) & 4.761(89) \\ \hline
\end{tabular}
\end{center}
\caption{$\SU(7)$ first excited state meson masses.\label{tab:su7table2}}
\end{table}

\begin{table}[!h]
\begin{center}
\begin{tabular}{|c|c|c|c|c|c|c|}\hline
$\kappa$&$N_s^3 \times N_t$&$m_{\pi^\star}/\sqrt{\sigma}$&$m_{\rho^\star}/\sqrt{\sigma} $&$m_{a_1^\star}/\sqrt{\sigma} $&$m_{b_1^\star}/\sqrt{\sigma} $&$m_{a_0^\star}/\sqrt{\sigma} $\\ \hline
0.15298 & $ 12 ^3\times 24 $ & 4.530(48) & 4.660(59) & 5.56(10) & 5.625(85) & 5.344(53) \\\hline
0.15521 & $ 12 ^3\times 24 $ & 4.137(62) & 4.292(76) & 5.199(64) & 5.29(14) & 4.949(81) \\\hline
0.15684 & $ 12 ^3\times 24 $ & 3.865(82) & 4.033(88) & 4.976(86) & 5.07(20) & 4.67(12) \\\hline
0.15755 & $ 12 ^3\times 24 $ & 3.76(10) & 3.928(93) & 4.90(10) & 4.99(25) & 4.57(18) \\\hline
\end{tabular}
\end{center}
\caption{$\SU(17)$ first excited state meson masses.\label{tab:su17table2}}
\end{table}

\begin{table}[!h]
\begin{center}
\begin{tabular}{|c|c|c|c|c|}\hline
State $X$& $A_{X,1}$ & $A_{X,2}$& $B_{X,1}$& $B_{X,2}$ \\\hline
$a_0$ & 2.402(34) & 4.25(62) &  2.721(53) & -6.84(96) \\ 
$a_1$ & 2.860(21) &0.84(36)& 2.289(35)  &  -2.02(61) \\
$b_1 $ &2.901(23) &1.07(40)&  2.273(38)  &  -2.83(72) \\
$\pi^\star $&  3.392(57)  & 1.0(1.1)& 2.044(80)   & -1.2(1.6) \\
$\rho^\star $& 3.696(54)  &0.23(55)&1.782(67)  &  -1.30(54) \\
$a_0^\star $ & 4.356(65)  &1.8(1.4)&1.902(98)  &  -2.9(2.1) \\
$a_1^\star $ & 4.587(75)  &1.2(1.2)& 1.76(12)  & -2.1(19) \\
$b_1^\star $ &  4.609(99)  &1.7(1.5)& 1.87(15) & -2.5(2.2) \\  \hline 
\end{tabular}
\end{center}
\caption{Coefficients for the expansion of the particle masses as $m_X/\sqrt{\sigma} = \left( A_{X,1} + A_{X,2}/N^2 \right)+  \left( B_{X,1} + B_{X,2}/N^2 \right) \mpcac/\sqrt{\sigma} $. \label{tab:otherParticlesFit}}
\end{table}

\begin{table}[!h]
\begin{center}
\begin{tabular}{|c|c|c|c|c|c|}\hline
$\kappa$&$N_s^3 \times N_t$&$\hat{F}_{\pi}^{\mathrm{lat}}/\sqrt{\sigma}$&$ \hat{F}_{\pi}/\sqrt{\sigma}$&$\hat{f}_{\rho}^{\mathrm{lat}}/\sqrt{ \sigma} $&$\hat{f}_{\rho}/\sqrt{\sigma} $\\ \hline
0.14581 & $ 24 ^3\times 48 $ & 0.4033(62) & 0.350(35) & 0.900(14) & 0.693(70) \\
0.14827 & $ 24 ^3\times 48 $ & 0.3598(65) & 0.313(32) & 0.883(15) & 0.681(69) \\
0.15008 & $ 24 ^3\times 48 $ & 0.3159(79) & 0.274(28) & 0.861(17) & 0.664(68) \\
0.15096 & $ 24 ^3\times 48 $ & 0.291(10) & 0.253(27) & 0.844(21) & 0.650(67) \\
0.151959 & $ 24 ^3\times 48 $ & 0.275(20) & 0.239(30) & 0.845(36) & 0.651(71) \\
0.152496 & $ 24 ^3\times 48 $ & 0.24(16) & 0.21(14) & 0.74(11) & 0.57(10) \\ \hline
\end{tabular}
\end{center}
\caption{Decay constants for the $\SU(2)$ theory.\label{tab:su2table3}}
\end{table}

\begin{table}[!h]
\begin{center}
\begin{tabular}{|c|c|c|c|c|c|}\hline
$\kappa$&$N_s^3 \times N_t$&$\hat{F}_{\pi}^{\mathrm{lat}}/\sqrt{\sigma}$&$ \hat{F}_{\pi}/\sqrt{\sigma}$&$\hat{f}_{\rho}^{\mathrm{lat}}/\sqrt{ \sigma} $&$\hat{f}_{\rho}/\sqrt{\sigma} $\\ \hline
0.15002 & $ 24 ^3\times 48 $ & 0.4132(58) & 0.349(35) & 0.9402(83) & 0.686(69) \\
0.1522 & $ 24 ^3\times 48 $ & 0.3727(51) & 0.315(32) & 0.9143(85) & 0.667(67) \\
0.1538 & $ 24 ^3\times 48 $ & 0.3325(64) & 0.281(29) & 0.8835(91) & 0.644(65) \\
0.15458 & $ 24 ^3\times 48 $ & 0.3083(78) & 0.261(27) & 0.865(10) & 0.631(64) \\
0.155638 & $ 24 ^3\times 48 $ & 0.2742(78) & 0.232(24) & 0.849(18) & 0.619(63) \\
0.15613 & $ 24 ^3\times 48 $ & 0.246(14) & 0.208(24) & 0.893(55) & 0.652(77) \\ \hline
\end{tabular}
\end{center} 
\caption{Decay constants for the $\SU(3)$ theory.\label{tab:su3table3}}
\end{table}

\begin{table}[!h]
\begin{center}
\begin{tabular}{|c|c|c|c|c|c|}\hline
$\kappa$&$N_s^3 \times N_t$&$\hat{F}_{\pi}^{\mathrm{lat}}/\sqrt{\sigma}$&$ \hat{F}_{\pi}/\sqrt{\sigma}$&$\hat{f}_{\rho}^{\mathrm{lat}}/\sqrt{ \sigma} $&$\hat{f}_{\rho}/\sqrt{\sigma} $\\ \hline
0.15184 & $ 24 ^3\times 48 $ & 0.4305(37) & 0.360(36) & 0.9807(64) & 0.701(70) \\
0.154 & $ 24 ^3\times 48 $ & 0.3857(39) & 0.323(32) & 0.9526(67) & 0.681(68) \\
0.15559 & $ 24 ^3\times 48 $ & 0.3417(46) & 0.286(29) & 0.9206(77) & 0.658(66) \\
0.15635 & $ 24 ^3\times 48 $ & 0.3227(34) & 0.270(27) & 0.8998(92) & 0.643(65) \\
0.157173 & $ 24 ^3\times 48 $ & 0.2934(49) & 0.245(25) & 0.870(14) & 0.622(63) \\
0.15764 & $ 24 ^3\times 48 $ & 0.2710(71) & 0.227(23) & 0.855(25) & 0.611(64) \\ \hline
\end{tabular}
\end{center}
\caption{Decay constants for the $\SU(4)$ theory.\label{tab:su4table3}}
\end{table}

\begin{table}[!h]
\begin{center}
\begin{tabular}{|c|c|c|c|c|c|}\hline
$\kappa$&$N_s^3 \times N_t$&$\hat{F}_{\pi}^{\mathrm{lat}}/\sqrt{\sigma}$&$ \hat{F}_{\pi}/\sqrt{\sigma}$&$\hat{f}_{\rho}^{\mathrm{lat}}/\sqrt{ \sigma} $&$\hat{f}_{\rho}/\sqrt{\sigma} $\\ \hline
0.15205 & $ 24 ^3\times 48 $ & 0.4350(31) & 0.363(36) & 0.9702(50) & 0.688(69) \\
0.15426 & $ 24 ^3\times 48 $ & 0.3909(32) & 0.326(33) & 0.9372(52) & 0.665(67) \\
0.15592 & $ 24 ^3\times 48 $ & 0.3484(37) & 0.290(29) & 0.8984(57) & 0.637(64) \\
0.15658 & $ 24 ^3\times 48 $ & 0.3291(42) & 0.274(28) & 0.8788(65) & 0.623(62) \\
0.157548 & $ 24 ^3\times 48 $ & 0.2951(39) & 0.246(25) & 0.845(10) & 0.600(60) \\
0.158355 & $ 24 ^3\times 48 $ & 0.269(14) & 0.224(25) & 0.815(55) & 0.578(70) \\ \hline
\end{tabular}
\end{center}
\caption{Decay constants for the $\SU(5)$ theory.\label{tab:su5table3}}
\end{table}

\begin{table}[h!]
\begin{center}
\begin{tabular}{|c|c|c|c|c|c|}\hline
$\kappa$&$N_s^3 \times N_t$&$\hat{F}_{\pi}^{\mathrm{lat}}/\sqrt{\sigma}$&$ \hat{F}_{\pi}/\sqrt{\sigma}$&$\hat{f}_{\rho}^{\mathrm{lat}}/\sqrt{ \sigma} $&$\hat{f}_{\rho}/\sqrt{\sigma} $\\ \hline
0.15264 & $ 24 ^3\times 48 $ & 0.4374(25) & 0.364(36) & 0.9903(36) & 0.698(70) \\
0.15479 & $ 24 ^3\times 48 $ & 0.3944(26) & 0.328(33) & 0.9600(35) & 0.677(68) \\
0.15636 & $ 24 ^3\times 48 $ & 0.3536(30) & 0.294(30) & 0.9248(37) & 0.652(65) \\
0.15712 & $ 24 ^3\times 48 $ & 0.3302(35) & 0.275(28) & 0.9031(42) & 0.637(64) \\
0.158051 & $ 24 ^3\times 48 $ & 0.2987(34) & 0.248(25) & 0.8458(95) & 0.597(60) \\
0.158845 & $ 24 ^3\times 48 $ & 0.2676(91) & 0.222(24) & 0.776(36) & 0.548(60) \\ \hline
\end{tabular}
\end{center}
\caption{Decay constants for the $\SU(6)$ theory.\label{tab:su6table3}}
\end{table}

\begin{table}[!h]
\begin{center}
\begin{tabular}{|c|c|c|c|c|c|}\hline
$\kappa$&$N_s^3 \times N_t$&$\hat{F}_{\pi}^{\mathrm{lat}}/\sqrt{\sigma}$&$ \hat{F}_{\pi}/\sqrt{\sigma}$&$\hat{f}_{\rho}^{\mathrm{lat}}/\sqrt{ \sigma} $&$\hat{f}_{\rho}/\sqrt{\sigma} $\\ \hline
0.152816 & $ 24 ^3\times 48 $ & 0.4386(23) & 0.364(36) & 0.9918(43) & 0.697(70) \\
0.154967 & $ 24 ^3\times 48 $ & 0.3949(24) & 0.328(33) & 0.9592(44) & 0.675(68) \\
0.156547 & $ 24 ^3\times 48 $ & 0.3529(27) & 0.293(29) & 0.9214(48) & 0.648(65) \\
0.157339 & $ 24 ^3\times 48 $ & 0.3273(31) & 0.272(27) & 0.8961(55) & 0.630(63) \\
0.158273 & $ 24 ^3\times 48 $ & 0.2906(47) & 0.241(24) & 0.8602(84) & 0.605(61) \\
0.159062 & $ 24 ^3\times 48 $ & 0.249(16) & 0.206(25) & 0.814(28) & 0.573(60) \\ \hline
\end{tabular}
\end{center}
\caption{Decay constants for the $\SU(7)$ gauge theory.\label{tab:su7table3}}
\end{table}

\begin{table}[!h]
\begin{center}
\begin{tabular}{|c|c|c|c|c|c|}\hline
$\kappa$&$N_s^3 \times N_t$&$\hat{F}_{\pi}^{\mathrm{lat}}/\sqrt{\sigma}$&$ \hat{F}_{\pi}/\sqrt{\sigma}$&$\hat{f}_{\rho}^{\mathrm{lat}}/\sqrt{ \sigma} $&$\hat{f}_{\rho}/\sqrt{\sigma} $\\ \hline
0.15298 & $ 12 ^3\times 24 $ & 0.4456(31) & 0.369(37) & 1.0055(92) & 0.704(71) \\
0.15521 & $ 12 ^3\times 24 $ & 0.4045(35) & 0.335(34) & 0.986(10) & 0.690(69) \\
0.15684 & $ 12 ^3\times 24 $ & 0.3610(44) & 0.299(30) & 0.958(11) & 0.670(68) \\
0.15755 & $ 12 ^3\times 24 $ & 0.3379(53) & 0.280(28) & 0.939(13) & 0.657(66) \\
0.158531 & $ 12 ^3\times 24 $ & 0.307(12) & 0.254(27) & 0.902(17) & 0.631(64) \\
0.15931 & $ 12 ^3\times 24 $ & 0.296(28) & 0.245(34) & 0.851(48) & 0.595(68) \\ \hline
\end{tabular}
\end{center}
\caption{Decay constants for the $\SU(17)$ theory.\label{tab:su17table3}}
\end{table}


\end{document}